\documentclass[12pt,a4paper]{article}

\usepackage{epsf}
\usepackage{amsmath}
\usepackage{bbm}
\usepackage{rotate}
\usepackage{rotating}
\usepackage[hypertex]{hyperref}
\renewcommand{\baselinestretch}{1.5}
\makeatletter \@addtoreset{equation}{section} \makeatother

\newcommand{\be}{\begin{equation}}
\newcommand{\ee}{\end{equation}}
\newcommand{\bea}{\begin{eqnarray}}
\newcommand{\eea}{\end{eqnarray}}
\newcommand{\cO}{\mathcal{O}}
\newcommand{\cN}{\mathcal{N}}
\newcommand{\cL}{\mathcal{L}}

\newcommand{\cV}{{\mathbbm V}}
\newcommand{\cD}{\mathcal{D}}
\newcommand{\cW}{\mathcal{W}}
\newcommand{\cM}{\mathcal{M}}

\newcommand{\unit}{{\mathbbm{1}}}
\newcommand{\imag}{i}
\newcommand{\rmd}{\mbox{\rm{d}}}
\newcommand{\e}{\mbox{\rm{e}}}
\newcommand{\ub}{\bar{u}}
\newcommand{\vb}{\bar{v}}
\newcommand{\jb}{\bar{\jmath} }
\newcommand{\ib}{\bar{\imath} }

\newcommand{\lb}{\bar{l}}

\newcommand{\qd}{\dot{q}}
\newcommand{\pd}{\dot{\phi}}
\newcommand{\ad}{\dot{\alpha}}
\newcommand{\bd}{\dot{\beta}}
\newcommand{\nd}{\dot{\nu}}

\newcommand{\Ud}{\dot{U}}

\newcommand{\mscr}[1]{\mbox{\scriptsize #1}}
\newcommand{\ft}[2]{{\textstyle\frac{#1}{#2}}}

\begin{document}

\begin{titlepage}
\begin{center}

\hfill hep-th/0310174\\
\hfill FSU-TPI-09/03

\vskip 1cm {\large \bf  M-theory cosmologies from singular Calabi-Yau compactifications
}\footnote{Work supported by the `Schwerpunktprogramm Stringtheorie' of the DFG.}

\vskip .5in

{\bf Laur J\"arv, Thomas Mohaupt and Frank Saueressig }  \\

{\em Institute of Theoretical Physics,
Friedrich-Schiller-University 
Jena, \\ 
 Max-Wien-Platz 1, D-07743 Jena, Germany}\\
{\tt L.Jaerv, T.Mohaupt, F.Saueressig@tpi.uni-jena.de}
%\vskip 0.2cm

\end{center}
%\vskip 1.2cm

\begin{center} {\bf ABSTRACT } \end{center}
\vspace{-2mm}

\noindent 
We study five-dimensional Kasner cosmologies in a time-dependent Calabi-Yau compactification
of M-theory undergoing a flop transition. The dynamics of the additional states, which become massless at the transition point and give rise to a scalar potential, are taken into account using a recently constructed gauged supergravity action. Due to the dynamics of these states the moduli  do not show the usual run-away behavior 
but oscillate around the transition region. Moreover, the solutions
typically exhibit short periods of accelerated expansion.
We also analyze the interplay between the geometries of moduli space and space-time.
%
%We also analyze the behavior of Kasner solutions at the boundary of the moduli space, and we investigate whether cosmological solutions can become singular. In addition, we show for any Calabi-Yau compactification which becomes singular through a type I contraction (flop transition) or type III contraction ($SU(2)$ enhancement) that Kasner solutions remain smooth.

\end{titlepage}

\tableofcontents

\begin{section}{Introduction}

The derivation of cosmology from a fundamental theory of gravity
and particle interactions is a very interesting and challenging 
problem. In particular, it turns out to be rather hard to understand the
origin of primordial inflation, which is a key ingredient of the most compelling cosmological models
\cite{L,LL}.
Moreover, recent astronomical observations \cite{WMAP,SN,SN2} indicate that our universe currently undergoes a modest accelerated expansion. Both cases are typically explained by postulating a suitable scalar potential.
As long as one has no or minimal ($\cN=1$) supersymmetry, this
potential is an independent function, and there are many choices
which lead to inflation. Even though there has been considerable progress in constructing such potentials from string theory \cite{KKLT,KKLMMT,BKQ}, inflation is non-generic and involves some degree of 
`functional fine tuning'.

The situation becomes even more involved when one considers effective theories which preserve more than the minimal supersymmetry ($\cN \geq 2$) or live in higher dimensions ($D \geq 5$):
here the potential is not an independent function, but is determined
by a so-called gauging, i.e., by the action of the gauge group on
the manifold parametrized by the scalar fields. This reduces the
options for fine-tuning and it becomes difficult
to find potentials which allow for inflation at all. 
%
%Further problems arise 
%in the context of string and M-theory compactifications, because
%there are no-go theorems which rule out accelerated expansion
%for a time-independent, non-singular and compact internal manifold \cite{Gib,MalNun}.
%
So far two mechanisms are known for 
deriving potentials from string and M-theory compactifications which are of interest for getting accelerated cosmological expansion
\cite{Tow}: (i) 
compactifications on time-dependent hyperbolic spaces \cite{hyper_comp_1,hyp_1,hyp_2,hyper_comp_2,emp,hyp_3,hyper_comp_3a,hyper_comp_3,hyper_comp_4} and (ii) compactifications with fluxes \cite{flux_c_1,flux_c_2,emp,hyper_comp_3,hyper_comp_4} or, in a broader sense, brane world cosmologies (we refer to \cite{Que} for a review and more references). Generically, the amount of expansion in these models is not sufficient for primordial inflation, but may be relevant for explaining the acceleration observed today.

In this paper we investigate a third mechanism to obtain accelerated expansion from M-theory compactifications: the inclusion of extra light modes arising from the internal Calabi-Yau (CY) manifold $X$ becoming singular. In this case 
some cycles of $X$ are contracted 
to zero volume.
The additional states, 
which we will call `transition states' (following \cite{flop}) are 
the winding modes of strings or branes wrapping the contracted cycles. This intrinsically stringy phenomenon has profound consequences.
Although the manifold $X$ becomes singular in this limit, the full
string or M-theory physics is non-singular. This also applies to the corresponding low energy effective action (LEEA),  provided that all
relevant, i.e., light, degrees of freedom are taken into 
account \cite{Str,Witten1}. 
Moreover,  degenerations of the internal manifold 
can often be smoothed in more than one way, implying the existence
of so-called topological phase transitions, where one passes from
a smooth family of compactification spaces to another one, with different 
topology \cite{GreFlop,WitD=2,Str2,Witten1}.\footnote{We refer
to \cite{GreRev} for a review and further references.}

We study the effect of the transition states
arising in an M-theory flop transition on the dynamics of
Kasner cosmological solutions. Thereby, we extend earlier work by
Br\"andle and Lukas \cite{flop} where the hypermultiplet scalar manifold was 
taken to be flat. This assumption is compatible with rigid, but 
not with local supersymmetry. To describe the flop transition, we use the gauged supergravity
action constructed in \cite{model}, which has $\cN = 2$ {\it local} supersymmetry.
Our work continues the study of transition states in the 
framework of low energy effective actions along the lines of 
\cite{GMMS,GSS,KMS,SU2,TMProc,SU2/4}.

We find that the inclusion of the dynamical transition states has two important effects:
the first is that the scalar fields no longer show the usual run-away
behavior but are dynamically attracted to the flop region, 
where they oscillate around the transition locus.
The second is that we indeed find solutions with several, though very short,  
periods of accelerated expansion. Yet, in our specific model 
we do not realize slow-roll inflation and the net effect of the
accelerating periods on cosmic expansion is not significant.

The stabilization of moduli in the vicinity of the flop line
was already observed in \cite{flop}. Part of the motivation of our work is to reinvestigate this interesting phenomenon using a locally supersymmetric action.  Although it is natural to expect that 
the coupling to supergravity does not modify the properties of a 
scalar potential in an essential way, we need to point out
that this is actually more subtle. In theories with more than
four supercharges the scalar potential is not an independent
function, but is determined by the gauging of the scalar manifold. Also, the conditions on the geometries
of these manifolds are different for local and for rigid supersymmetry.
Therefore it is not guaranteed that the properties of a rigidly supersymmetric
matter sector carry over to supergravity.   
One illustrative example was given in \cite{Kal}: while it is possible 
to construct potentials suitable for hybrid inflation in rigid
four-dimensional ${\cal N}=2$ supersymmetry, their extension to
${\cal N}=2$ supergravity did not work in this case.

We now give an outline of the paper and summarize our main  
results.
 Section 2 contains
the necessary background material. First we 
review the properties of the particular CY space that 
we use for compactification, the $\mathbbm{F}_1$-model described in detail in \cite{Jan,phasetransitions}.
Then we specify two different LEEA for the compactified theory. The usual LEEA is obtained by dimensional reduction on a smooth
CY manifold and only contains states which are generically
massless, while the transition states are left out. We call this
description the `Out-picture'. The `In-picture' is 
obtained by including the transition states as dynamical fields in the LEEA.
We then derive the equations of motion for Kasner cosmological
solutions,
\be
ds^2 = - e^{2 \nu(t)} \, d t^2 + e^{2 \alpha(t)} d \vec{x}^2 + e^{2 \beta(t)}
d y^2 \;.
\ee
Since our space-time is five-dimensional, we allow different scale factors
of the three-dimensional part, with coordinates $\vec{x}$, and of the
extra dimension, with coordinate $y$. This ansatz also includes de Sitter spaces. In this case the lapse function $\nu(t)$ is set to zero and the logarithmic scale factors $\alpha$ and/or $\beta$ increase linearly in time. Our main reason
for considering a five-dimensional instead of a four-dimensional
cosmology is that we can treat the sector which controls the flop
transition exactly, as we will explain in detail below. 
%
%We hope to
%study an analogous four-dimensional model in the future. 
%
Of course,
it would also be interesting to integrate our five-dimensional model 
into a brane world setup.

In section 3
we consider cosmological solutions of the Out-picture equations of motion.
We show that Kasner cosmological solutions are smooth when the internal 
space passes through a topological phase transition. The proof is independent 
of the
choice of the internal manifold, and applies to  flops as well as
to transitions with $SU(2)$ gauge symmetry enhancement.
We also show that an accelerated expansion is 
possible along the extra $y$-direction, but not in the three-dimensional
$\vec{x}$-space. 
Then we illustrate the dynamics of the moduli, using
representative numerical solutions for the $\mathbbm{F}_1$-model. 
These show the usual run-away
behavior, i.e., they roll through the moduli space until they reach 
its boundary. One remarkable feature is that the solutions do not
become singular as long as the moduli take values in the interior
of the moduli space, which fits nicely with observations made in the
context of BPS solutions and adds further evidence to the idea
that the K\"ahler cone acts as a cosmic censor \cite{KMS,TMProc}. 
The behavior of solutions at the boundaries is studied
analytically, and we give examples of solutions which connect
any two boundary components of the moduli space in finite time.

In section 4 we analyze cosmological solutions in the In-picture. First we 
review the properties of the scalar potential which is induced by
the transition states. Then we show that the equations of motion
have a family of non-hyperbolic fixed points, which is parametrized
by the flat directions of the scalar potential. `Non-hyperbolic' means
that the fixed manifold is neither attractive nor repulsive, so that 
linearized solutions exhibit oscillations. 
Therefore it is difficult to make an 
analytic statement about 
the asymptotic
behavior of the full non-linear system. However, the numerical solutions
behave uniformly. They are attracted towards the region of the flop where 
they oscillate around the transition locus. 
We neither see an attractor behavior where all the scalars approach
constant values nor a run-away behavior.

The second new feature of the In-picture is that 
accelerated expansion of the three-dimensional $\vec{x}$-space
is now possible. 
In order to investigate whether one can generate a sufficient
amount of inflation we use generalized slow-roll conditions 
suitable for non-linear sigma models (see for example \cite{Stewart}). A numerical investigation 
shows that these conditions cannot be satisfied in our model.
We also study explicit examples
of numerical solutions with an accelerating scale factor 
of the three-dimensional $\vec{x}$-space. These solutions  oscillate rapidly between between 
accelerated and decelerated expansion, leaving the
net effect of the inflationary episodes negligible.

In the last section we discuss our  results and 
explain the dynamics of the In-picture
in terms of the scalar potential. We also consider which modifications
are needed in order to get realistic moduli stabilization and inflation before we conclude with an outlook on future directions of research.

\end{section}

\begin{section}{Review of the  model}
In this section we collect all the material which is needed
later. First we 
review flop transitions in M-theory and their description
in the Out- and In-picture. We then
give the details of the LEEA for the particular CY
space we use to set up our cosmological model. Finally we derive the
equations of motion for Kasner space-times.

\begin{subsection}{Scalar manifolds: Out-picture vs In-picture}
The compactification of eleven-dimensional supergravity on 
a smooth CY threefold $X$ with Hodge numbers $h^{p,q}$ results in five-dimensional
minimal supergravity coupled to $n_V = h^{1,1} -1$ abelian vector 
multiplets and $n_H = h^{2,1} + 1$ neutral hypermultiplets
\cite{CCDF}. The theory has
no scalar potential, and all the scalars are moduli. The
vector multiplet scalars parametrize deformations of
the K\"ahler form of $X$ at fixed total volume, while 
the hypermultiplet scalars parametrize the volume of $X$,
deformations of its complex structure, and deformations of
the eleven-dimensional three-form gauge field. 

Supersymmetry imposes restrictions on the geometries
of the scalar manifolds. The vector multiplet scalars are coordinates of
 a so-called very special real manifold
${\cal M}_{\mscr{VM}}$, which can be realized as the level set
of a homogeneous, real, cubic polynomial, called the
prepotential \cite{GST}:
\be
{\cal V}(\phi) = C_{IJK} h^I(\phi) h^J(\phi) h^K(\phi) = 1 \;.
\ee
Here $\phi=(\phi^x)$, $x=1,\ldots, n_V$ denote the scalars
which parametrize ${\cal M}_{\mscr{VM}}$, while $h^I$,
$I=0,\ldots, n_V$ are the embedding coordinates. The
coefficients $C_{IJK}$ of the prepotential ${\cal V}$
determine all couplings in the vector multiplet sector. 
In CY compactifications they are given by
the triple intersection numbers of $X$.

The hypermultiplet manifold ${\cal M}_{\mscr{HM}}$ must be
a quaternion-K\"ahler manifold with a negative Ricci scalar \cite{BagWit}.
There is no simple relation between the topological
data of $X$ and ${\cal M}_{\mscr{HM}}$. In fact, not much
is known about ${\cal M}_{\mscr{HM}}$ in general. The so-called
universal hypermultiplet, which is the subspace of $\cM_{\mscr{HM}}$ containing the volume of $X$ and no 
 complex structure deformations, is known.
At the classical level the
corresponding quaternion-K\"ahler space is 
$X(1) = U(2,1) / (U(2) \times U(1))$ \cite{CCDF}.\footnote{It has been 
shown recently that there is a non-trivial one-loop correction \cite{AMTV}.
We will work with the tree level manifold for simplicity.}

The K\"ahler moduli space of $X$ has the structure of a cone
and is called the K\"ahler cone ${\cal K}_X$ \cite{Wilson} (see also 
\cite{Huebsch}). If the boundary 
of ${\cal K}_X$ is approached, certain cycles within $X$ are contracted to zero volume and
one obtains a singular space $\hat{X}$. The most simple degeneration
is a type I contraction, where a finite number $N$ of
holomorphic curves is contracted. In this case it is possible
to make a flop transition and 
to obtain a new family of smooth CY spaces $\tilde{X}$,
parametrized by a new K\"ahler cone ${\cal K}_{\tilde{X}}$. 
The extended K\"ahler cone is constructed by joining
the K\"ahler cones of all CY spaces related by a flop transition along their common boundaries. The spaces $X$ and 
$\tilde{X}$ have different topologies, in particular they have
different triple intersection numbers $C_{IJK}$ and
$\tilde{C}_{IJK}$. In terms of the LEEA the changes in the couplings of
the vector multiplet sector can be explained as the threshold
corrections arising from integrating out the $N$ charged hypermultiplets, which 
become massless at the transition point \cite{Witten1}.
Moreover, the winding states of M2-branes around the $N$ contracted curves
indeed account for these $N$ additional charged hypermultiplets.

One way to describe the dynamics near the flop transition is 
to work with the standard Out-picture LEEA. In this case the flop transition
manifests itself in a discontinuous change of the triple
intersection numbers, $C_{IJK} \rightarrow \tilde{C}_{IJK}$.
However, configurations where the transition states
are excited cannot be described. But
a reliable description of M-theory low energy dynamics needs to include all
the light states. Therefore it is preferable to work with the extended
In-picture LEEA. The vector multiplet
sector of this action can be found exactly: the 
Lagrangians in the Out-picture are determined by the
triple intersection numbers of $X$ and $\tilde{X}$, and the 
threshold corrections induced by integrating out $N$ charged
hypermultiplets are also known exactly \cite{Witten1}. The
prepotential in the In-picture is given by the
`averaged triple intersection numbers'\footnote{This `orbit sum
rule' does not only apply to flops, but also to transitions
with $SU(2)$ gauge symmetry enhancement \cite{SU2}.}
\be
\hat{C}_{IJK} = \ft12 ( C_{IJK} + \tilde{C}_{IJK} ) \, .
\label{OSR}
\ee

The situation in the hypermultiplet sector is more complicated:
it is  not known how to compute the metric of ${\cal M}_{\mscr{HM}}$
in string or M-theory, whether in the Out-picture or in the
In-picture. However,  in \cite{model} we constructed a
toy model
for ${\cal M}_{\mscr{HM}}$ which (i) gives a consistent supergravity
action, (ii) has the correct physical properties to describe a flop,
and (iii) is simple enough to allow for explicit calculations. 
The hypermultiplet manifold in the In-picture contains $N+1$ hypermultiplets and is taken to be
\be
X(1+N) = \frac{U(1+N,2)}{U(1+N) \times U(2)} \;,
\ee
where $N$ of the
hypermultiplets carry the charges appropriate for the
transition states of a flop. This charge assignment
uniquely fixes the 
gauging of isometries, which in turn determines the 
gauged supergravity Lagrangian and in particular the scalar potential.
The remaining neutral hypermultiplet parametrizes the space $X(1)$ and is identified with the universal
hypermultiplet.\footnote{As discussed in \cite{model} the relation between the hypermultiplet manifolds in the In- and the Out-picture may be more complicated due to threshold corrections which we neglect in this paper.}

\subsection{The vector multiplet sector of the $\mathbbm{F}_1$-model}
We consider a particular CY compactification, where the internal 
manifold is the elliptic fibration over the first Hirzebruch
surface $\mathbbm{F}_1$ \cite{Jan,phasetransitions}. We will refer to this choice as the
$\mathbbm{F}_1$-model. It has two vector multiplets and
consists of two K\"ahler cones which, following  \cite{phasetransitions}, will be called region II and region III. These regions are conveniently parametrized by the three real variables  $h^I= T,U,W$, and their prepotentials 
are given by
\bea\label{2.1}
\nonumber {\cal V}_{\rm (II)} & := & \frac{3}{8} \, U^3 + \frac{1}{2} \, U \, T^2 - \frac{1}{6} \, W^3 = 1 \, , \\
{\cal V}_{\rm (III)} & := & \frac{5}{24} \, U^3 + \frac{1}{2} \, U^2 W - \frac{1}{2} \, U \, W^2 + \frac{1}{2} \, T^2 \, U = 1 \, .
\eea
The boundaries of the regions II and III are located at 
\be\label{2.2}
\begin{array}{llll}
 W = 0 \, , \quad & U = W \, , \quad & T = \frac{3}{2} U \, , & \mbox{and} \\
 U = 0 \, , \quad & U = W \, , \quad & T = \frac{1}{2} U + W \, , &
\end{array}
\ee
respectively. The flop transition occurs at the boundary $U=W$ where a single holomorphic curve shrinks to zero volume. The extended K\"ahler cone of this model is obtained by joining the two regions at this boundary. Here the discontinuity of the prepotentials 
(\ref{2.1}), 
\be\label{2.3}
{\cal V}_{\rm (II)} - {\cal V}_{\rm (III)} = \frac{1}{6} (U - W)^3 \, ,
\ee
corresponds precisely to the threshold correction arising from
integrating out one hypermultiplet with unit charge under the
$U(1)$ associated with the direction $U-W$ \cite{Witten1,phasetransitions}.

In order to find the metrics on the two K\"ahler cones, we first calculate 
the coupling matrix $a_{IJ}$ of the gauge fields,\footnote{Here and in the following
we use the standard formulae for five-dimensional vector multiplets \cite{GST}.}
\be\label{2.4}
%\begin{split}
a_{IJ} :=   - \, \frac{1}{3} \partial_I \, \partial_J \ln( {\cal V}) 
\left|_{ {\cal V} = 1} \right. \, \\
=   -2 \, C_{IJK} \, h^K + 3 \, C_{IKL} \, C_{JMN} \, h^K h^L h^M h^N \, ,
%\end{split}
\ee
where the index $I$ enumerates the fields $h^I = T, U,W$. To construct the
vector multiplet scalar metric, we single out two coordinates which will
serve as the vector multiplet scalar fields: $\phi^x =  U,W $. Solving the 
constraints (\ref{2.1}) in terms of the remaining coordinate $T$, we
explicitly obtain $T(U,W)$ in the regions II and III as

\be\label{2.5}
\begin{split}
T_{\rm (II)}(U,W) & =  \, \left( \frac{24 + 4 W^3 - 9 U^3 }{12 U} \right)^{1/2} \qquad  \mbox{and} \\
T_{\rm (III)}(U,W) & =  \,\left( \frac{24 + 12 U W^2 - 12 U^2 W - 5 U^3}{12 U} \right)^{1/2} \,  ,
\end{split}
\ee
respectively. The metric $g_{xy}$ on the vector multiplet scalar manifold is 
 proportional to the pull-back of $a_{IJ}$ by the immersion
\be\label{2.6}
g_{xy} = h^I_x \, h^J_y \, a_{IJ} \, , \quad \mbox{where} \quad h^I_x := - \sqrt{\frac{3}{2}} \partial_x \, h^I(\phi) \, .
\ee
Here ``$\partial_x$'' denotes the partial derivative with respect to the vector multiplet scalar $\phi^x$ and $T$ is taken as $T(U,W)$. The prepotentials (\ref{2.1}) give rise to the following scalar manifold metrics:
\bea\label{2.7}
g^{\rm (II)}_{xy}  &  = & \left[ \begin{array}{cc} 
\frac{9U^3W^3 + 54U^3 - 12 W^3 - 36 - W^6}{2 U^2 \left( 9U^3 - 4 W^3 - 24 \right)} & 
- \, \frac{W^2 \left( 9 U^3 - W^3 - 6 \right)}{2 U \left( 9U^3 - 4 W^3 - 24 \right)} \\ 
- \, \frac{W^2 \left( 9 U^3 - W^3 - 6 \right)}{2 U \left( 9U^3 - 4 W^3 - 24 \right)} &
\frac{W \left( 9 U^3 - W^3 - 24 \right)}{2 \left( 9U^3 - 4 W^3 - 24 \right)}
\end{array} \right] \, , \\
\nonumber g^{\rm (III)}_{xy} & = & \left[ \begin{array}{cc} 
\frac{18 - 15 U^3 - 18 U^2 W + 12 U W^2 - 4 U^4 W^2 }{U^2 \left( 24 + 12 U W^2 - 12 U^2 W - 5 U^3 \right)} &
\frac{6 W - 9 U + 4 U^3 W }{24 + 12 U W^2 - 12 U^2 W - 5 U^3} \\ 
\frac{6 W - 9 U + 4 U^3 W }{24 + 12 U W^2 - 12 U^2 W - 5 U^3} &
\frac{4 U \left( 3 - U^3 \right) }{24 + 12 U W^2 - 12 U^2 W - 5 U^3}
\end{array} \right] \, .
\eea
These metrics are non-degenerate at the flop line $U=W$, 
and can be connected continuously.
The properties of the other boundaries of the extended K\"ahler cone shown in the left diagram of Fig. \ref{eins} can be found in \cite{phasetransitions}
and are summarized in Table \ref{t.1}.
\begin{figure}[t]
\renewcommand{\baselinestretch}{1}
\epsfxsize=0.45\textwidth
\begin{center}
\leavevmode
\epsffile{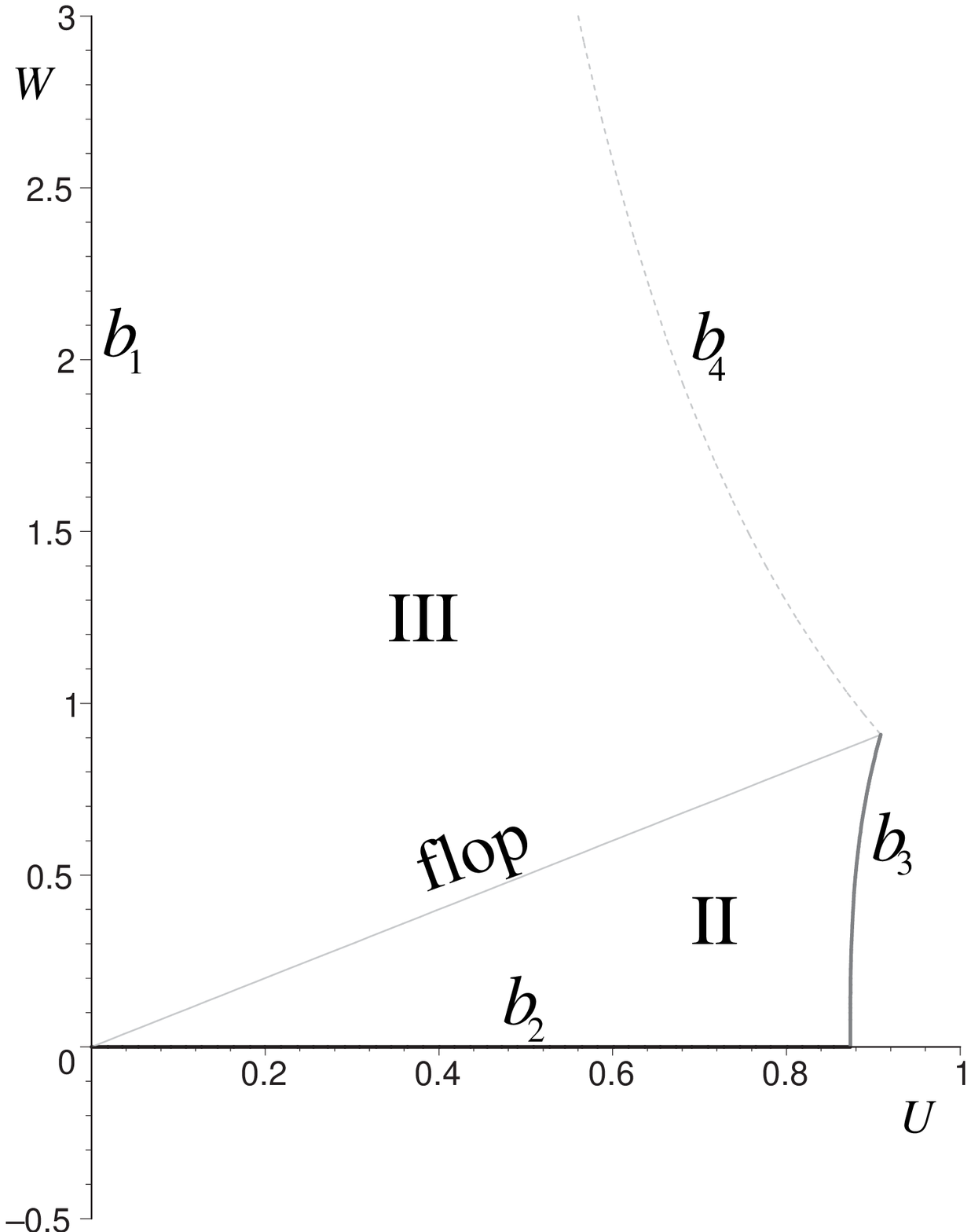} \, \, \, \, \,
\epsfxsize=0.45\textwidth
\epsffile{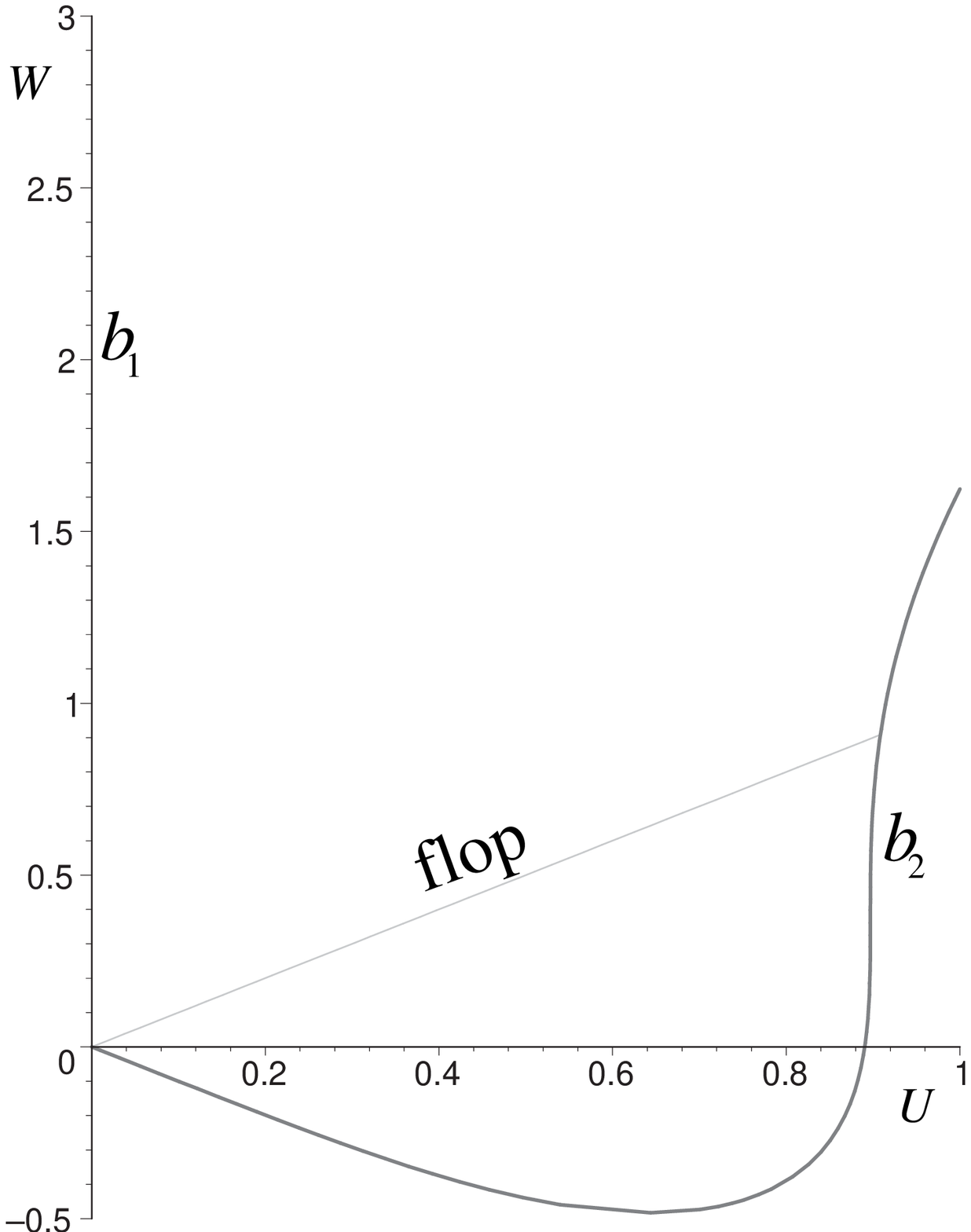}
\end{center}
\parbox[c]{\textwidth}{\caption{\label{eins}{\footnotesize Comparison between the vector multiplet scalar manifolds in the Out- (left)  and In-picture (right). The gray line labeled ``flop'' indicates the locus of the flop transition $U=W$. The labels ``$b_1$'', ``$b_2$'', ``$b_3$'' and ``$b_4$'' indicate the other boundaries of the scalar manifolds. The location of these boundaries is different in the Out- and In-picture.}}}
\end{figure}
\begin{table}[t]
\begin{tabular*}{\textwidth}{@{\extracolsep{\fill}} cccc} \hline \hline
boundary & location &  $\det(g_{xy})$ & microscopic picture \\ \hline
$b_1$ & $U = 0$ & infinite &  zero CY volume   \\ 
$b_2$ & $W = 0$ & zero & tensionless strings \\
%loss of the K\"ahler modulus $W$  \\
%
$b_3$ & $T_{\rm (II)} = \frac{3}{2} U$ & zero & 
tensionless strings  \\ 
$b_4$ &  $T_{\rm (III)} = \frac{1}{2} U + W$ & finite & $SU(2)$-enhancement  \\
\hline \hline
\end{tabular*}
\renewcommand{\baselinestretch}{1}
\parbox[c]{\textwidth}{\caption{\label{t.1}{\footnotesize
Boundaries of the Out-picture K\"ahler cones. Here $T_{\rm (II)}$ and $T_{\rm (III)}$ are given in eq. (\ref{2.5}) and $g_{xy}$ is short for $g_{xy}^{\rm (II)}$ or $g_{xy}^{\rm (III)}$ depending on the region. The last column  gives the interpretation of the boundaries in terms 
of microscopic M-theory physics.}}}
\end{table}  
%
%This table is organized as follows. The first column ``boundary'' gives the label of the boundary as in Fig. \ref{eins} and the second column ``location'' gives its location in terms of $U,W$. In the third column ``$\det(g_{xy})$'' 
%The boundaries are labeled "$b_1$", \ldots, "$b_4$"
%and we give the value of the determinant of $g_{xy}$ for each boundary. 
%Here $g_{xy}$ should be understood as being $g_{xy}^{II}$ and $g_{xy}^{III}$
%in region II and III, respectively. 

%
%At generic points of 
%the boundary of the K\"ahler cone corresponding to 
%$b_1$ the volume of the internal space becomes zero. 
%However, the scalar parametrizing the total volume sits in a hypermultiplet,
%not a vector multiplet. The degeneration found at $b_1$, which is the
%intersection of the K\"ahler cone with the cubic hypersurface (\ref{2.1})
%corresponds to shrinking some two-cycles, while others are decompactified
%in a such a way that the total volume is kept fixed \cite{TMProc}.
%At $b_2$  the extended K\"ahler cone extends beyond this boundary but the 
%corresponding vector multiplet manifold is parameterized by $U$ only. 
%At $b_2, b_3$  we get tensionless strings and
%there is an infinite number of transition states, which become massless at this boundary.\footnote{The boundary component $b_2$ corresponds to 
%a smooth CY compactification on the elliptic fibration over $\mathbbm{CP}^2$.} 
%Finally at $b_4$  one obtains two additional vector multiplets which become massless and form an $SU(2)$ triplet with one of the vector multiplets already present.

Let us now consider the vector multiplet sector of the In-picture LEEA. 
Application of the `orbit sum rule' (\ref{OSR}) to the case at hand gives
\be\label{2.8}
\begin{split}
{\cal V}_{\rm (In)} = & \frac{1}{2} \left( {\cal V}_{\rm (II)} 
+ {\cal V}_{\rm (III)} \right) \\
= & \frac{7}{24} U^3 + \frac{1}{2} U T^2 - \frac{1}{12} W^3 + \frac{1}{4} U^2 W - \frac{1}{4} U W^2 = 1 \, .
\end{split}
\ee
The metric on the vector multiplet scalar manifold is obtained in the same way as the metrics (\ref{2.7}). Explicitly, we find
\be\label{2.9}
 g_{xy}^{\rm (In)}  =  
\left[ \begin{array}{cc} 
 g_{UU} & g_{WU}  \\ g_{WU} & g_{WW}
\end{array} \right] \, ,
\ee 
where the entries are given by
\bea\label{2.10}
\nonumber
g_{UU} & = & \frac{1 }{8 U^2 \, K} \, \bigg( 144 - 6 W^4 U^2 + 48 U W^2 + 4 W^5 U - 72 U^2 W \,  \\
\nonumber && \qquad \qquad  - 14 U^3 W^3 - 17 U^4 W^2 - 168 U^3 + 24 W^3 + W^6 \bigg) \, , \\
 g_{WU} & = &  \frac{1}{8 U \, K} \, \bigg( 24 U W - 36 U^2 - 12 W^2 - W^5 - 4 W^4 U \\
\nonumber && \qquad \qquad  + 6 U^2 W^3 +  14 U^3 W^2 + 17 U^4 W \bigg)  \, ,  \\
\nonumber
g_{WW} & = & \frac{1}{8 \, K} \, \left( 48 (U+W) + 4 U W^3 - 6 U^2 W^2 - 14 U^3 W + W^4 - 17 U^4 \right) \,  ,
\eea
with
\be\label{2.11}
K :=  24 - 7 U^3 + 2 W^3 - 6 U^2 W + 6 U W^2 \, .
\ee
 The boundaries
of the vector multiplet manifold are defined 
by the loci where the metric $g_{xy}^{\rm (In)}$ degenerates. 
This manifold is shown in the right diagram of Fig. \ref{eins}. Here we 
distinguish between the two boundaries $b_1$ and $b_2$ which correspond to 
the lines where $\det(g_{xy}^{\rm (In)}) \rightarrow \infty$ and  
$\det(g_{xy}^{\rm (In)}) = 0$, respectively. Note that the 
boundaries of the vector multiplet scalar manifolds of the Out- and the 
In-picture have slightly different locations.

Combined with the hypermultiplet
sector discussed below the prepotential (\ref{2.8}) defines  a consistent supergravity Lagrangian. In order to interpret this Lagrangian correctly it
is important to realize that it can only be expected to be a good
description of M-theory low energy physics as long as the masses of
the transition states are smaller than those of all other massive
M-theory states. This is true in the vicinity of the flop line. 
If one moves far away from this line, there are many other states which have 
masses comparable to those of the transition states. One should then 
either include these other massive states in the LEEA (which so far
has not been done) or one should
integrate out the transition states and work in the Out-picture 
(valid at energies less than the mass of the lightest massive
excitation).

We are of course free to study the In-picture LEEA (\ref{2.8}) as 
an interesting  gauged supergravity model in its own right.
Therefore we will later consider solutions within
the full scalar manifold, including the boundaries. However, 
only those properties which are determined by the behavior
close to the flop line are guaranteed to capture the M-theory
physics correctly.

\subsection{The hypermultiplet sector of the $\mathbbm{F}_1$-model}
Let us now turn to the hypermultiplet sector in the Out-picture, which can be 
truncated consistently   to the universal hypermultiplet. 
This multiplet contains the CY volume $V$ and three  real
scalars $\sigma, \theta, \tau$ arising from the dimensional reduction of the 
three-form field. These scalars parametrize the coset
\be\label{2.12}
X(1) = \frac{U(2,1)}{U(2) \times U(1)} \, .
\ee
The explicit metric obtained from CY compactifications of M-theory is \cite{UHM_2}
\be\label{2.13}
ds^2 = \frac{1}{4V^2} \rmd V^2 + \frac{1}{4 V^2} \left( \rmd \sigma + 2 \theta \rmd \tau - 2 \tau \rmd \theta \right)^2 + \frac{1}{V} \left( \rmd \theta^2 + \rmd \tau^2 \right) \, .
\ee
This metric completely determines the hypermultiplet sector of the Out-picture LEEA.

The hypermultiplet sector of the In-picture has in 
addition a second hypermultiplet, 
the transition states. Following \cite{model} we take the 
hypermultiplet manifold to be
\be\label{2.14}
X(2) = \frac{U(2,2)}{U(2) \times U(2)} \, .
\ee
The metric on this space was constructed in \cite{model}
by employing the superconformal quotient construction 
\cite{SCQ,SCgauging,SCQ2} resulting in\footnote{Note that the metrics obtained by the superconformal quotient construction are {\it negative} definite. Their relation to the positive definite metrics appearing in the Lagrangian (\ref{2.22}) is given by $ ds^2( g_{XY} )  = -  ds^2( G_{XY} )$.}
\be\label{2.15}
 G_{XY}  = \left[ \begin{array}{cccc}
 0 & G_{v \vb} & 0 & G_{v \ub} \\ 
 G_{\vb v} & 0 & G_{\vb u} & 0 \\
 0 & G_{u \vb} & 0 & G_{u \ub} \\
 G_{\ub v} & 0 & G_{\ub u} & 0 \\
\end{array} \right] \, ,
\ee
with the entries:
\bea\label{2.16}
\nonumber G_{u_i \ub_{ \jb }}  & = & \frac{1}{2 \phi_-}  \left( \eta^{i \jb} + \vb^{\jb} \, v^i \right) 
- \frac{1}{2 \phi_-^2}
\left( \eta^{\jb l} u_l + \vb^{\jb} \left( v^l u_l \right) \right)  
\big( \eta^{i \lb} \ub_{\lb} + v^i \big( \vb^{\lb} \ub_{\lb} \big) \big) \, , \\
G_{\vb^{\ib} u_{j}} & = &  \frac{1}{2 \phi_-^2} \, \left( \ub_{\ib} v^j \, \left( 1 + \eta^{k \lb} u_k \ub_{\lb} \right) - \ub_{\ib} \eta^{j \lb} \ub_{\lb} \left( v^l u_l \right) \right)  \, , \\
\nonumber G_{v^i \vb^{\jb}} & = & 
\frac{1}{2 \phi_+} \, \eta_{i \jb} 
- \frac{1}{2 \phi_+^2} \, \big( \eta_{i \lb} \, \vb^{\lb} \big) \, \big( \eta_{\jb l} \, v^{l} \big) \, - \, \frac{1}{\phi_+ \phi_-} \, u_i \ub_{\jb} \\
\nonumber && + \, \frac{1}{2 \phi_-} \, u_i \, \ub_{\jb} \, - \, \frac{1}{2 \phi_-^2} \, u_{i} \, \ub_{\jb} \, \big( v^l \, u_l \big) \big( \vb^{\lb} \ub_{\lb} \big) \, .
\eea
Here the index $i = 1,2$ enumerates the hypermultiplets parametrized by the complex fields $ v^i, u_i $ and $X,Y = \left\{ v^i, \vb^{\ib}, u_i, \ub_{\ib} \right\}$. Further $\eta_{i \lb} = {\rm diag}\left[-1,-1 \right]$ and $\eta^{i \lb}$ is its inverse. Additionally $\phi_+$ and $\phi_-$ are defined by:
\be\label{2.16a}
 \phi_+ := 1 + \eta_{i \bar{\jmath}} \, v^i \bar{v}^{\bar{\jmath}}  \, , \quad 
\phi_- :=  1 + \eta^{i \bar{\jmath}} \, u_i \bar{u}_{\bar{\jmath}} + \left( v^i u_i \right) \left( \bar{v}^{\bar{\imath}} \, 
\bar{u}_{\bar{\imath}} \right)  \, .
\ee
The other non-vanishing entries of the matrix can be obtained from the relations $G_{v^i \vb^{\jb}} = G_{\vb^{\jb} v^i}$, $G_{u_i \ub_{\jb}} = G_{\ub_{\jb} u_i}$, $G_{v^i \ub_{\jb}} = G_{\ub_{\jb} v^i}$ and $G_{v^i \ub_{\jb}} = \left( G_{\vb^{\ib} u_j} \right)^*$, where ``$*$'' denotes complex conjugation.

We take $ v^1, u_1 $ to be neutral while the 
second hypermultiplet $ v^2, u_2 $ carries the charge required 
to describe the transition states. This is achieved 
by gauging the particular $U(1)$ isometry corresponding to the
Killing vector
\be\label{2.17}
k_{\rm gauge} =  - \, \imag \left[ \, 0 \, , \, v^2 \, , \, 0 \, , \,  - \vb^2 \, , \,  0 \, , \,  - u_2 \,  , \, 0, \,  \ub_2 \, \right]^{\rm T} \, , 
\ee
given with respect to the basis
$
 \left\{ \partial_{v^1}, \partial_{v^2},\partial_{\vb^1},\partial_{\vb^2},\partial_{u_1},\partial_{u_2},\partial_{\ub_1},\partial_{\ub_2} \right\} \, .
$
The masses of these states are encoded in the scalar potential of the Lagrangian. We will come back to this point in the next subsection.

Since we identify 
the neutral hypermultiplet of the In-picture with the universal
hypermultiplet, we would like to know how 
the coordinates $ v^1, u_1 $ of $X(2)$ are explicitly related
to the fields 
$V, \sigma, \theta, \tau $.
We first observe that restricting the
metric on $X(2)$ to the subspace of vanishing transition states, $v^2 = u_2 =
0$,
%, \rmd v^2 = \rmd u_2 = 0$ 
provides us with a metric on $X(1)$. Explicitly,
this metric is given by 
\bea\label{2.19}
\nonumber G_{u \ub} & = & - \, \frac{1}{2 \phi_-^2} \, \left( 1 - v \vb \right) \, ,  \\
G_{u \vb} & = & \frac{1}{2 \phi_-^2} \, \ub v \, , \\
\nonumber G_{v \vb} & = & - \, \frac{1}{2 \, \phi_{+}^{2} \, \phi_-^2} \, \left(1 - u \ub \left( 1 - v \vb \right)^2 \right)  \, ,
\eea
which is the metric of $X(1)$ obtained in \cite{SCQ,model}. 
Note that here and in the rest of this subsection, we drop the index $1$ from the fields $ v^1, u_1 $.

%Integrating out the charged 
%hypermultiplet should a priori result in threshold corrections which 
%modify the selfcouplings of the neutral hypermultiplet. Following
%\cite{model} we assmue (self-consistently) that 
%these can be ignored in a first approximation, 
%so that the two hypermultiplet manifolds are related
%by a simple truncation. In particular the $X(1)$ subspace
%of $X(2)$ is identified with the universal hypermultiplet.

Taking into account that $ ds^2( g_{XY} )  = -  ds^2( G_{XY} )$, the transformation between $ v, u $ and $V,\sigma, \theta,\tau
$  can then be found by composing the coordinate transformations between 
the different parametrizations of the universal hypermultiplet given in 
\cite{UHM,SCQ}. This leads to the relations
\bea\label{2.20}
\nonumber
V & = & \frac{1}{L} \, \left( 1 - u \ub \left( 1 - v \vb \right)^2 - v \vb \right) \, , \\ \nonumber
\sigma & = & \frac{\imag}{L} \left( u - \ub \right) \, \left( 1 - v \vb \right) \, , \\ \nonumber
\theta & = & \frac{1}{2 L} \left( \vb \left( 1 + \ub \left( 1 - v \vb \right) \right) + v \left( 1 + u \left( 1 - v \vb \right) \right) \right) \, , \\  
\tau & = & \frac{\imag}{2 L} \left( \vb \left( 1 + \ub \left( 1 - v \vb \right) \right) - v \left( 1 + u \left( 1 - v \vb \right) \right) \right) \, ,
\eea
where $L$ is given by
\be\label{2.21}
L := \left( 1 + u \left( 1 - v \vb \right) \right) \left( 1 + \ub \left( 1 - v \vb \right) \right) \, .
\ee
It is straightforward to check that this coordinate transformation
indeed relates the metrics (\ref{2.13}) and (\ref{2.20}) for the universal
hypermultiplet given above. 
Therefore we recover the eleven-dimensional meaning of the
parameters in the subspace $X(1)$ and 
in particular their relation to the  CY volume $V$,
which we will take to be time-dependent in the cosmological solutions 
of the following sections. For later use we note that taking $v = 0$ 
and $u = \ub$ corresponds to setting $\sigma = \theta = \tau = 0$ and truncating the universal hypermultiplet to the volume scalar $V$. 
\end{subsection}
\begin{subsection}{The low energy effective Lagrangians}
After discussing the scalar manifolds in detail, we can now specify the LEEA.
The general form of a $\cN = 2, D = 5$ gauged supergravity Lagrangian is known from \cite{gs,UHM} 
and the concrete form of the LEEA modeling a flop transition was explicitly worked out in \cite{model}. We refer
to these papers for further details. 

The bosonic part of the Lagrangian is \cite{model}
\bea\label{2.22}
\nonumber \sqrt{-g}^{-1} \cL^{\cN = 2}_{\rm bosonic} & = &  - \frac{1}{2} R - \frac{1}{4} a_{IJ} F^I_{\mu\nu} F^{J~\mu\nu} \\ 
&& - \frac{1}{2} g_{XY} \cD_{\mu} q^X \cD^{\mu} q^Y - \frac{1}{2} g_{xy} \partial_{\mu} \phi^x \partial^{\mu} \phi^y \\ 
\nonumber && + \frac{1}{6 \sqrt{6}} C_{IJK} \sqrt{-g}^{-1} \epsilon^{\mu\nu\rho\sigma\tau} F^I_{\mu \nu} F^J_{\rho \sigma} A^K_\tau - {\rm g}^2 \cV(\phi, q) \, .  
\eea
Here $\phi^x$ and $q^X$ denote the vector and hypermultiplet scalars
parametrizing the scalar manifolds introduced in the previous subsection. 
Additionally, we have the five-dimensional metric $g_{\mu \nu}$ with Ricci scalar
$R$ and three vector fields $A^I_\mu$, $I = T, U,W$, with 
field strengths $F^I_{\mu \nu} = \partial_\mu A^I_\nu -
\partial_\nu A^I_{\mu}$. The $C_{IJK}$ are determined by the cubic polynomials
(\ref{2.1}) and (\ref{2.8}), and 
$a_{IJ}$ is given in eq. (\ref{2.4}). The covariant derivative appearing in the hypermultiplet kinetic term is given by $\cD_{\mu} q^X := \partial_{\mu} q^X + {\rm g} A^I_\mu K^X_{I}(q)$. This Lagrangian is general enough to describe  both the Out- and the In-picture.

The  Out-picture Lagrangian is obtained by setting ${\rm g} = 0$ and taking the scalar metrics 
to be  (\ref{2.7}) and (\ref{2.13}). To get the In-picture Lagrangian we have to use
 the scalar metrics (\ref{2.9}) and (\ref{2.16}) instead. Moreover, we now have a non-trivial
gauging of the hypermultiplet isometry corresponding to the Killing vector (\ref{2.17}).
The explicit form of the covariant derivative is 
\be\label{2.24}
\cD_\mu \, q^X =  \partial_\mu \, q^X + {\rm g} \left( A^U_\mu - A^W_\mu \right) k_{\rm gauge}^X(q) \, . 
\ee
The gauge coupling ${\rm g}$ is uniquely fixed in terms of microscopic M-theory data \cite{flop,model}, 
\be\label{2.25}
{\rm g} = 
\sqrt{\frac{2}{3}} \, T_{(2)} (6 {\rm v})^{1/3} =
\sqrt{\frac{2}{3}} ( 48 \pi )^{1/3}
\, ,
\ee
where $T_{(2)}$ denotes the tension of the M2-brane and ${\rm v} = \ft{ \kappa_{(11)}^2 }{\kappa_{(5)}^2}$ relates the
eleven-dimensional and the five-dimensional gravitational couplings. In the second step we used
$T_{(2)} = ( \ft{8 \pi}{\kappa_{(11)}^2} )^{1/3}$ and set $\kappa_{(5)}=1$.

The gauging determines the scalar potential 
\be\label{2.26}
\cV( \phi, q) = -6 \cW^2 + \frac{9}{2} g^{\Lambda \Sigma} \partial_{\Lambda} \cW \partial_{\Sigma} \cW \, ,
\ee
where $g^{\Lambda \Sigma} = g^{XY} \oplus g^{xy}$ is the direct sum of the inverse hypermultiplet and vector multiplet scalar metrics. The superpotential $\cW$ is given by
\be\label{2.27}
\cW = 6^{-5/6} \, \left( \frac{1}{2 \phi_-} \, \left( \ub_2 u_2 \right) + \frac{1}{2 \phi_+} \left( \vb^2 v^2 \right) \right)  \, \left( U - W \right) \, .
\ee
\end{subsection}
\begin{subsection}{Kasner space-times}
We now turn to the cosmological solutions of the Lagrangian (\ref{2.22}). 
For the five-dimensional space-time metric we make
the following Kasner ansatz: 
\be\label{2.28}
ds^2_5 = - \e^{2 \nu(t)} \rmd t^2 + \e^{2 \alpha(t)} \rmd \vec{x}^2 + \e^{2 \beta(t)} \rmd y^2  \, .
\ee
Here $\vec{x} = (x^1, x^2, x^3 )$ are three space-like coordinates,
parametrizing the macroscopic dimensions, while $y$ is the coordinate of
the fifth, extra dimension. Note that we include a non-trivial 
lapse function $\e^{\nu(t)}$ in the ansatz. This will play a crucial role 
in solving the Einstein equations in the Out-picture analytically.
We also impose that 
all fields are homogeneous in the four space-like directions, i.e., they do not depend on the spatial coordinates.

Moreover, we restrict ourselves to the case where the vector fields $A^I_\mu$ can be consistently  set to zero, as this will considerably simplify the later analysis. 
In the Out-picture the equations of motion are always solved 
by $A^I_\mu = 0$. The In-picture Lagrangian, however, contains a non-trivial source term, 
which arises from the covariant derivative in the hypermultiplet kinetic term,
 \be\label{2.30}
j^{\rm source}_{\mu} := \sqrt{-g} \, {\rm g} \, g_{XY}(q) \partial_\mu q^X k^Y_{\rm gauge}(q) \, .
\ee
This term vanishes, if the complex  hypermultiplet scalar fields $v^i, u_i$ are restricted to be real. Therefore we will use this truncation in the rest of the paper.

Under these assumptions we get the 
following non-trivial equations of motion from the Lagrangian (\ref{2.22}): \\
Einstein equations:
\bea\label{2.31}
\nonumber 3 \left( \ad^2 + \ad \, \bd \right) & = & T + {\rm g^2} \, \e^{2 \nu} \, \cV \, , \\
2 \ddot{\alpha} + \ddot{\beta} + 2 \ad \bd + 3 \ad^2 + \bd^2 - 2 \nd \ad - \nd \bd &=& - T + {\rm g^2} \, \e^{2 \nu} \, \cV \, , \\
\nonumber 3 \left( \ddot{\alpha} + 2 \ad^2 - \nd \ad\right) & = & - T + {\rm g^2} \, \e^{2 \nu} \, \cV \, ,
\eea
vector multiplet sector:
\be\label{2.32}
\ddot{\phi}^x + \gamma^x_{~yz} \, \pd^y \, \pd^z + \left( 3 \ad + \bd - \nd \right) \pd^x + \e^{2 \nu} {\rm g^2} g^{xy} \frac{\partial \cV}{\partial \phi^y } = 0 \, ,
\ee
hypermultiplet sector:
\be\label{2.33}
\ddot{q}^X + \Gamma^X_{~YZ} \, \qd^Y \, \qd^Z + \left( 3 \ad + \bd - \nd \right) \qd^X + \e^{2 \nu} {\rm g^2} g^{XY} \frac{\partial \cV}{\partial q^Y }= 0 \, .
\ee
Here the ``overdot'' indicates a derivative with respect to the time coordinate $t$, i.e., $\qd^X := \frac{\partial}{\partial t} q^X$, etc. We also introduced the kinetic energy $T$ as
\be\label{2.34}
T := \frac{1}{2} \, g_{XY} \, \qd^X \, \qd^Y + \frac{1}{2} \, g_{xy} \, \pd^x \pd^y \, .
\ee
The $ \gamma^x_{~yz}$ and $ \Gamma^X_{~YZ}$ denote the Christoffel symbols of the vector and hypermultiplet scalar metrics $g_{xy}$ and $g_{XY}$, respectively. We note that $T$ and $\cV(\phi,q)$ are both positive semi-definite.

To obtain the equations of motion valid in the Out- or In-picture, we make
the appropriate substitutions for the potential and the scalar field metrics. When studying numerical solutions we will
fix ${\rm g}$ to the value determined by M-theory, eq. (\ref{2.25}).

\end{subsection}
\end{section}

\begin{section}{Solutions in the Out-picture}

In this section we study cosmological solutions in the Out-picture.
First we derive some of their general properties which can be established without
specifying the vector and hypermultiplet manifolds. Then we
focus upon the $\mathbbm{F}_1$-model.

\begin{subsection}{General properties of cosmological solutions}
%
%Before turning to our particular model, let us discuss the properties of the
%functions appearing in the Kasner metric (\ref{2.28}) when the internal
%manifold undergoes a generic topological phase transition. 

We start our investigation by considering an arbitrary CY threefold and 
cosmological solutions which pass through a
 topological phase transition involving a finite number
of transitions states. This means that we do admit both
flop transitions with an arbitrary number of charged hypermultiplets,
and type III contractions which lead to 
$SU(2)$ gauge symmetry enhancement \cite{Wilson,KMP,Witten1,SU2}. In the latter case one 
has two charged vector multiplets together with a number
of charged hypermultiplets depending on the details of the
contraction.
We will show that $ \nu(t), \alpha(t)$ and $
\beta(t) $ are {\it smooth} across any topological phase transition satisfying
these assumptions.\footnote{This was also shown in \cite{flop} for
the flop of a single curve.}

%Let us comment on our assumptions. The Out-picture Lagrangian for a generic
%topological phase transition does not contain any charged matter fields. Hence
%it is always possible to consistently take the vector fields to be
%zero. Further integrating out the transition states only affects the vector
%multiplet sector of the theory. Therefore taking the hypermultiplet scalar
%metric to be continuous is also well justified. 
The only source of discontinuities in the Out-picture 
is the jump in the triple intersection numbers, which can be understood as a
threshold effect. When integrating out the
transition states one finds that the Out-picture prepotentials
in the two K\"ahler cones differ by the amount \cite{Witten1,KMP,SU2} 
\be\label{3.1}
\Delta {\cal V} = \frac{1}{6} \left( \delta n_H - \delta n_V \right) \, (h^\star)^3 .
\ee 
Here $h^\star = q_I h^I$ is proportional to the volume of the collapsing 
cycle\footnote{The $C^I$ are a basis of $H_2(X,\mathbbm{Z})$, 
therefore $q_I \in \mathbbm{Z}$.} $C^\star = q_I C^I$, while $\delta n_V$ and
$ \delta n_H$ count the vector and hypermultiplets which
become massless in the transition. The transition locus corresponds to
$h^\star = 0$. By virtue of eq. (\ref{2.4}) we find that $\Delta {\cal V}$ does not
contribute to the metric $a_{IJ}$ at the transition point. Hence $a_{IJ}$ and
also $g_{xy}$ will be continuous. 
But the first derivative of
$a_{IJ}$ with respect to $h^\star$ is not continuous due to the jump in the triple intersection numbers  indicated by eq. (\ref{3.1}). This implies that the
derivative of $g_{xy}$ and therefore the Christoffel connection
$\gamma^x_{~yz}$ is discontinuous at the transition locus.

Next we impose that the function $T$ defined in (\ref{2.34}) is constant. 
As we will see, this corresponds to a 
specific choice of the lapse function in our ansatz and leads
to a consistent solution.
If $T$ is constant  the Einstein equations decouple from the matter 
equations and can be solved analytically:
\bea\label{3.2}
\nonumber \alpha(t) & = & c_1 \, t + c_2 \, , \\
\beta(t) & = & - \, \frac{1}{3 \, c_1} \left( 3 \, c_1^2 - T \right) \, t + c_3 \, , \\
\nonumber \nu(t) & = & \frac{1}{3 \, c_1} \left( 6 \, c_1^2 + T \right) \, t + c_4 \, , 
\eea
where the $c_i$ are constants of integration. This solution does not depend on the choice of the vector and hypermultiplet scalar metric. It satisfies
\be\label{3.3}
3 \ad + \bd - \nd = 0 \, .
\ee
We read this as a condition which fixes the lapse function such that 
$T$ is constant with respect to the corresponding time variable.
Now we have to check whether this is consistent with the
scalar equations of motion. 
Substituting in our result,
these reduce to the standard geodesic equations with respect to the vector and
hypermultiplet scalar metrics:
\be\label{3.4}
\ddot{\phi}^x + \gamma^x_{~yz} \, \pd^y \, \pd^z  =  0 \, , \qquad 
\ddot{q}^X + \Gamma^X_{~YZ} \, \qd^Y \, \qd^Z  =  0 \, .
\ee
By taking the time derivative of $T$ and using the geodesic equations
above we find that $T$ is conserved, $\ft{d}{dt}T=0$, so that we indeed have a consistent
solution of the equations of motion.

%
%\be\label{3.5}
%\frac{d}{dt} T = 0 \, \qquad \mbox{for geodesic motion.}
%\ee
%
%This result justifies the assumption that $T$ is conserved. 

In order to prove  that the functions $\alpha(t), \beta(t)$ and $\nu(t)$ given by (\ref{3.2}) are smooth,
we only need to show that the piecewise constant function $T$ is continuous
at the transition point. Looking at the definition (\ref{2.34}), we find that $T$ contains the vector
and hypermultiplet scalar metrics as well as $\dot{\phi}^x$ and
$\dot{q}^X$. The metrics have already been shown to be continuous. 
To establish the continuity of the scalar fields, we observe that their
dynamics is governed by the geodesic equations (\ref{3.4}). These
differential equations are different at both sides of the transition
line because of the  discontinuity in the Christoffel symbols
$\gamma^x_{~yz}$. But since eq. (\ref{3.4}) is of second order, we can choose
our free constants of integration in such a way that the scalar fields
$\phi^x, q^X$ as well as their first derivatives $\pd^x, \dot{q}^X$ are
continuous at the transition locus. Then $T$ is continuous at
the transition line. This completes the proof of the statement given above.

Note that we only get a smooth solution if 
we choose a particular lapse function.
This choice is distinguished by the fact that 
the gravitational and scalar equations of motion decouple. Moreover, the time variable corresponding to our lapse function is the affine parameter of the geodesic equations on the moduli spaces. When studying cosmological solutions, however, we will use the standard cosmological time $\tau$, which amounts
to setting the lapse function to unity. 
In this parametrization the second derivatives of $\alpha(\tau)$ and
$\beta(\tau)$ are continuous, but have a kink when crossing the transition
locus. This frame dependence of the solutions resembles
the differences between the string frame and the Einstein frame 
familiar from string theory. 

\end{subsection}
\begin{subsection}{Cosmological solutions of the $\mathbbm{F_1}$-model }
We now turn to the cosmological solutions of the $\mathbbm{F}_1$-model. 
In order to have the standard parametrization used in cosmology
we now switch to the cosmological time $\tau$ so that the space-time metric (\ref{2.28}) becomes
\be\label{3.6}
\rmd s^2 = - \rmd \tau^2 + \e^{2 \alpha(\tau)} \rmd \vec{x}^2 + \e^{2 \beta(\tau)} \rmd y^2 \,  .
\ee 
Here and henceforth it is understood that all fields and the functions
$\alpha$ and $\beta$ now depend on $\tau$ and we use the
``dot'' to indicate derivatives with respect to $\tau$.
The corresponding equations of motion can be obtained from
(\ref{2.31}), (\ref{2.32}) and (\ref{2.33}) by setting $\nu = \dot{\nu} =  0$
and replacing $t \rightarrow \tau$. Note that the 
Einstein and matter equations do not decouple anymore. By taking certain
linear combinations, Einstein equations may now be written as: 
\bea\label{3.7}
\nonumber 3 \left( \ad^2 + \ad \, \bd \right) & = & T(\tau) \, , \\
3 \, \left( \ddot{\beta} +  \bd^2 + 2 \ad \bd -  \ad^2 \right)   &=& -   T(\tau) \, , \\
\nonumber 3 \left( \ddot{\alpha} + 2 \ad^2  \right) & = & - T(\tau) \, .
\eea
In this parametrization $T$ is no longer conserved along the integral curves of (\ref{2.32}) and (\ref{2.33}) and therefore depends on $\tau$. 
%This feature considerably complicates studying the Einsteins equations analytically. 
%Nevertheless, we can understand some qualitative features of these equations. 
%
\subsubsection*{Possibilities for inflation}
In order to discuss whether  this model allows inflation,
we first introduce the following scale factors:
\be\label{3.8}
\begin{split}
a & = \e^{\alpha} \, ,  \quad \dot{a} = \ad \, \e^{\alpha} \,, \quad \ddot{a} = \left( \ddot{\alpha} + \ad^2 \right) \e^{\alpha} \, , \\
b & = \e^{\beta} \,,  \quad \dot{b} = \bd \,  \e^{\beta} \,, \quad \ddot{b} = \left( \ddot{\beta} + \bd^2 \right) \e^{\beta} \, .
\end{split}
\ee
An expansion of space-time in the $\vec{x}$- and $y$-dimensions is characterized by $\dot{a} > 0$ and $\dot{b} > 0$, 
while accelerated expansion corresponds to $\ddot{a} > 0$ and $\ddot{b} > 0$, respectively.

Rewriting Einstein equations (\ref{3.7}) in terms of $\ddot{a}$ and $\ddot{b}$ and taking appropriate linear 
combinations leads to
\be\label{3.9}
\ddot{a}  = - \, \left( \frac{1}{3} \, T + \ad^2 \right) \, \e^{\alpha}  \, , \qquad
\ddot{b}  = \left( - T + 3 \ad^2 \, \right) \e^{\beta} .
\ee
These equations show that the Out-picture does not allow for accelerated expansion in the $\vec{x}$-dimensions, since $\ddot{a}$ is negative semi-definite. An accelerating phase in the $y$-dimension is possible and requires $3 \ad^2 > T$. 
%This effect, however, is based on splitting the 5-dimensional space-time into a 4-dimensional and a 1-dimensional piece and 
%is not induced by the dynamics of the CY moduli. 
%
\subsubsection*{Numerical Examples}
We will now study numerical solutions of the eqs. (\ref{2.31}), (\ref{2.32}) and (\ref{2.33}) parametrized by the time $\tau$. In this course we restrict the universal hypermultiplet sector of our model to the volume scalar $V$, setting $\sigma, \theta, \tau$ to zero, which can be done consistently. In this case the eq. (\ref{3.3}) for the hypermultiplet sector becomes
\be\label{3.10}
\ddot{V} - \frac{1}{V} \, \dot{V}^2 + \left( 3 \ad + \bd \right) \dot{V} = 0 \, .
\ee

We first focus on the dynamics of the vector multiplet scalar fields $U, W$. Their typical dynamics is shown in Fig. \ref{zwei}. Here the arrows point towards increasing values of $\tau$. But the equations of motion (\ref{2.31}), (\ref{2.32}) and (\ref{2.33}) are invariant under time reversal $\tau \rightarrow -\tau$ so that each trajectory has a time-reversed conterpart.
\begin{figure}[t]
\renewcommand{\baselinestretch}{1}
\epsfxsize=0.45\textwidth
\begin{center}
\leavevmode
\epsffile{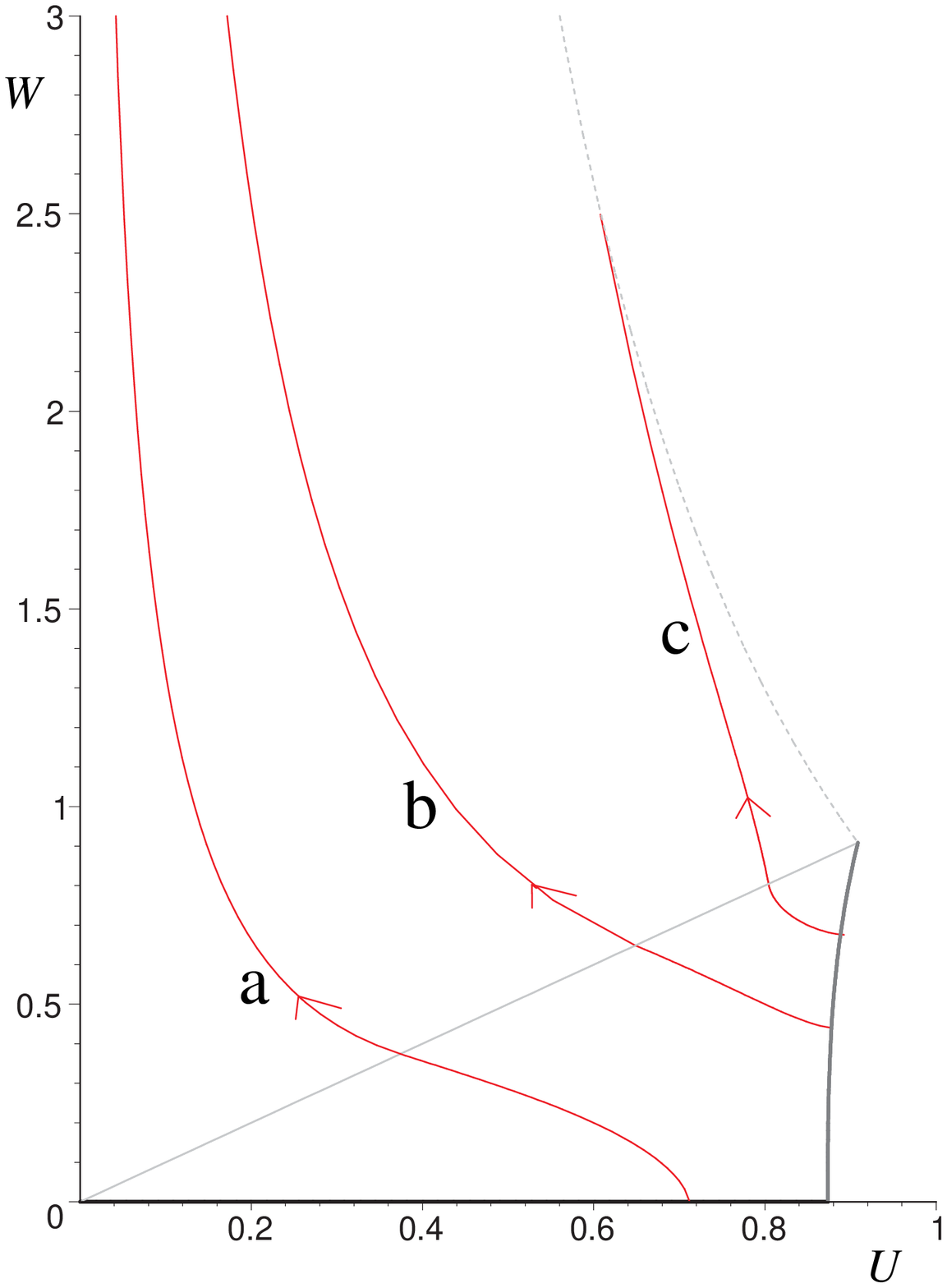} \, \, 
\epsfxsize=0.45\textwidth
\epsffile{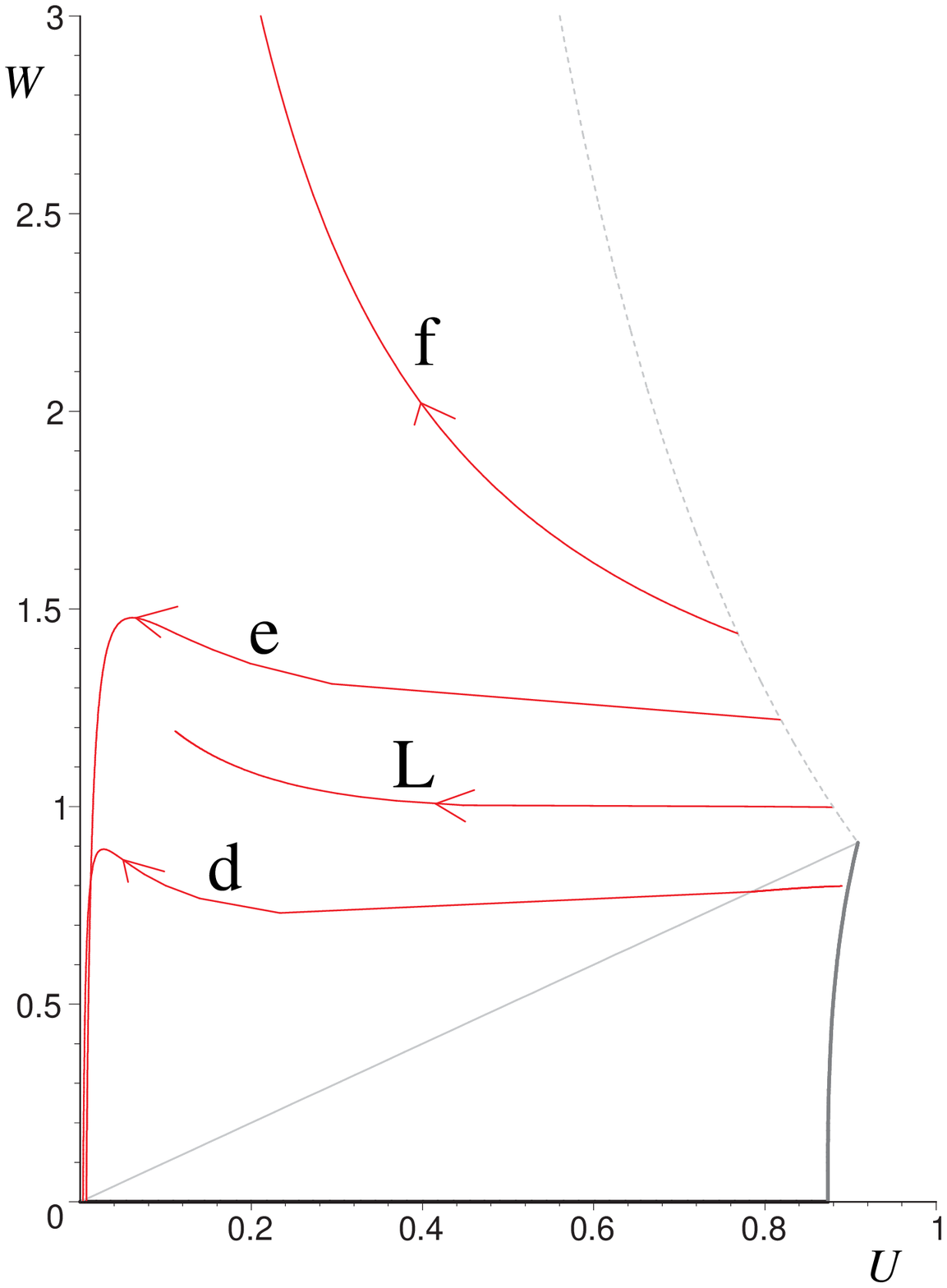}
\end{center}
\parbox[c]{\textwidth}{\caption{\label{zwei}{\footnotesize Illustrative examples for possible solutions of the Out-picture equations of motion with initial conditions given in Table \ref{t.2} in the Appendix A. The solutions ``a'' to ``f'' connect all possible boundaries of the extended K\"ahler cone. Solution ``L'' is an example for a solution that starts at the boundary and remains inside the moduli space for an exceptionally long time.}}}
\end{figure}

The vector multiplet scalar trajectories may be classified according to the boundaries where they start and end.  Fig. \ref{zwei} shows one example for every type of these trajectories. We find that for every possible pair of boundaries there is one class of trajectories which start at one of the boundaries and end at the other.\footnote{For the boundary $b_1$ this is not obvious from Fig. \ref{zwei}. The numerical analysis indicates that the trajectories which leave the plot at $W=3$ approach the boundary $b_1$ at large values of $W$. However, the numerical solutions are not conclusive about whether the boundary is reached at a finite of infinite value of $W$.} All boundaries can be reached in finite time which depends on the particular trajectory chosen. The solution ``L'' corresponds to a certain subclass of the types introduced above. Here the absolute value of the field derivatives decreases monotonically, so that the corresponding solution stays inside the vector multiplet moduli space for an exceptionally long time. None of the examples becomes singular while inside the extended K\"ahler cone. We further observe that all trajectories are one times differentiable at the flop line $U=W$. The behavior of the solutions at the boundaries of the vector multiplet scalar manifold will be discussed in more detail in the next subsection.

Let us now turn to the other fields appearing in the equations of motion. Examples of their characteristic behavior are illustrated in Fig. \ref{drei}. The initial conditions for these trajectories are given by the lines ``b'', ``d'' and ``L'' in Table \ref{t.2}.
\begin{figure}[p!]
\renewcommand{\baselinestretch}{1}
\mbox{\hspace*{0.14\textwidth} b \hspace*{0.28\textwidth} \, \,  d \hspace*{0.28\textwidth} \, \,  L \hspace*{0.14\textwidth}}
\begin{center}  
\leavevmode
\epsfxsize=0.30\textwidth
\epsffile{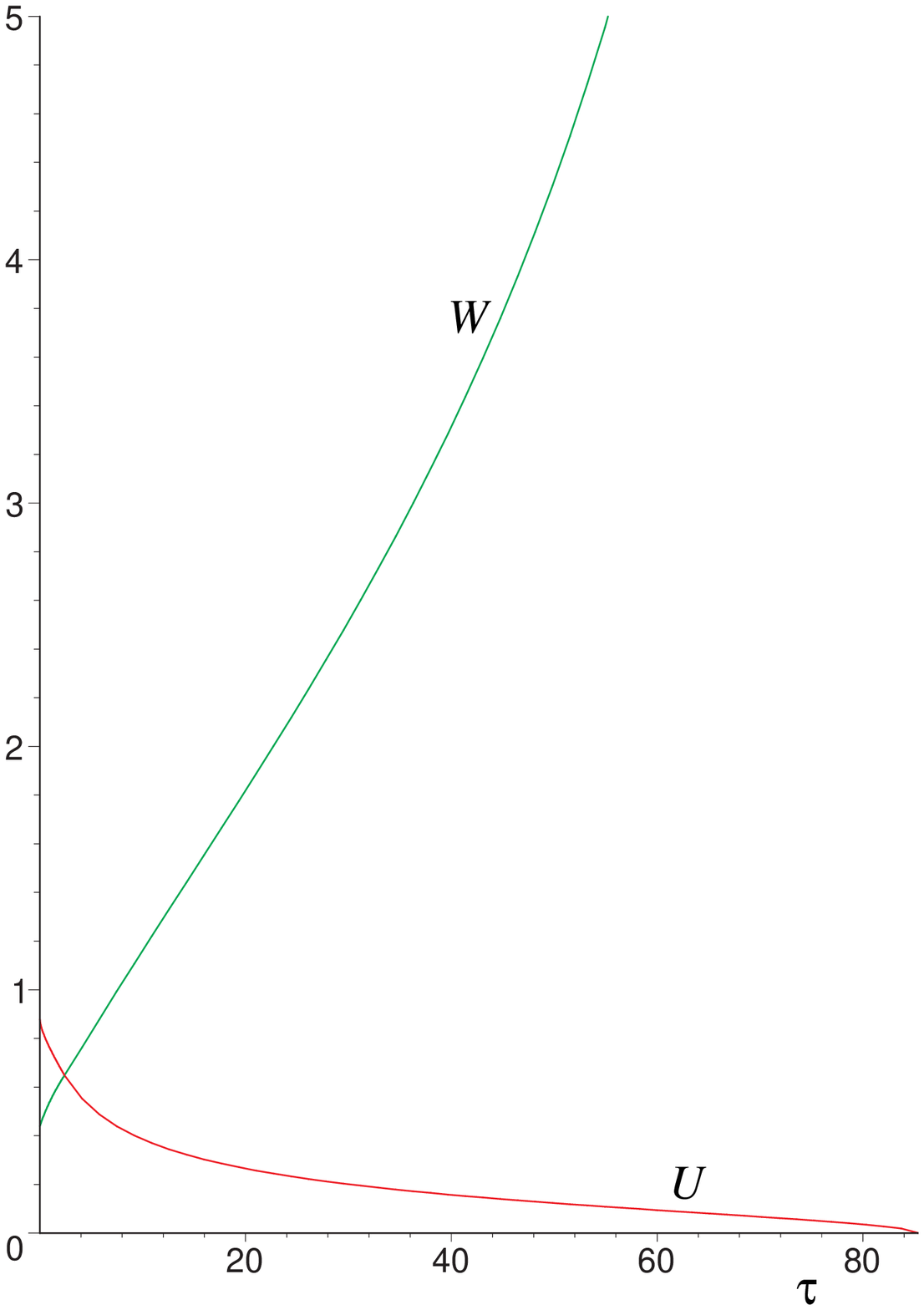} \, \, 
\epsfxsize=0.30\textwidth
\epsffile{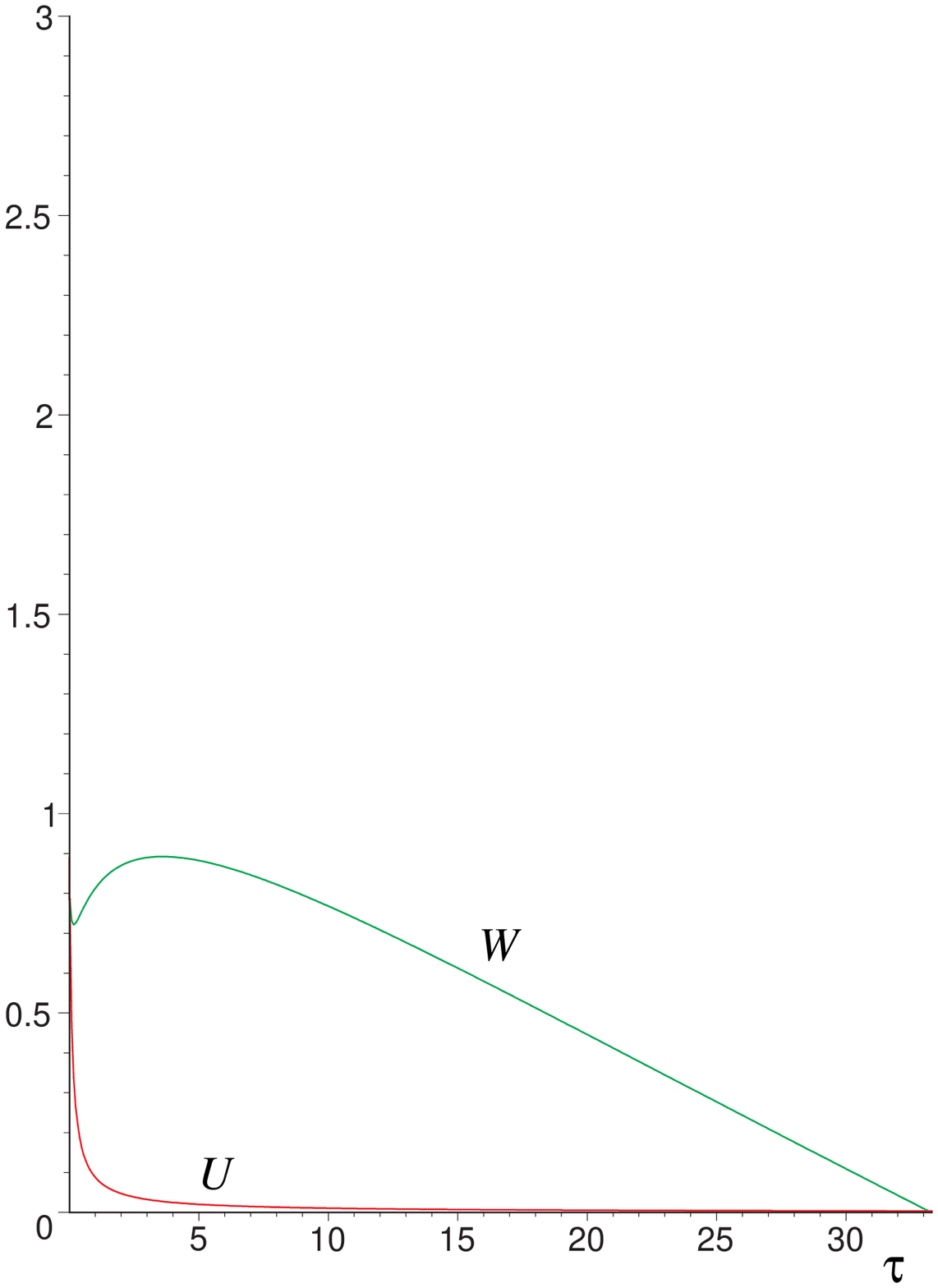} \, \, 
\epsfxsize=0.30\textwidth
\epsffile{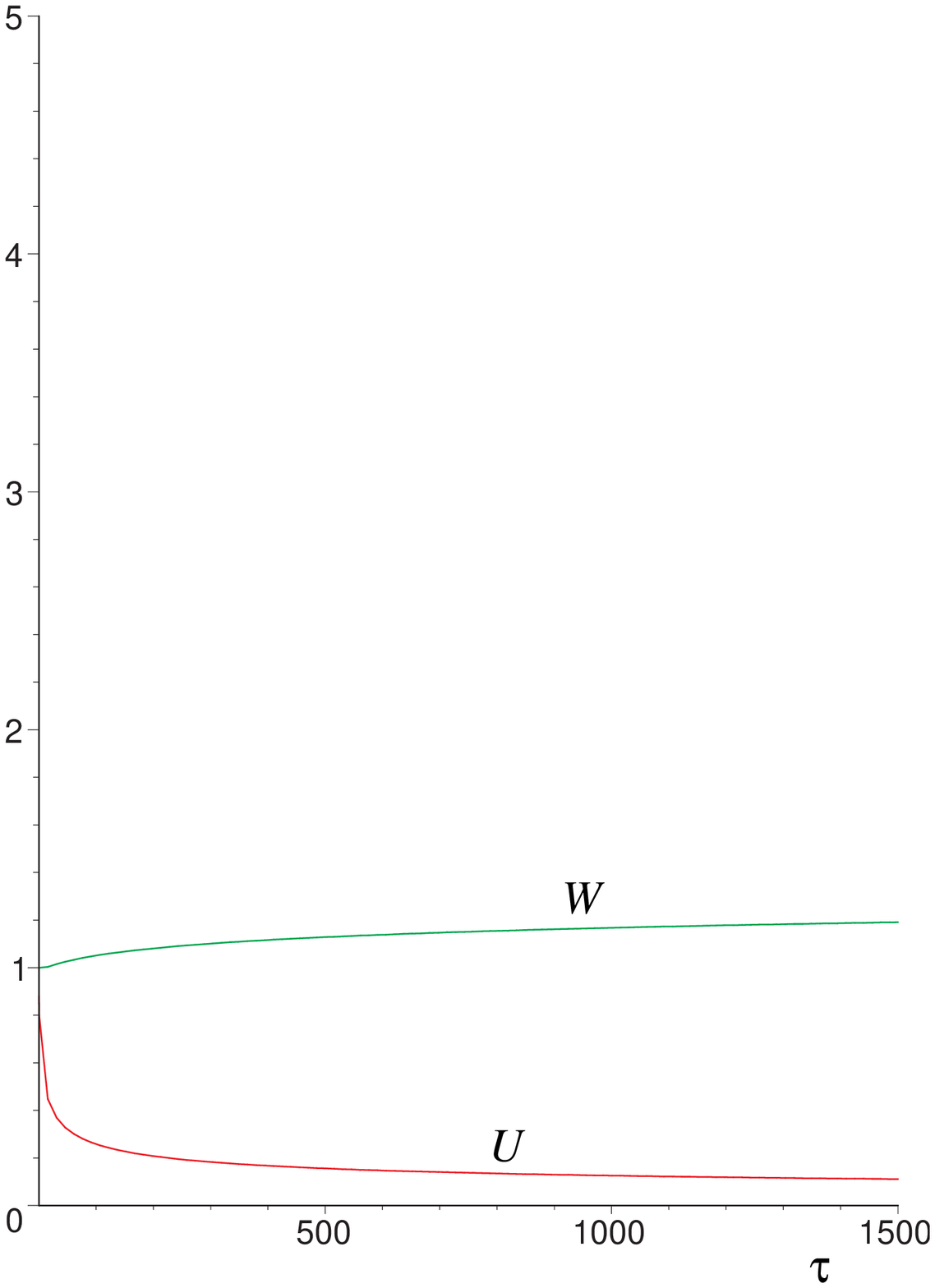}
%\end{center} 

%
\leavevmode
\epsfxsize=0.30\textwidth
\epsffile{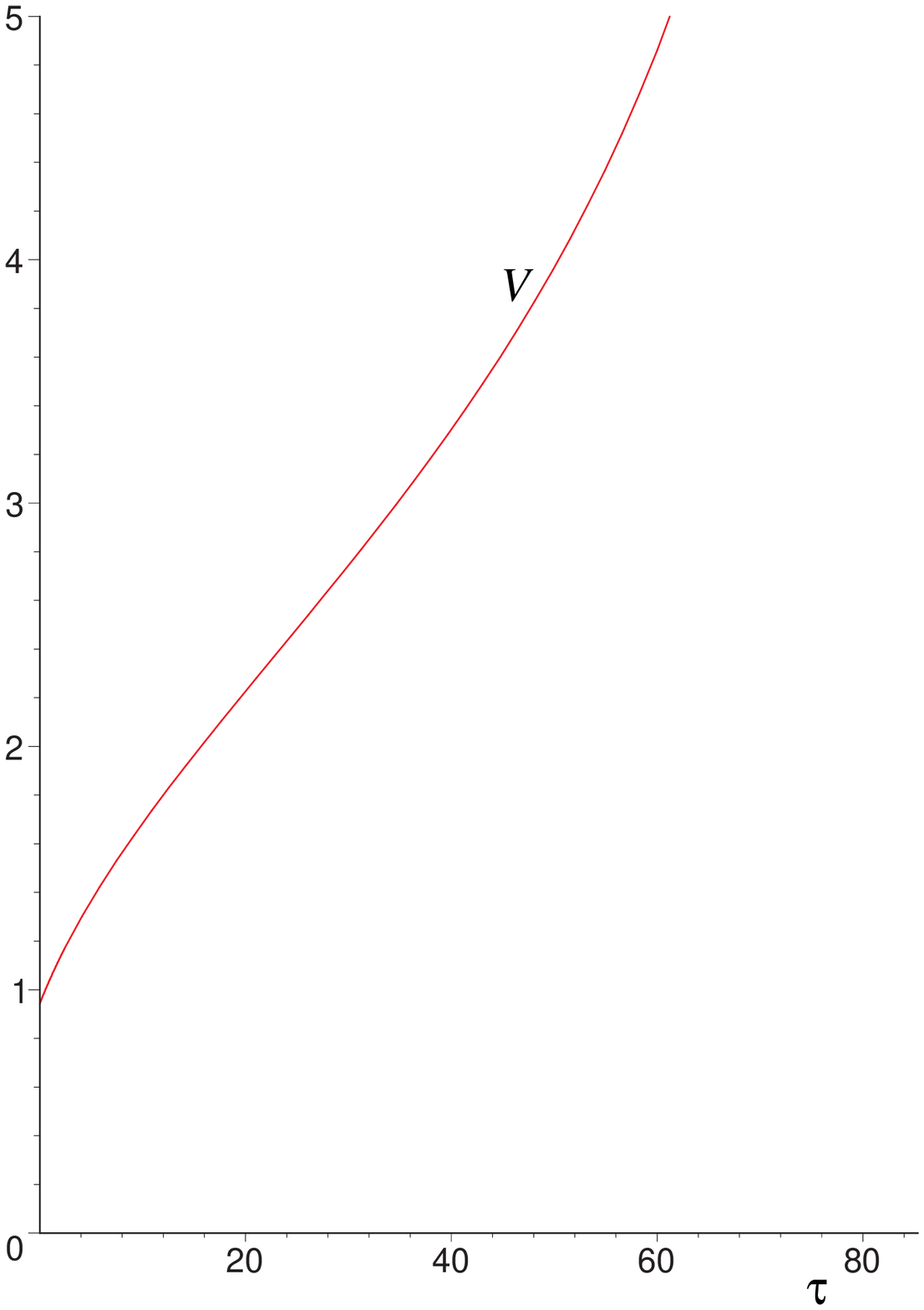} \, \, 
\epsfxsize=0.30\textwidth
\epsffile{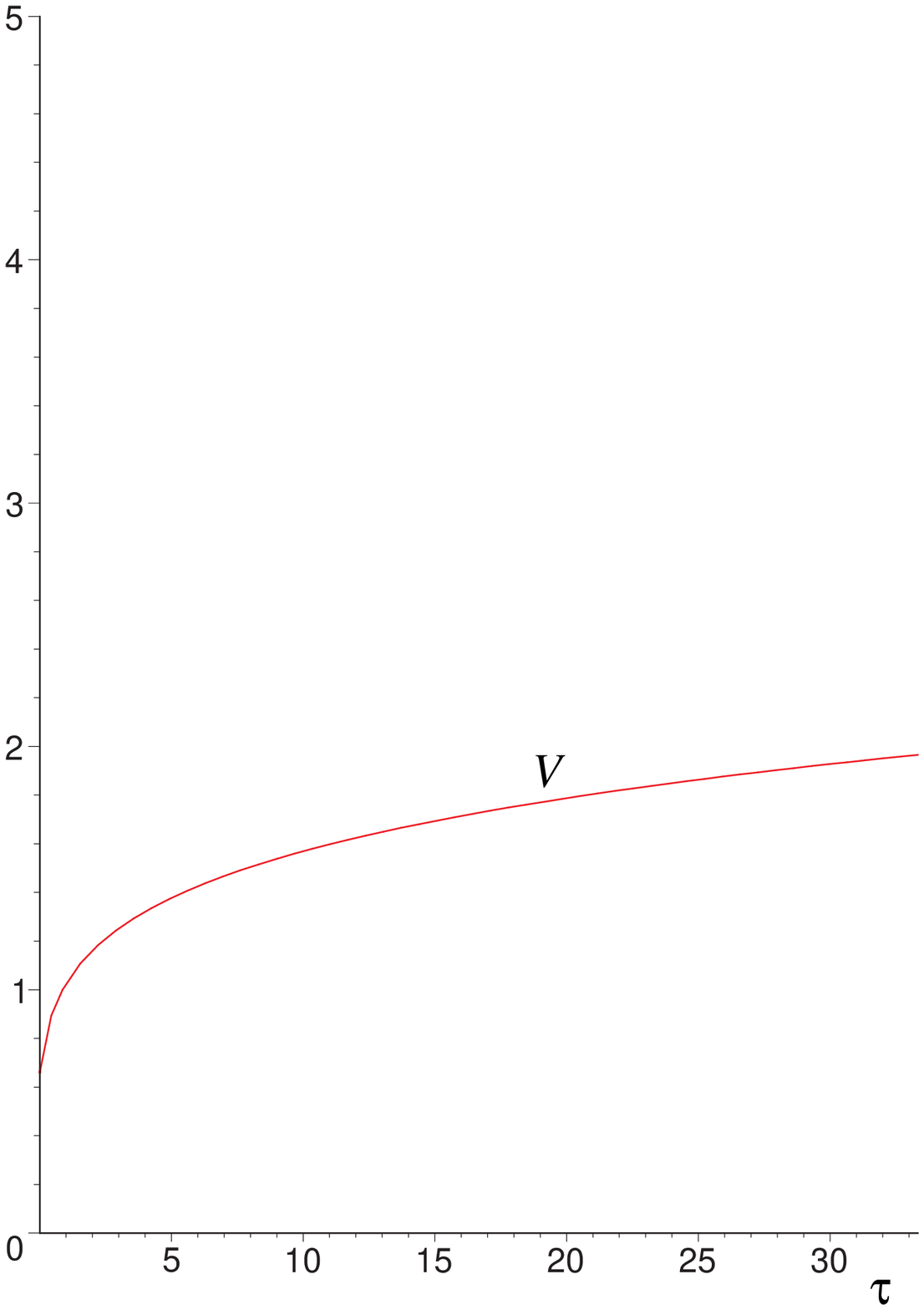} \, \, 
\epsfxsize=0.30\textwidth
\epsffile{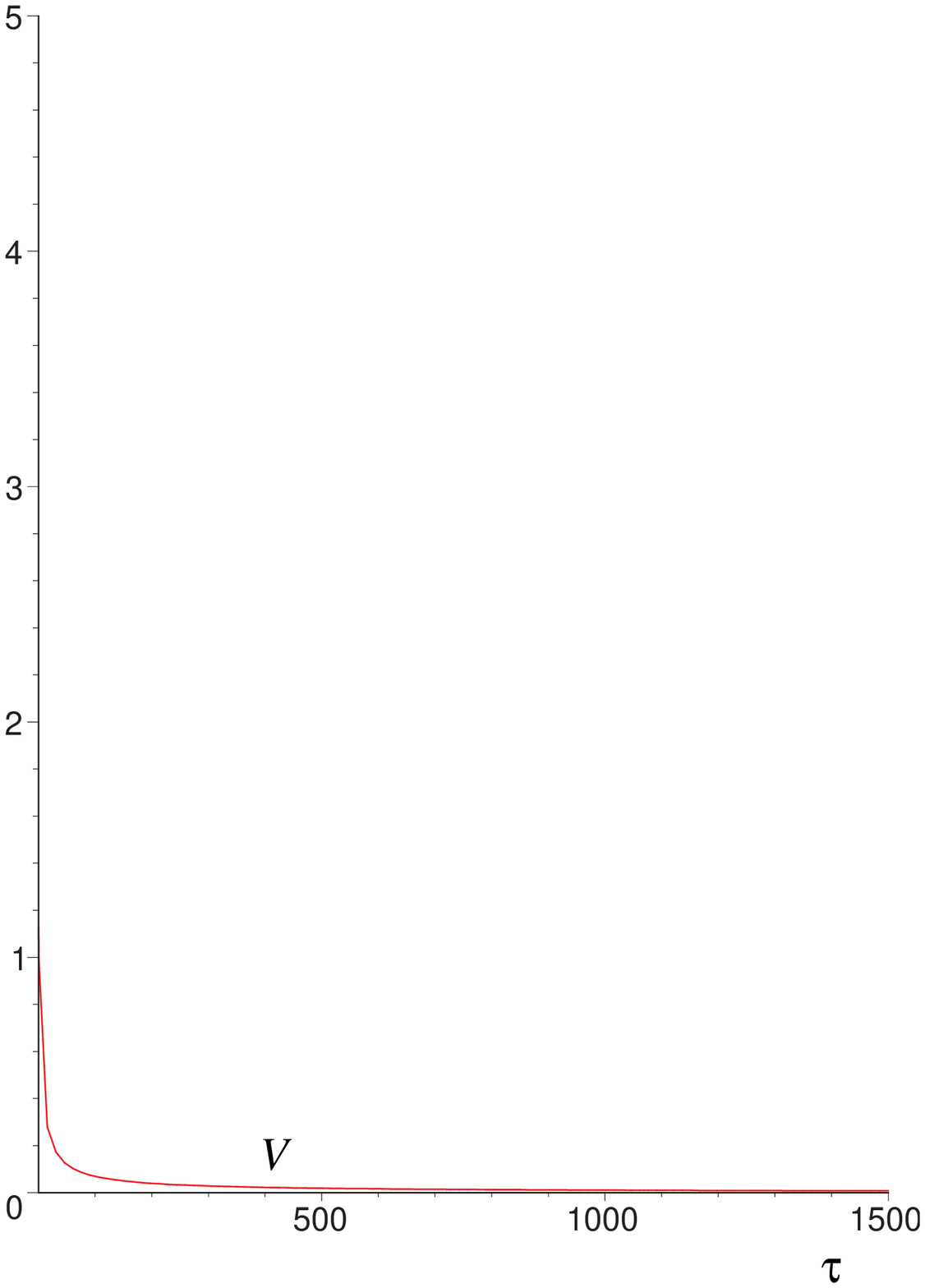}

%\begin{center}
\leavevmode
\epsfxsize=0.30\textwidth
\epsffile{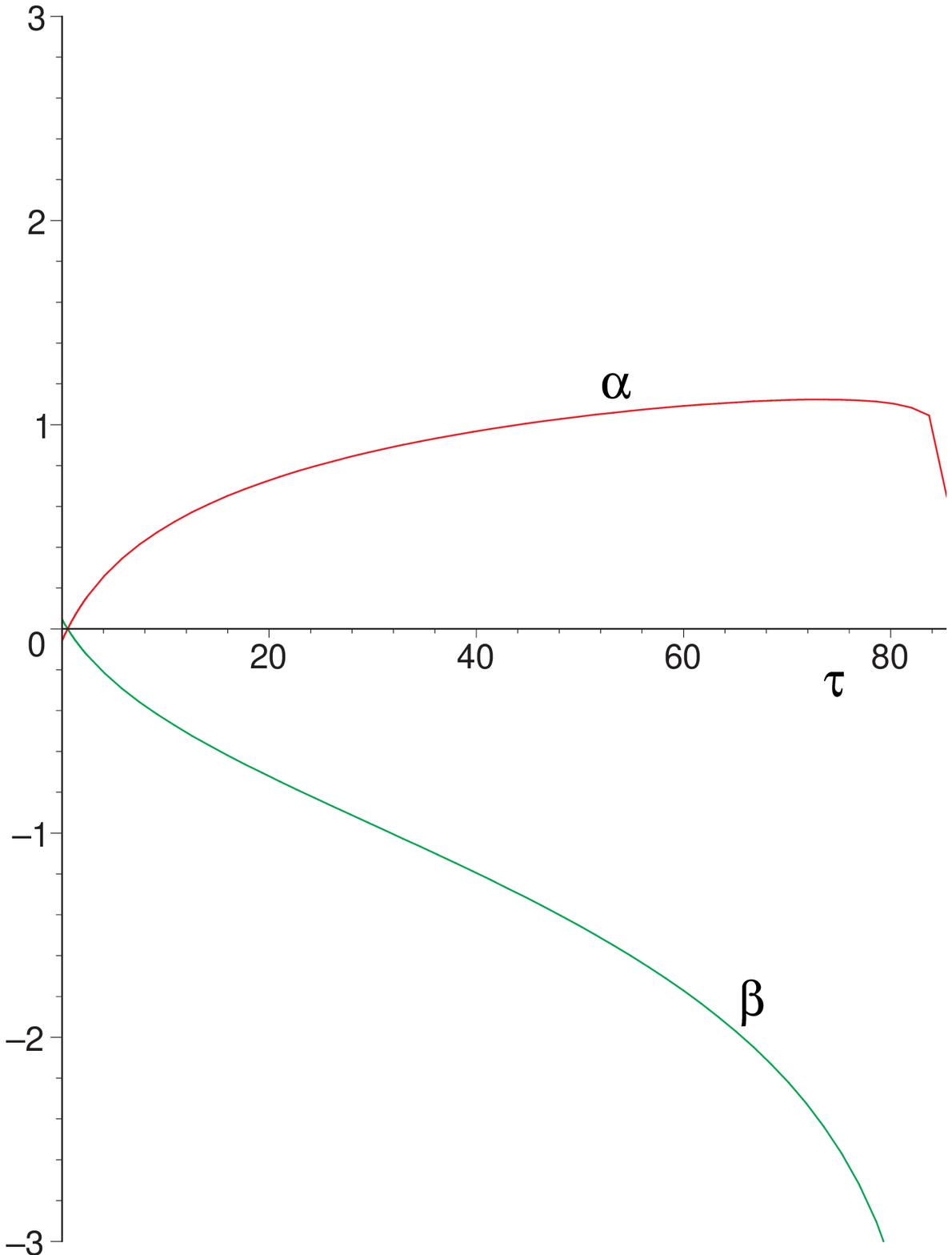} \, \, 
\epsfxsize=0.30\textwidth
\epsffile{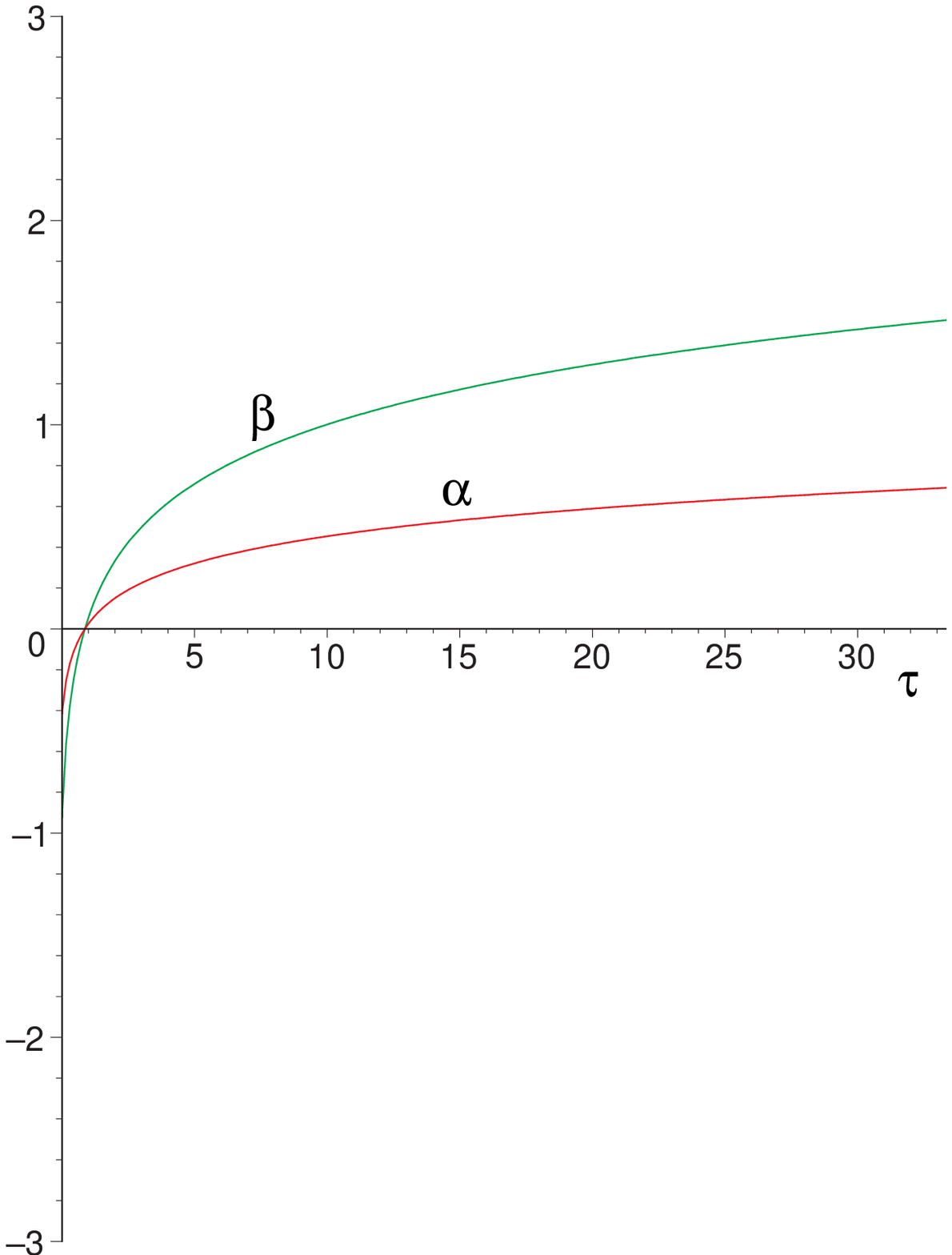} \, \, 
\epsfxsize=0.30\textwidth
\epsffile{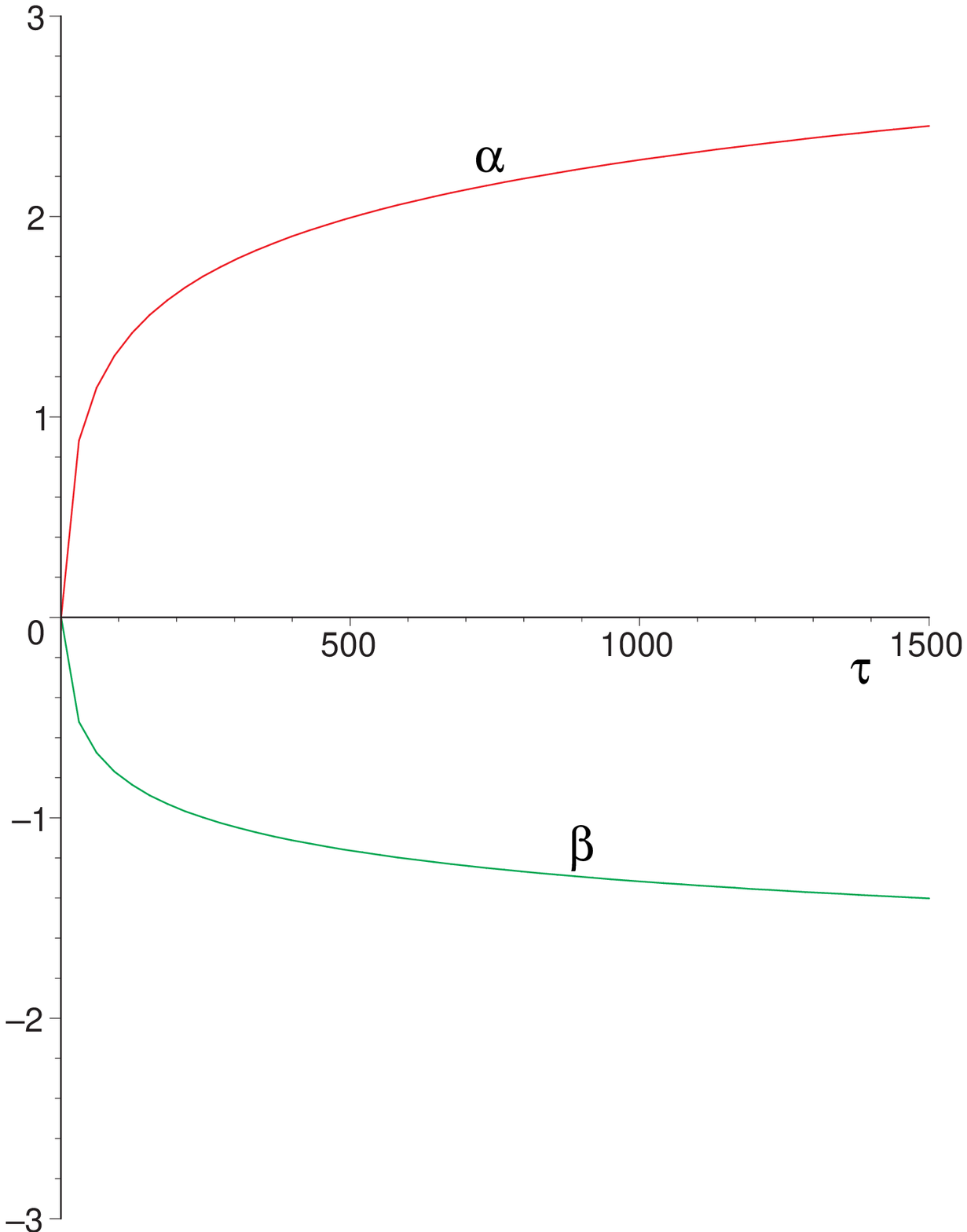}
%\end{center} 
%\begin{center}
%\leavevmode
\end{center} 
\parbox[c]{\textwidth}{\caption{\label{drei}{\footnotesize Three examples of solutions illustrating the typical behaviors of cosmological solutions in the Out-picture. The solutions ``b'' and ``d'' show the usual run-away behavior of the moduli fields while the ``L'' solution stays inside the extended K\"ahler cone for an exceptionally long time.}}}
\end{figure}

The first row of Fig. \ref{drei} shows the behavior of the vector multiplet scalar fields $U(\tau), W(\tau)$. The first two diagrams display the solutions ``b'' and ``d'' which both start at the boundary $b_3$ and end at $b_1$ and $b_2$, respectively.  
The second row shows the behavior of the CY volume $V$. Here we observe
that $V(\tau)$ either  increases or decreases  monotonically, depending on whether the
solution started with $\dot{V}(0) > 0$ or $\dot{V}(0) < 0$. Choosing
$\dot{V}(0) = 0$ leads to a constant volume. We find that $V(\tau)$ does not become
singular, $V(\tau)=0$ or $V(\tau) = \infty$, while the solution is inside the extended K\"ahler cone.
The evolution of the scale factors $\alpha(\tau)$ and $\beta(\tau)$ is shown
in the third row of Fig. \ref{drei}. All solutions have $\ddot{\alpha} \le 0$. Their behavior is governed almost entirely by their
initial conditions $\ad(0), \bd(0)$. For positive (negative) initial values
the solutions monotonically increase (decrease). The only exception to this
rule arises when the solution approaches the boundary $b_1$. In this case the
kinetic term $T$ becomes large due to the vector multiplet scalar metric
$g_{xy}^{\rm (III)}$ developing an infinite eigenvalue. This induces a rapid
decrease of both $\alpha(\tau)$ and $\beta(\tau)$. But the solution only
becomes singular at the boundary $b_1$. Therefore we do not
encounter any  space-time singularities, while the vector multiplet scalars
are {\it inside} the extended K\"ahler cone of the model.

\end{subsection}
\begin{subsection}{Behavior of solutions at the boundaries of moduli space}
After studying the general behavior of the vector multiplet scalar fields in
the previous subsections, we will now discuss their dynamics close to the
boundaries of the extended K\"ahler cone shown in the first diagram of
Fig. \ref{eins}. The case of an internal boundary (flop),
where the vector multiplet scalar metric $g_{xy}$ is regular and continuous,
was already discussed in subsection 3.1. Since boundaries where $\det(g_{xy})$ is infinite, zero, or regular, quite generally appear in
vector multiplet moduli spaces and are not limited to our particular model, we
will label such boundaries as being of type I, II, and III, respectively.\footnote{In comparison to \cite{Wilson} type I boundaries correspond to the cubic cone, type II boundaries are associated with type II contractions, and type III boundaries are related to type I or type III contractions.}  

In this subsection we will focus on the behavior
of our solutions close to the
boundaries $b_1$ and $b_2$, $b_3$ where $g_{xy}$ has an infinite or zero
eigenvalue, respectively. Here we observe that the Christoffel connection diverges when the solutions approach these boundaries. This singularity completely dominates the behavior of the solution in these regions.  We expand the connection piece around the singularity in a Laurent series and keep the leading order term only. The resulting differential equations can be solved analytically and the corresponding expressions properly describe the behavior of the solutions close to the boundary.
\subsubsection*{The boundary $b_1$}
We start with the type I boundary $b_1$ which is given by $ U = 0$. Here the metric $g_{xy}^{\rm (III)}$ has an infinite eigenvalue. Expanding the Christoffel symbols for small $U$ we obtain
\be\label{1.1}
\begin{array}{ll}
\gamma^U_{~~UU} = - \, \frac{1}{U} + \cO(U) \, , \quad &
\gamma^W_{~~WW} = \cO(U^0) \, , \\
\gamma^U_{~~WW} = \cO(U^5) \, , \quad &
\gamma^W_{~~UU} = - \, \frac{W}{4 \, U^2} + \cO(U^{-1}) \, , \\
\gamma^U_{~~WU} = \cO(U^4) \, , \quad &
\gamma^W_{~~WU} = \frac{1}{2 U} + \cO(U^0) \, .
\end{array}
\ee
Assuming that $W$ and $\dot{W}$ are of order one, the equation for
$U(\tau)$ decouples, 
\be\label{1.2}
\ddot{U}(\tau) - \frac{1}{U(\tau)} \dot{U}(\tau)^2 = 0 \, ,
\ee
and we can determine the behavior of $U(\tau)$ close to $U = 0$:
\be\label{1.3}
U(\tau) = c_1 \, \e^{c_2 \tau} \, .
\ee
Here $c_1$ and $c_2$ are the constants of integration, which are fixed by the initial conditions. In order for the solution to be inside region III and $U(0) \ll 1$ we need $0 < c_1 \ll 1$. The condition that $U(\tau)$ approaches the boundary fixes $c_2 < 0$. 
Substituting this result into the equation for $W(\tau)$, we find 
\be\label{1.4}
\ddot{W}(\tau) + \frac{c_2}{2} \dot{W}(\tau) + \frac{(c_2)^2}{4} W(\tau) = 0 \, .
\ee
This is the equation for a damped harmonic oscillator, which has the solution
\be\label{1.5}
W(\tau) = 
c_3 \, \e^{- \frac{c_2}{4} \tau} \sin \left( \frac{\sqrt{3} c_2}{4} \, \tau \right)
+ c_4 \, \e^{- \frac{c_2}{4} \tau} \cos \left( \frac{\sqrt{3} c_2}{4} \, \tau \right) \, .
\ee
Here $c_3$ and $c_4$ are determined by the initial conditions for $W(\tau)$, while $c_2$ enters from the equation for $U(\tau)$. Since $c_2 < 0$ this solution describes harmonic oscillations with exponentially growing amplitude. According to the initial values $c_3$ and $c_4$ we can distinguish 3 different types of behavior:
\begin{enumerate}
\item The values $c_2$, $c_3$ and $c_4$ are such that the solution monotonically decreases and reaches $W = 0$ in a finite time.
\item The solution first oscillates away from $W = 0$ before running into the boundary $W = 0$.
\item The solution grows rapidly and leaves the region where $W(\tau)$ and $\dot{W}(\tau)$ are small compared to $\frac{1}{U(\tau)}$. Here our approximation breaks down and the solutions (\ref{1.3}) and (\ref{1.5}) are no longer valid.
\end{enumerate}
This implies that the solutions 1) and 2) do not run into the boundary $b_1$, since eq. (\ref{1.3}) shows they cannot reach this boundary in a finite time. Instead they run into the boundary $b_2$ where $W = 0$. For the solutions 3) the analytic expression allows no conclusions. 
\begin{figure}[t]
\renewcommand{\baselinestretch}{1}
\epsfxsize=0.3\textwidth
\begin{center}
\leavevmode
\epsffile{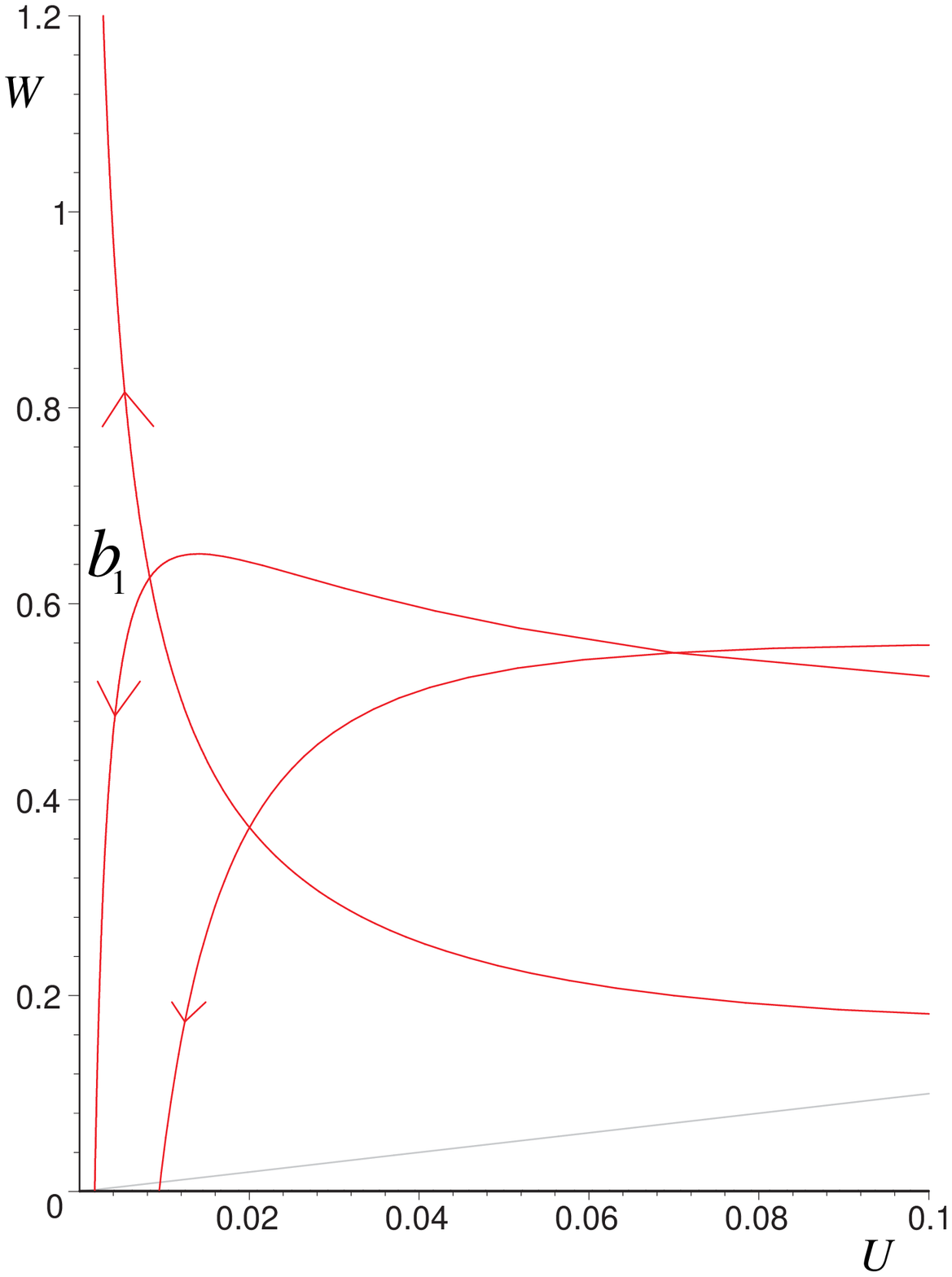} \, \, 
\epsfxsize=0.3\textwidth
\epsffile{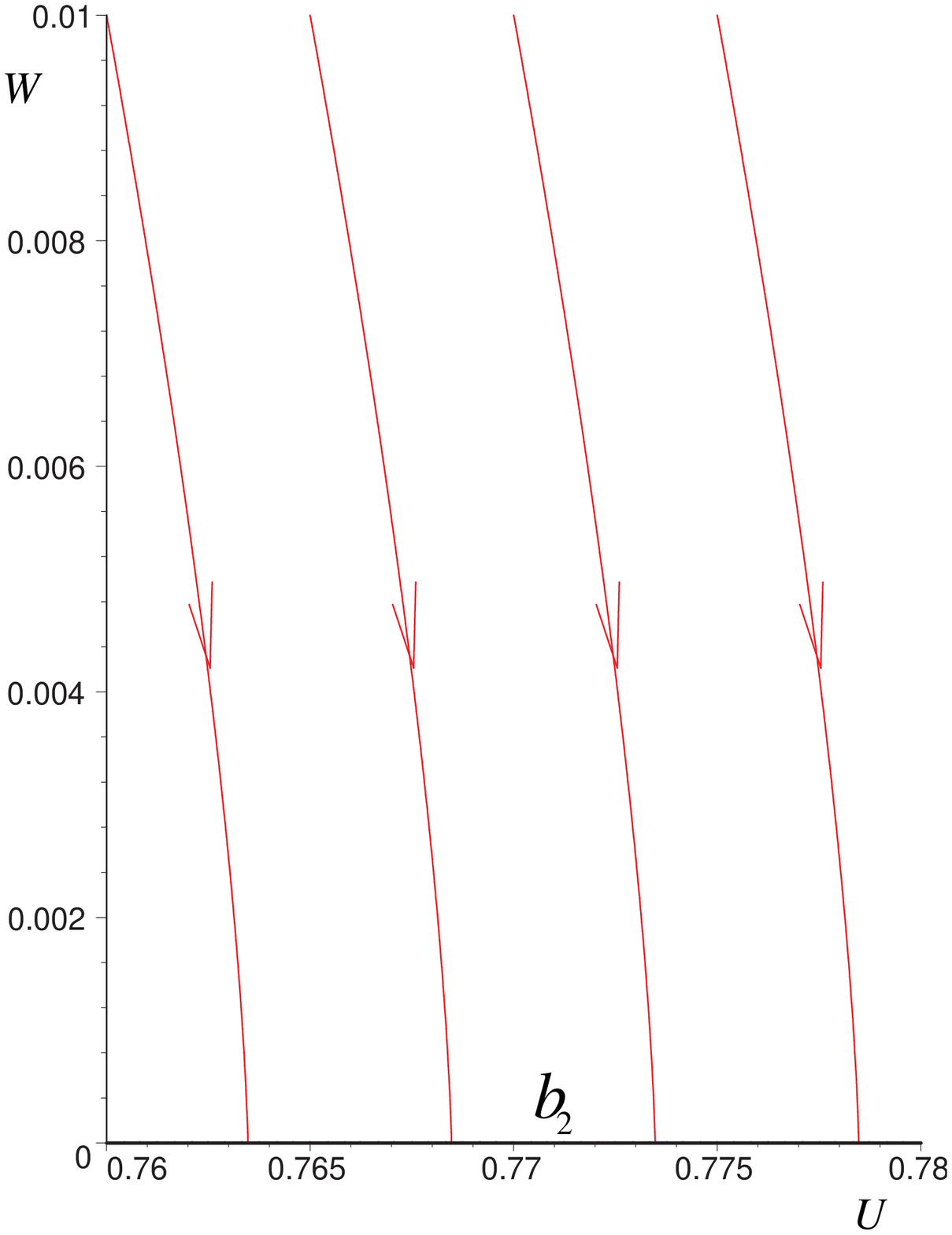} \, \, 
\epsfxsize=0.3\textwidth
\epsffile{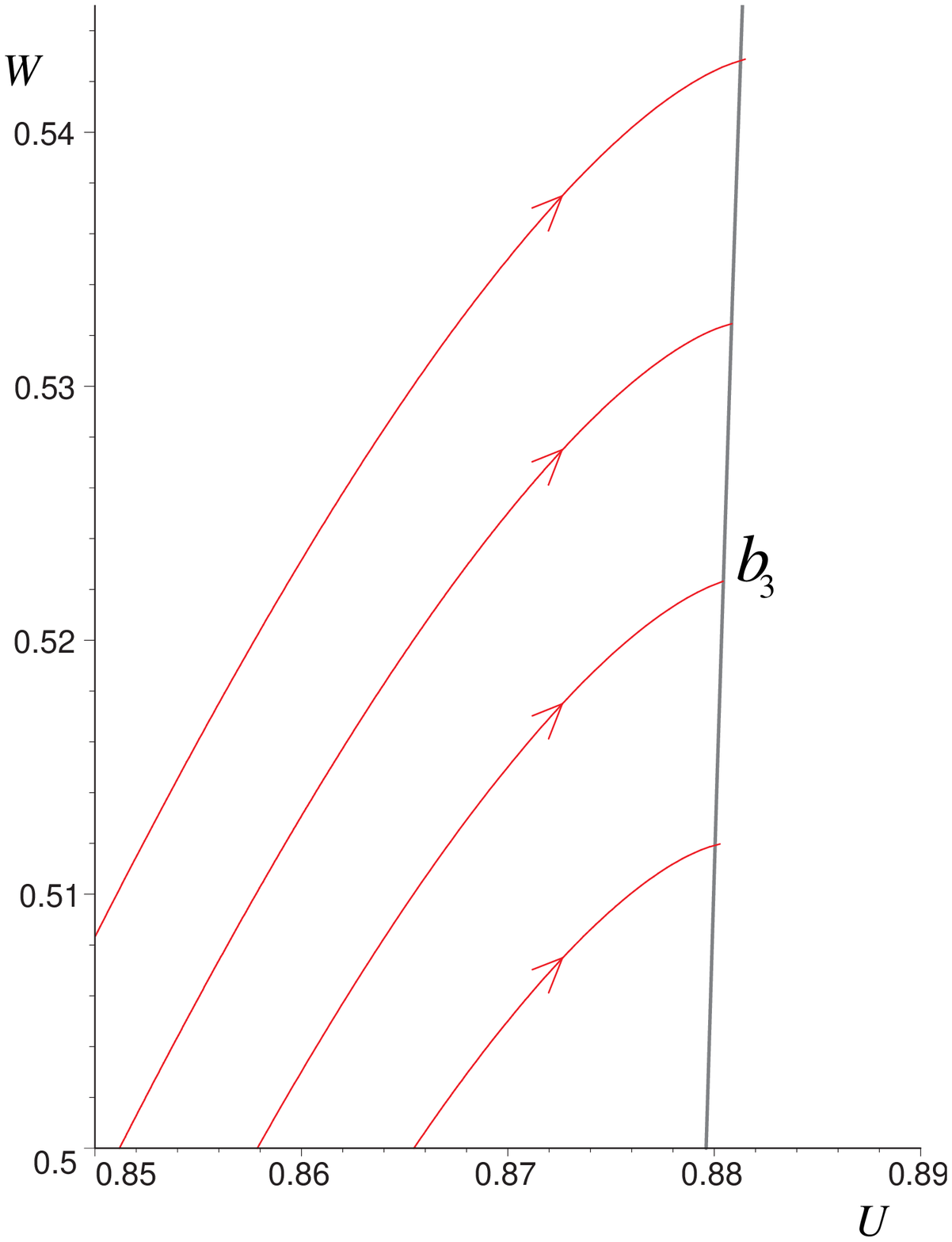}
\end{center}
\parbox[c]{\textwidth}{\caption{\label{vier}{\footnotesize Numerical investigation of the behavior of solutions close to the boundaries ``$b_1$'', ``$b_2$'' and ``$b_3$''. The initial conditions for the trajectories are given in Table \ref{t.4} in Appendix A. The arrow points in the direction of increasing value of $\tau$. These example solutions confirm the qualitative features found analytically.}}}
\end{figure}

Numerical examples for the solutions of the full equations of motion close to $b_1$ are shown in the first diagram of Fig. \ref{vier}. Here we give one example for all 3 types of solutions discussed above, confirming the qualitative behavior found in our analytic investigation.
\subsubsection*{The boundary $b_2$}
Let us now turn to the boundary $b_2$ which is of type II and given by the line $ 0 < U < \left( \frac{2}{3} \right)^{1/3}$, $W = 0$. At this line the metric $g^{\rm (II)}_{xy}$ has a zero eigenvalue along the $W$-direction. Expanding the corresponding Christoffel symbols around $W = 0$, we obtain
\be\label{1.6}
\begin{array}{ll}
 \gamma^U_{~~UU}  = - \, \frac{9 U^6 - 3 U^3 + 16}{U \left( 3 U^3 - 8 \right) \left( 3 U^3 - 2 \right)} + \cO(W^3) \, , \quad  & 
\gamma^W_{~~WW} = \frac{1}{2 W} + \cO(W^2)  \, , \\
 \gamma^U_{~~WW}  = - \, \frac{U}{4} W + \cO(W^4) \, ,  &
\gamma^W_{~~UU}  =  \cO(W^{1}) \, , \\

\gamma^U_{~~WU} = \cO(W^2)  \, ,  &  
\gamma^W_{~~WU}  = \cO(W^3) \, .
\end{array}
\ee
Assuming that $\dot{U}(\tau)$ is of order one, we can write down the equation of motion for $W(\tau)$ close to $W=0$:
\be\label{1.7}
\ddot{W}(\tau) + \frac{1}{2 W(\tau)} \, \dot{W}(\tau)^2 = 0 \, .
\ee
This equation has the solution
\be\label{1.8}
W(\tau) = \left( \frac{3}{2} \left( c_1 \tau + c_2 \right) \right)^{2/3} \, , \quad \dot{W}(\tau) = \frac{c_1}{\sqrt{W(\tau)}} \, .
\ee
In order for $W(\tau)$ to run inside the region II and to approach the boundary, we take $c_2 > 0$ and $c_1 < 0$. Then eq. (\ref{1.8}) indicates that the solution reaches $W(\tau) = 0$ in a finite time $\tau_b = - \frac{c_2}{c_1} > 0$. Looking at $\dot{W}(\tau)$ we see that $\dot{W}(\tau) \rightarrow - \infty$ as $\tau \rightarrow  \tau_b$. Substituting the solution for $W(\tau)$ into the equation of motion for $U(\tau)$ gives
\be\label{1.10}
\ddot{U}(\tau) = \frac{9 U(\tau)^6 - 3 U(\tau)^3 + 16}{U(\tau) \left( 3 U(\tau)^3 - 8 \right) \left( 3 U(\tau)^3 - 2 \right)} \, \dot{U}(\tau)^2 + \frac{(c_1)^2}{4} U(\tau) \, . 
\ee
Since the RHS of this equation does not become singular as $W(\tau) \rightarrow 0$,  $\dot{U}(\tau)$ is finite. The result that  $ \dot{U}(\tau)$ is finite while $\dot{W}(\tau) \rightarrow -  \infty$ as $\tau \rightarrow \tau_b$ implies that the corresponding trajectories become orthogonal to the  line $b_2$ when $W(\tau) \rightarrow 0$. There is no mechanism preventing the solution from reaching the boundary in a finite time.

This behavior is also found when studying numerical solutions of the full equations of motion. Examples of such solutions which run into the boundary $b_2$ are shown in the second diagram of Fig. \ref{vier}. These solutions exactly match the analytic behavior found above.
\subsubsection*{The boundary $b_3$}
The type II boundary $b_3$ is given by the line $6 + W^3 - 9 U^3 = 0$. Here the metric $g_{xy}^{\rm (II)}$ has a zero eigenvalue in the $U$-direction. 
In order to analyze the behavior of solutions at this boundary analytically, we use coordinates $ W$ and 
\be\label{1.12}
\epsilon = 6 + W^3 - 9 U^3,
\ee
where $\epsilon \ge 0$ measures the distance from the boundary $b_3$. The metric $g_{xy}^{\rm (II)}$ then becomes
\be\label{1.13}
g_{xy}^{\rm (II)}\left( \epsilon, W \right) = \left[ 
\begin{array}{cc} 
g_{\epsilon \epsilon} & g_{\epsilon W} \\ 
g_{W \epsilon} & g_{WW} 
\end{array} \right] \, ,
\ee
where
\bea\label{1.14}
\nonumber
g_{\epsilon \epsilon} & = & \frac{\epsilon \left( 6 + W^3 \right)}{18 \left( 6 + W^3 - \epsilon \right)^2 \,  \left( 18 + 3 W^3 + \epsilon \right) } \, , \\
g_{\epsilon W} & = & - \, \frac{\epsilon^2 \,  W^2 }{18 \left( 6 + W^3 - \epsilon \right)^2 \,  \left( 18 + 3 W^3 + \epsilon \right) } \, , \\
\nonumber 
g_{WW} & = & \frac{W \left( 648 + 216 W^3 + 18 W^3 - 30 W^3 \epsilon + 6 \epsilon^2 + \epsilon^3 \right)}{18 \left( 6 + W^3 - \epsilon \right)^2 \,  \left( 18 + 3 W^3 + \epsilon \right) } \, .
\eea
Calculating the Christoffel symbols and expanding around $\epsilon = 0$, we obtain
\be\label{1.15}
\begin{array}{ll}
 \gamma^\epsilon_{~~\epsilon \epsilon}  =  \frac{1}{2 \epsilon} + \cO(\epsilon^2) \, , \quad & 
\gamma^W_{~~WW} = - \, \frac{W^3 - 3}{W \left( 6 + W^3 \right)} + \cO(\epsilon) \, , \\ 
 \gamma^\epsilon_{~~WW} = \cO(\epsilon^0) \, , &
\gamma^W_{~~\epsilon \epsilon}  =  - \, \frac{W}{108 \left( 6 + W^3 \right)^2 } \epsilon + \cO(\epsilon^2) \, , \\
\gamma^\epsilon_{~~W \epsilon} = \cO(\epsilon^0) \, , &
 \gamma^W_{~~W \epsilon} = \cO(\epsilon^2)  \, .
\end{array}
\ee
Assuming that $\dot{W}(\tau)$ is of order one and $W(\tau) > 0$ gives the following leading order differential equation for $\epsilon(\tau)$ close to the boundary:
\be\label{1.16}
\ddot{\epsilon}(\tau) + \frac{1}{2 \epsilon(\tau)} \, \dot{\epsilon}(\tau)^2 = 0 \, ,
\ee
solved by
\be\label{1.17}
\epsilon(\tau) =  \left( \frac{3}{2} \left( c_1 \tau + c_2 \right) \right)^{2/3} \, , \quad \dot{\epsilon}(\tau) = \frac{c_1}{\sqrt{\epsilon(\tau)}} \, .
\ee
Imposing that we start inside region II and run towards the boundary fixes $c_2 > 0$ and $c_1 < 0$. Note that this is exactly the same solution as we obtained for $W(\tau)$ at the boundary $b_2$. This in particular implies that $\epsilon = 0$ can be reached in finite time $ \tau_b = - \, \frac{c_2}{c_1} > 0 $ and $\dot{\epsilon}(\tau) \rightarrow - \infty$ when the solution approaches the boundary. 

Substituting this solution into the differential equation for $W(\tau)$ yields
\be\label{1.18}
\ddot{W}(\tau) = \frac{W(\tau)^3 - 3}{W(\tau) \left(6 + W(\tau)^3 \right)} \dot{W}(\tau)^2 + \frac{(c_1)^2 W(\tau)}{108 \left(6 + W(\tau)^3 \right)^2} \, .
\ee
As in the $b_2$ case we observe that the RHS of this equation does not become singular at \mbox{$\epsilon(\tau) = 0$}. This indicates that it is the part of $\dot{\epsilon}(\tau)$ pointing in the $U$-direction which becomes infinite as we approach the boundary, while $\dot{W}(\tau)$ stays finite. Hence the trajectory $\{ U(\tau), W(\tau) \}$ turns parallel to the $U$-direction when reaching the boundary $b_3$.

This behavior of the solutions of the full equations of motion can also be observed numerically. The third diagram of Fig. \ref{vier} shows a family of trajectories running into the boundary $b_3$.  Their behavior is in exact agreement with the analytic analysis above.

Summarizing the results of  this subsection, we find that the boundaries of
type II, $b_2$ and $b_3$, can be reached in finite time. At these boundaries
the metric develops a zero eigenvalue. The derivatives of the scalar field
associated with this zero eigenvalue diverge while the derivatives of the
other scalar fields remain finite.  As a consequence, the trajectories become
parallel to the direction associated with the zero eigenvalue when reaching
the boundary. Furthermore, the divergence of the scalar field derivatives exactly cancels the zero eigenvalue of the scalar field metric, so that the kinetic energy $T$ is finite at these boundaries.
For the type I boundary $b_1$, associated with a
space-time singularity,\footnote{Here we assume that the boundary is approached with finite $\Ud(\tau)$.} we observe that there is a mechanism preventing the
solutions from running into this boundary due to a
singularity in the Christoffel symbols. However, this analysis does not cover trajectories like ``a'' and ``b'' shown in Fig. \ref{eins}, as these correspond to the case 3) found for the boundary $b_1$.

These results imply that in all cases which could be treated analytically, no space-time singularity occurs.  Since the behavior of our solutions is universally determined by the type of degeneracy of the scalar field metric, we conjecture that these results will hold for any type I or type II boundary.
\end{subsection}

\end{section}
\begin{section}{Solutions in the In-picture}
After analyzing the behavior of cosmological solutions in the Out-picture, we
now turn to the In-picture. The new ingredient 
is the non-trivial scalar potential induced by the transition states.
As we will show in the following, it is this new feature
that  can give rise to an accelerating phase in our cosmological
solutions and leads to the dynamical stabilization of the moduli
close to the flop.\footnote{Since the In-picture LEEA is smooth, it is clear that the corresponding Kasner solutions are smooth at flop transitions and also at any other transition involving finitely many transition states.} Thus we start the discussion of the In-picture by examining
the properties of this potential.

\begin{subsection}{The scalar potential}

In \cite{model} it was found that the potential arising in the In-picture description of a flop transition is of the form (\ref{2.26}). It is completely determined by the vector and hypermultiplet metrics (\ref{2.9}), (\ref{2.16}) 
and the superpotential (\ref{2.27}). 
The potential is positive semi-definite, $\cV(\phi, q) \ge 0$.  
Its minima correspond to supersymmetric Minkowski vacua which 
are parametrized by a subset $\cM_C$ of the scalar manifold, 
characterized by vanishing transition states:
\be\label{9.2}
\partial_\Lambda \cV \left|_{\cM_C} \right. = 0 \, , \quad \cM_C = \left\{ 
\begin{array}{l} 
v^2  = u_2 = 0 \\
v^1 , u_1 , U, W \quad \mbox{undetermined .}
\end{array} 
\right. 
\ee
As explained in \cite{ACDP,model}, these minima are the only critical points of the potential. 
They are also critical points of the superpotential: $\partial_\Lambda \cW |_{\cM_C} = 0 = \cW |_{\cM_C}$. 

Fig. \ref{fuenf} shows the potential $\cV(\phi,q)$ for some non-trivial but fixed values of the hypermultiplet scalar fields.
\begin{figure}[t]
\renewcommand{\baselinestretch}{1}
\epsfxsize=0.45\textwidth
\begin{center}
\leavevmode
\epsffile{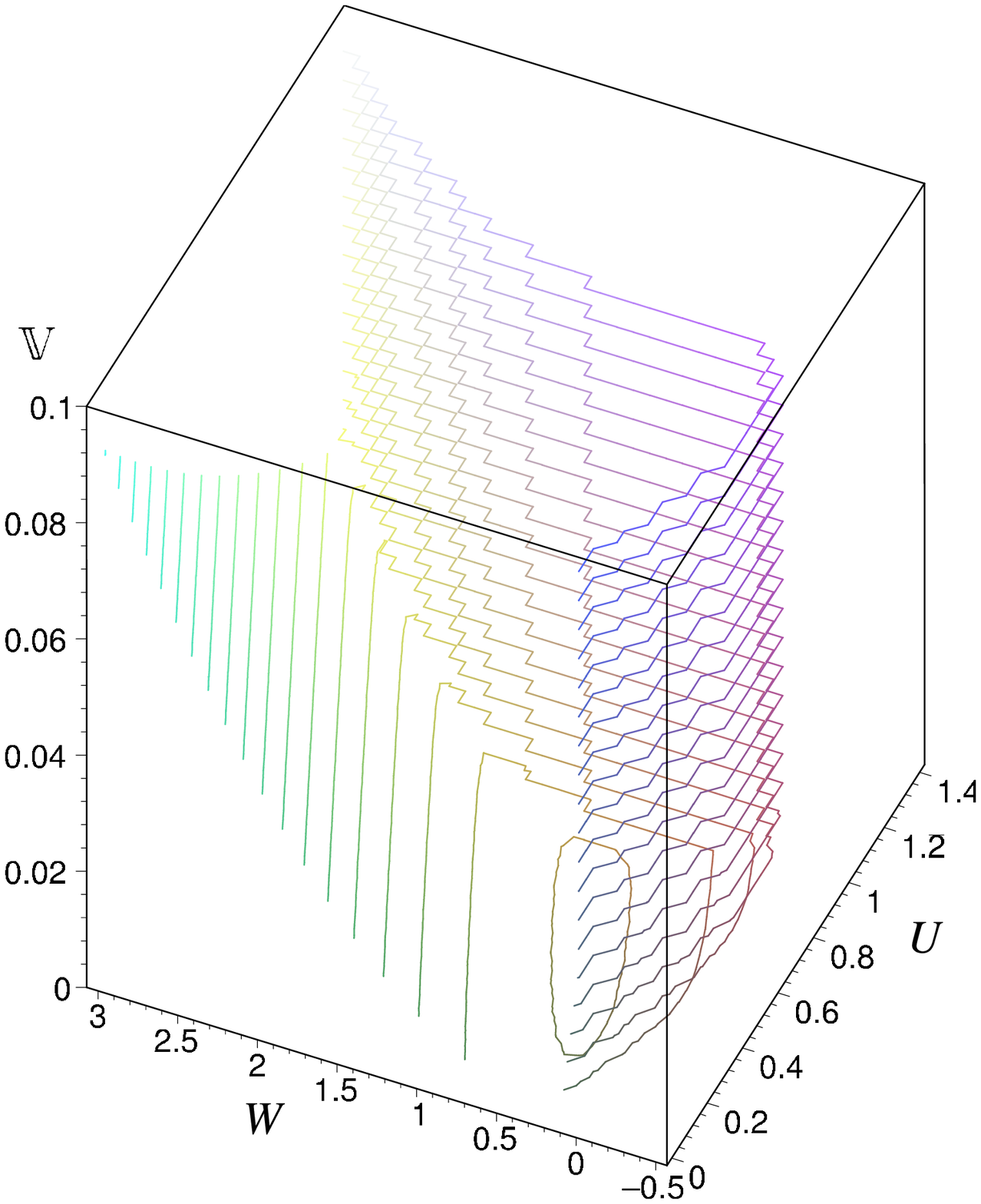} \, \, \, \, \,
\epsfxsize=0.45\textwidth
\epsffile{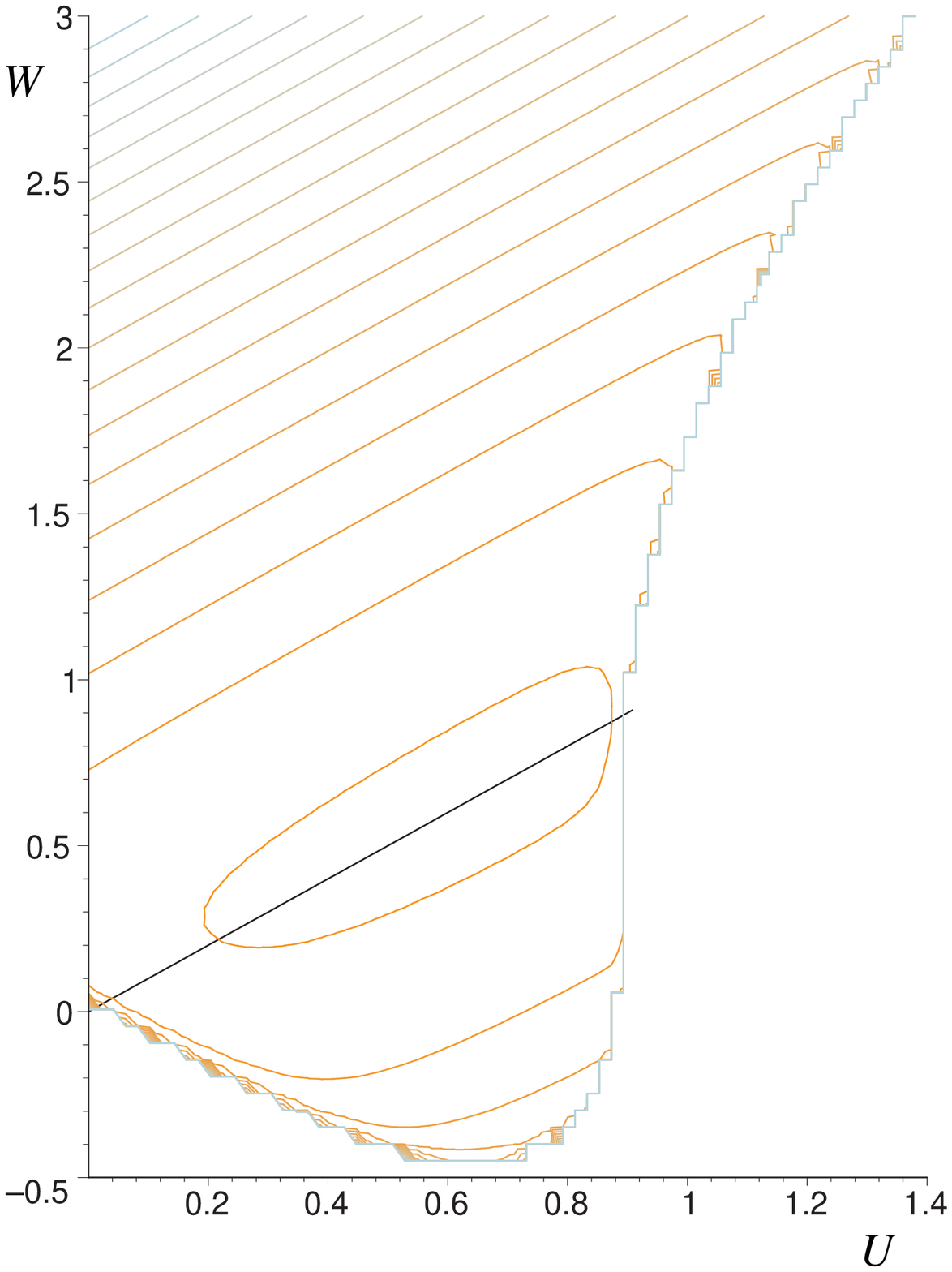}
\end{center}
\parbox[c]{\textwidth}{\caption{\label{fuenf}{\footnotesize The potential $\cV(\phi, q)$ at fixed values of the 
hypermultiplet scalars $u_1 = u_2 = 0.2, v^1 = v^2 = 0$. The potential has its minimum in the vicinity of the flop line 
$U=W$. At the boundary ``$b_2$'', where the metric $g_{xy}^{\rm (In)}$ has a zero eigenvalue, the potential diverges, while at the boundary 
``$b_1$'', where $\det(g_{xy}^{\rm (In)})$ becomes infinite, it is finite.}}}
\end{figure}
The potential is positive definite and finite as long as we are inside the vector multiplet scalar manifold. While Fig. 
\ref{fuenf} clearly shows that the value of the potential is small in the vicinity of the flop line $U=W$, an explicit 
calculation reveals that its actual minimum for fixed non-zero values of the transition states is {\it not located at} the flop line but slightly next to it.\footnote{Note that this point is not a critical point of the potential, since the derivatives with respect to the hypermultiplet scalars do not vanish.} The potential diverges at the boundary $b_2$ where the vector multiplet metric 
$g_{xy}^{\rm (In)}$ has a zero eigenvalue. At the boundary $b_1$, where $\det(g_{xy}^{\rm (In)})$ is infinite, the potential is finite. This 
feature can be traced back to the second term of the scalar potential (\ref{2.26}) which contains the inverse metric 
$g^{{\rm (In)}~xy}$. Finally we observe that in the limit $W \rightarrow \infty$ the 
potential diverges quadratically, $\cV \propto W^2$.

After considering the properties of the potential at the boundaries of the vector multiplet scalar manifold, let us comment on the 
boundaries appearing in the hypermultiplet sector. These are given by the loci where the hypermultiplet metric (\ref{2.16}) 
has an infinite eigenvalue. This occurs for either $\phi_+$ or $\phi_-$ defined in (\ref{2.16a}) becoming zero. The potential diverges at all boundaries of the hypermultiplet moduli space.

Fig. \ref{sechs} illustrates the dependence of the potential on the transition states. Taking $v^2 = p$, $u_2 = q$ both real and $v^1 = u_1 = 0$ and substituting this restriction into $\phi_+$ and $\phi_-$, we obtain
\be\label{9.7}
\phi_+ = 1 - p^2 \, , \quad \phi_- = 1 - q^2 + p^2 \, q^2 \, . 
\ee
This indicates that $p$ is bounded and takes values $-1 < p < 1$, while $q$ is unbounded. The potential blows up when $\phi_+$ or $\phi_-$ given above vanish.
\begin{figure}[t]
\renewcommand{\baselinestretch}{1}
\begin{center}
\leavevmode
\epsfxsize=0.45\textwidth
\epsffile{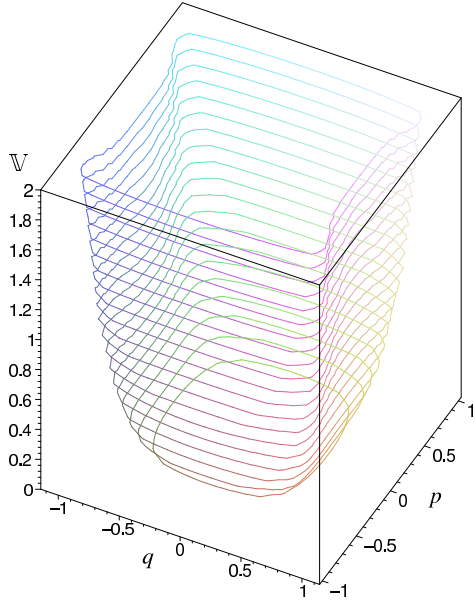}
\end{center}
\parbox[c]{\textwidth}{\caption{\label{sechs}{\footnotesize The scalar potential restricted to the submanifold $U=W=0.6$, $v^1 = u_1 = 0$ and $v^2 = p, u_2 =q$ taken to be real. The potential diverges at the loci where $\phi_+$ or $\phi_-$ become zero.}}}
\end{figure}

In summary we find that the potential diverges at the 
boundaries of the moduli space where the vector multiplet scalar metric develops a zero or the hypermultiplet scalar metric an infinite eigenvalue. At the boundaries where $\det(g_{xy}^{\rm (In)})$ becomes infinite, the potential is finite. These features can also be deduced from eq. (\ref{2.26}).
However, as explained before, it is not clear to which extent the In-picture LEEA at the boundaries really captures the
microscopic M-theory physics. 
\end{subsection}
\begin{subsection}{Fixed points of the equations of motion}
After discussing the properties of the scalar potential, we now turn to the In-picture equations of motion. Before investigating their properties numerically, let us first obtain some analytic results by studying  the fixed point properties of these equations. We begin by rewriting the system of second order equations (\ref{2.31}), (\ref{2.32}) and (\ref{2.33}) as a set of coupled autonomous first order equations. This is done in the standard way by introducing the momentum variables:
\be\label{9.8}
p^X := \dot{q}^X \, , \quad \rho^x := \dot{\phi}^x \, , \quad \gamma := \ad \, , \quad \delta := \bd \, .
\ee
Substituting in these variables, the equations of motion can be written as
\be\label{9.9}
\dot{q}^X = \beta_q^X \, , \, \dot{p}^X = \beta_p^X \, , \, \dot{\phi}^x = \beta_\phi^x \, , \, \dot{\rho}^x = \beta_\rho^x \, , \, \dot{\gamma} = \beta_\gamma \, , \, \dot{\delta} = \beta_\delta \, , 
\ee
where the constraint arising in the Einstein equations takes the form
\be\label{9.10}
\delta = \frac{1}{\gamma} \left( \frac{1}{3} (T + {\rm g}^2 \, \cV) - \gamma^2 \right) \, .
\ee
In these expressions the functions $\beta_i$ are given by:
\bea\label{9.11}
\nonumber
\beta_q^X  & = & p^X \, , \qquad \beta_\phi^x  = \rho^x \, , \\
\nonumber
\beta_p^X & = &  - \left( 3 \gamma + \delta \right) p^X - \Gamma^X_{~YZ} \, p^Y \, p^Z - {\rm g}^2 \, g^{XY} \, \frac{\partial \cV}{\partial q^Y} \, , \\
\nonumber
\beta_\rho^x & = &  - \left( 3 \gamma + \delta \right) \rho^x - \gamma^x_{~yz} \, \rho^y \, \rho^z - {\rm g}^2 \, g^{xy} \, \frac{\partial \cV}{\partial \phi^y} \, , \\
\beta_\gamma & = &  - \gamma \, \delta - 3 \gamma^2 + \frac{2}{3} \,{\rm g}^2 \, \cV \, , \qquad 
\beta_\delta = - 3 \gamma \delta - \delta^2 + \frac{2}{3} \,{\rm g}^2 \, \cV \, .
\eea

The set of fixed points of these equations consists of the points where the functions $\beta_i$ vanish simultaneously. For the matter fields the only solution to the equations $\beta^X_q = 0, \beta^x_\phi = 0$ is given by $p^X = 0, \rho^x = 0$. Substituting this constraint into $\beta^X_p, \beta_\rho^x$, we find that these functions vanish iff $g^{\Lambda \Sigma} \, \frac{\partial \cV}{\partial \phi^\Sigma} = 0$. Since the metric $g^{\Lambda \Sigma}$ is non-degenerate, this condition requires $\frac{\partial \cV}{\partial \phi^\Sigma} = 0$. This is just the condition for a critical point of the potential, which has already been investigated in the previous subsection. Using the result (\ref{9.2}) we see that the fixed points of the matter equations are 
\be\label{9.12}
\cM_C^{\rm mat} = \left\{ \cM_C \, , \, p^X = 0 \, , \, \rho^x = 0 \, \right\} \, .
\ee
Hence the fixed point manifold of the matter equations of motion is parametrized
by the flat directions of the potential.

Concerning the Einstein equations, we first observe that under the condition (\ref{9.12})  $T$ and $\cV$ vanish identically. In this case $\beta_\gamma$, $\beta_\delta$ and the constraint (\ref{9.10}) simplify to
\be\label{9.13}
\left. \beta_\gamma \right|_{\cM_C^{\rm mat}} = - \gamma \, \delta - 3 \, \gamma^2 \, , \quad  
\left. \beta_\delta \right|_{\cM_C^{\rm mat}} = - 3 \, \gamma \, \delta -   \delta^2 \, , \quad \delta = - \gamma \, .
\ee
Applying the fixed point condition $\beta_\gamma = 0, \beta_\delta =0$, implies that $\gamma = 0$, $\delta = 0$ while the values of $\alpha$ and $\beta$ are not determined by the fixed point condition. Thus we find that the equations (\ref{9.9}) have an entire manifold of fixed points $\cM_C^{\rm FP}$, given by
\be\label{9.14}
\cM_C^{\rm FP} = \left\{ 
\begin{array}{l} 
\cM_C \, , \, p^X = 0 \, , \, \rho^x = 0 \, , \, \gamma = 0 \, , \, \delta = 0 \\
\alpha, \beta \; \mbox{undetermined.}
\end{array}  \right. 
\ee

Let us now discuss the properties of these fixed points. In this course we calculate the critical exponents arising from linearizing the equations (\ref{9.9}) in the vicinity of the fixed points. These exponents are given by the eigenvalues of the stability matrix
\be\label{9.15}
{\bf B}_{ij} := \left. \partial_j \, \beta_i \right|_{\cM_C^{\rm FP}} \, , \quad
\left\{ i,j \in \alpha, \beta, \delta, \rho, \phi^x , \rho^x , q^X , p^X \right\} \, .
\ee
In order to calculate the entries of this matrix, we observe that
\be\label{9.16}
\left. \frac{\partial \beta^\Lambda_\phi}{\partial \phi^\Sigma} \right|_{\cM_C^{\rm FP}} = \left. - \, {\rm g}^2 \, g^{\Lambda \Xi} \, \frac{\partial^2 \cV}{\partial \phi^\Sigma \, \partial \phi^{\Xi}} \right|_{\cM_C^{\rm FP}} \, 
\ee
is the negative of the mass matrix $\cM_{~\Sigma}^{\Lambda}$ computed in \cite{model}. There it was found that $\cM_{~\Sigma}^{\Lambda}$ is diagonal with entries
\be\label{9.17}
\cM_{~\Sigma}^{\Lambda} = \frac{3}{2}  \, 6^{-2/3} \, {\rm g}^2 \, (U - W)^2 \, {\rm diag} \left[  0 \, , \, 1 \, , \, 0 \, , \, 1 \, , \, 0 \, , \, 1 \, , \, 0 \, , \, 1 \, , \, 0 \, , \, 0 \, \right] \,.
\ee
The non-vanishing entries correspond to the transition states $\{v^2 , \vb^2, u_2 , \ub_2 \}$. 

With this information at hand, it is now straightforward to compute the entries of ${\bf B}_{ij}$ with respect to the basis (\ref{9.15}),
\be\label{9.18}
{\bf B}_{ij} = 
\left[ 
\begin{array}{cccccc}
0 & \unit_2 & 0 & 0 & 0 & 0 \\
0 & 0 & 0 & 0 & 0 & 0 \\
0 & 0 & 0 & \unit_2 & 0 & 0 \\
0 & 0 & 0 & 0 & 0 & 0 \\
0 & 0 & 0 & 0 & 0 & \unit_8 \\
0 & 0 & 0 & 0 & - \cM_{~X}^{Y} & 0 
\end{array}
\right] \, .
\ee
The eigenvalues of this matrix are either zero or purely imaginary. Here it is useful to distinguish between the directions corresponding to the fields $\alpha, \beta, \gamma, \delta, \phi^x , \rho^x, v^1, u_1, \dot{v}^1, \dot{u}_1 $ and the charged transition states $v^2, u_2, \dot{v}^2, \dot{u}_2 $. The corresponding eigenvalues for these fields are given by  
\be\label{4.19}
\theta^{\rm neutral} = 0  \, , \qquad \theta^{\rm trans} = \pm \imag \, \sqrt{\frac{3}{2}} \, 6^{-1/3} \, {\rm g} \, ( U - W ) \, ,
\ee
respectively. This result implies that the fixed plane is neutrally stable, i.e., the solutions are neither attracted nor repelled by the fixed points. In other words these fixed points are {\it non-hyperbolic} \cite{AAABGI,DS}.
Since in this case it is not guaranteed that the stability matrix encodes the behavior of the full non-linear system, we rely on numerical solutions. As we will see in the next subsection, the transition states indeed oscillate 
 around their value at the fixed point $ v^2 = u_2 = 0 $. The
frequency of these oscillations depends on the actual values of the vector
multiplet scalar fields. This is the same qualitative behavior as indicated in eq. (\ref{4.19}). 

\end{subsection}
\begin{subsection}{Numerical solutions}
We now turn to the numerical solutions of the equations of motion
(\ref{2.31}), (\ref{2.32}) and (\ref{2.33}).
In order to be able to work with a trivial vector field background, we restrict the theory to the case where the hypermultiplet scalar fields are real, i.e.,
\be\label{8.1}
Q_v = {\rm Re}(v^1) \, , \quad q_v = {\rm Re}(v^2) \, , \quad Q_u = {\rm Re}(u_1) \, , \quad q_u = {\rm Re}(u_2) \, .
\ee
As discussed in subsection 2.5, this restriction provides a consistent truncation of the hypermultiplet equations of motion (\ref{2.33}).

To illustrate some characteristic features, we have picked a few examples of solutions whose initial conditions are given in Table \ref{t.3} in Appendix A. The solution labeled ``${\rm  b}^\prime$'' has the same initial conditions as the corresponding solution ``${\rm b}$'' in the Out-picture, the only difference being  one dynamical transition state for which we choose a non-vanishing initial value. This feature allows us to compare the qualitative behavior of the Out- and the In-picture solutions. Another example labeled ``i'' illustrates the behavior of a solution which initially starts far away from the flop line.

Let us first focus on the dynamics of the vector multiplet scalars $U,W$. The trajectories of the example solutions ``${\rm b}^{\prime}$'' and ``i'' projected to the vector multiplet scalar manifold are shown in Fig. \ref{sieben}.
\begin{figure}[t]
\renewcommand{\baselinestretch}{1}
\begin{center}
\leavevmode
\epsfxsize=0.45\textwidth
\epsffile{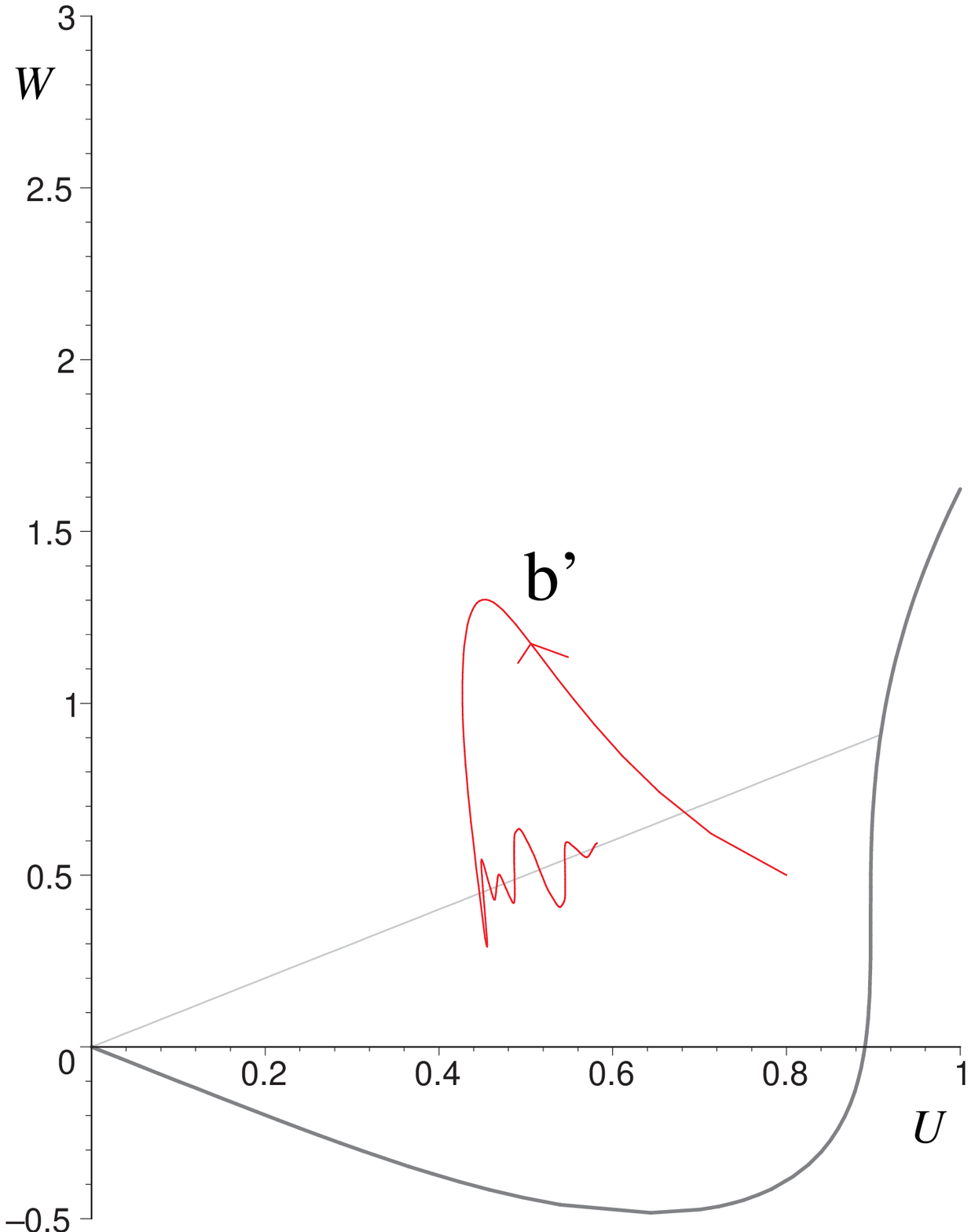} \, \, 
\epsfxsize=0.45\textwidth
\epsffile{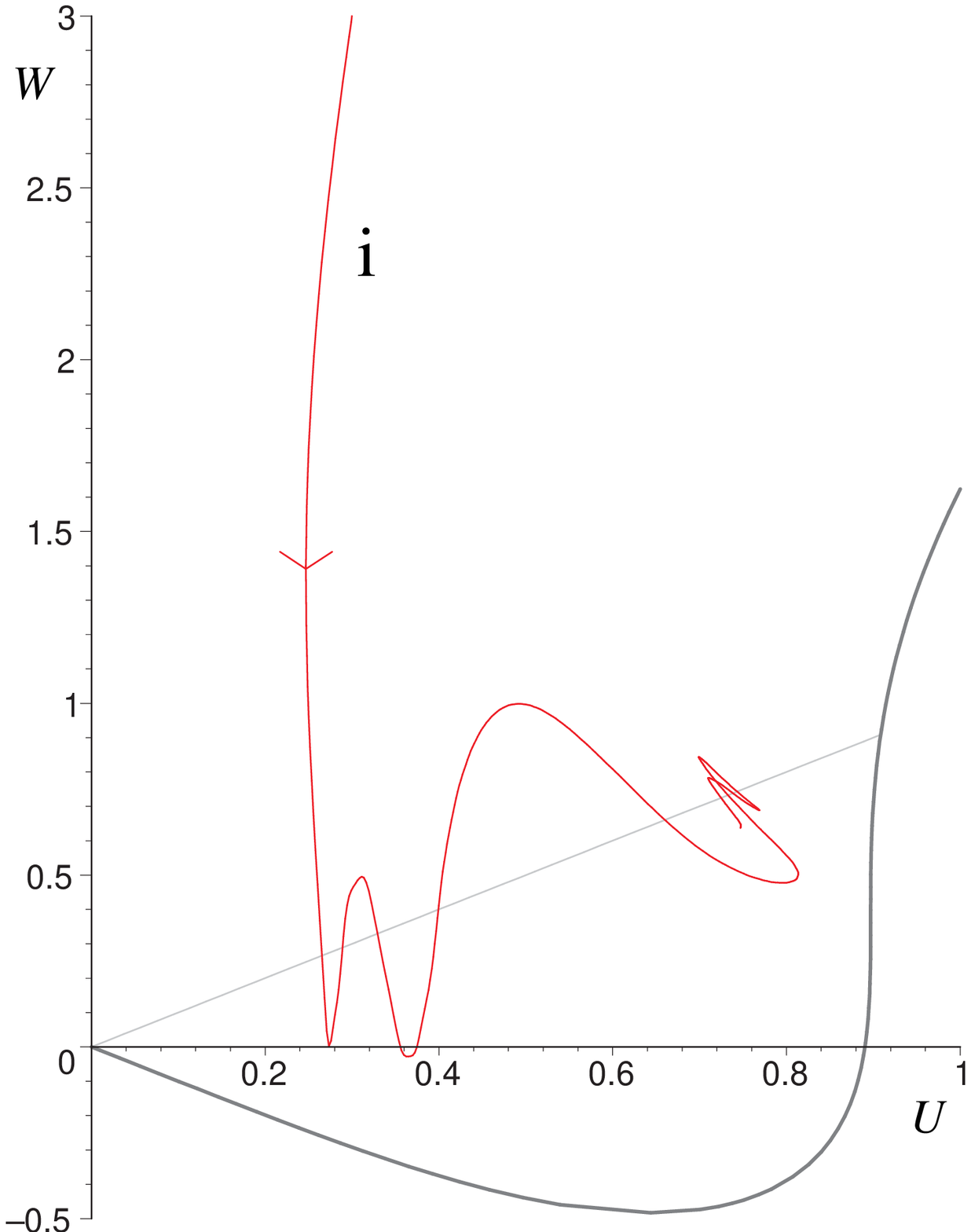}
\end{center}
\parbox[c]{\textwidth}{\caption{\label{sieben}{\footnotesize Numerical solutions of the In-picture equations of motion for the initial conditions ``${\rm b}^\prime$'' (left) and ``i'' (right) given in Table \ref{t.3}. The trajectories $U(\tau),W(\tau)$ stabilize in the vicinity of the flop line. The right diagram also illustrates how the solution is repelled by the boundary $b_2$.}}}
\end{figure}
The solutions first approach the flop line before starting to oscillate around  $U=W$. However, the flop line is
not an attractor: the solutions oscillate in the region close to the flop, but
do not settle down at a fixed point. 
This generic behavior of solutions differs from the corresponding behavior in the Out-picture, where the solutions crossed the flop line and did not stabilize. The In-picture solution ``${\rm b}^\prime$'' initially evolves analogously to its Out-picture cousin ``b'' but after crossing the flop line is turned back by the potential. A similar effect is shown in the second diagram, where one can also observe that the trajectory gets repelled when approaching the boundary $b_2$. Hence the In-picture potential provides a mechanism that prevents the solutions from running into the boundaries where $\det(g_{xy}^{({\rm In})}) = 0$. This is different from the Out-picture, where such boundaries are reached in finite time.

Let us now focus on the solution ``${\rm b}^\prime$'' more closely. The corresponding moduli are shown in Fig. \ref{acht}.
\begin{figure}[p!]
\renewcommand{\baselinestretch}{1}
\begin{center}
\leavevmode
\epsfxsize=0.45\textwidth
\epsffile{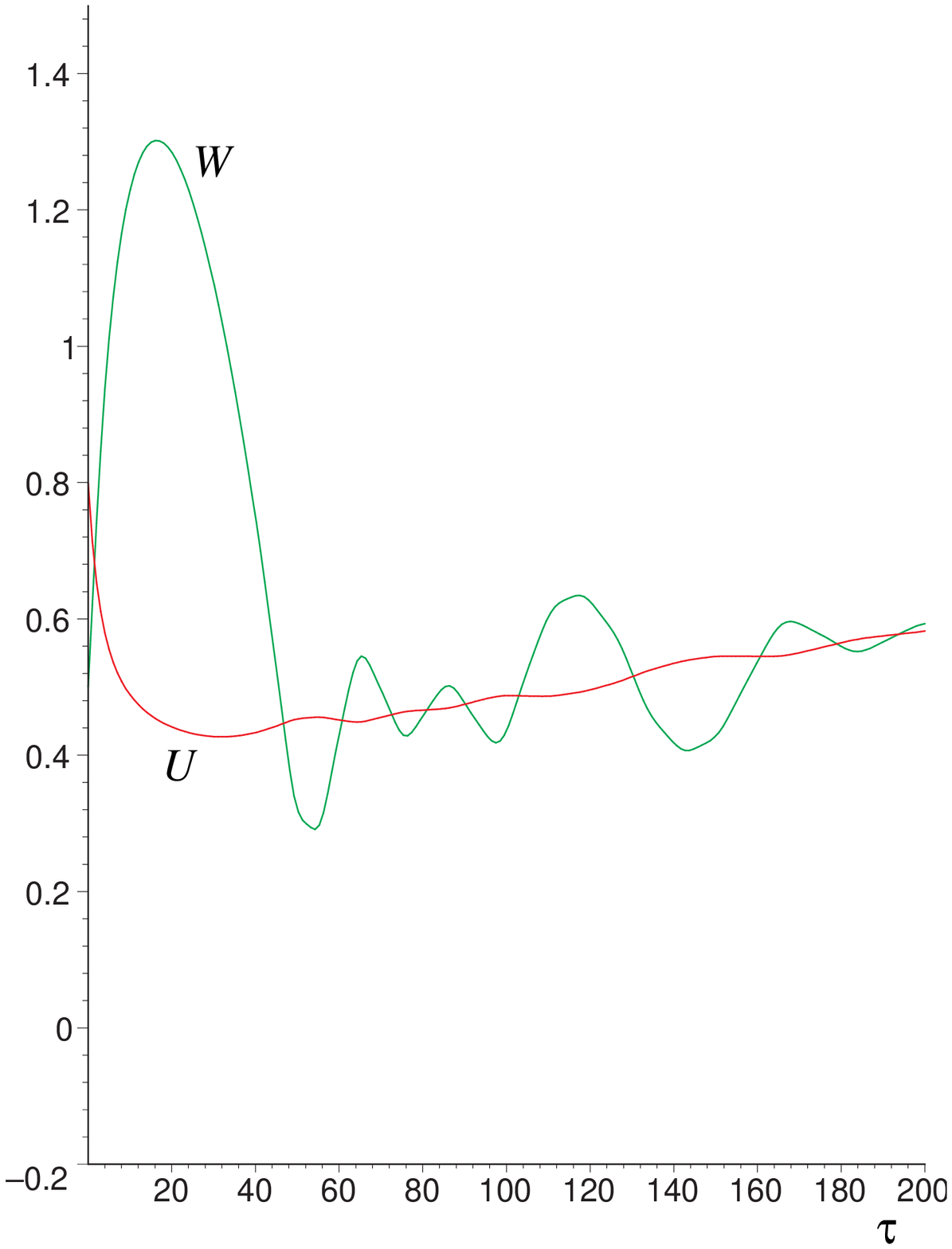} \, \, 
\epsfxsize=0.45\textwidth
\epsffile{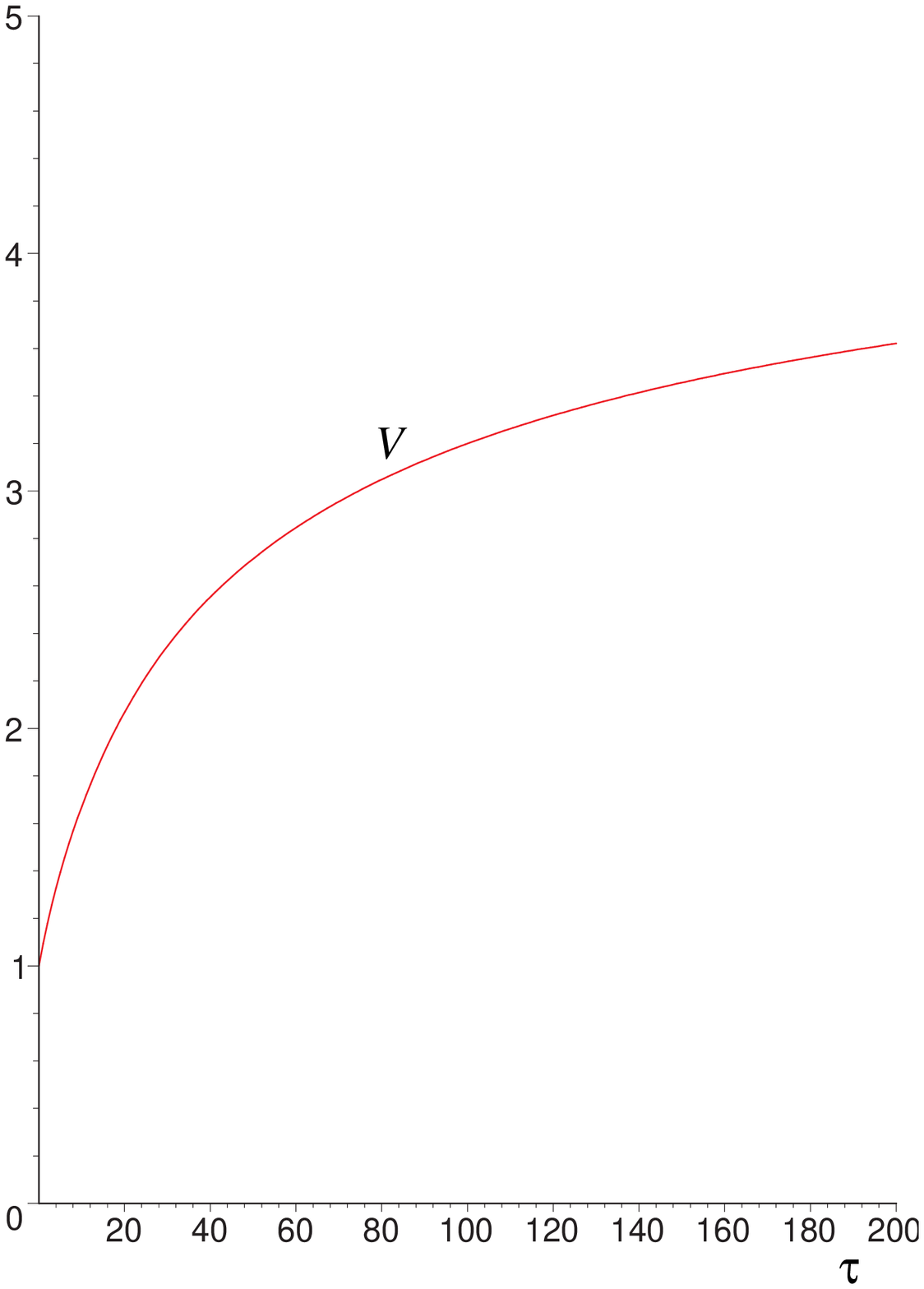}

\leavevmode
\epsfxsize=0.45\textwidth
\epsffile{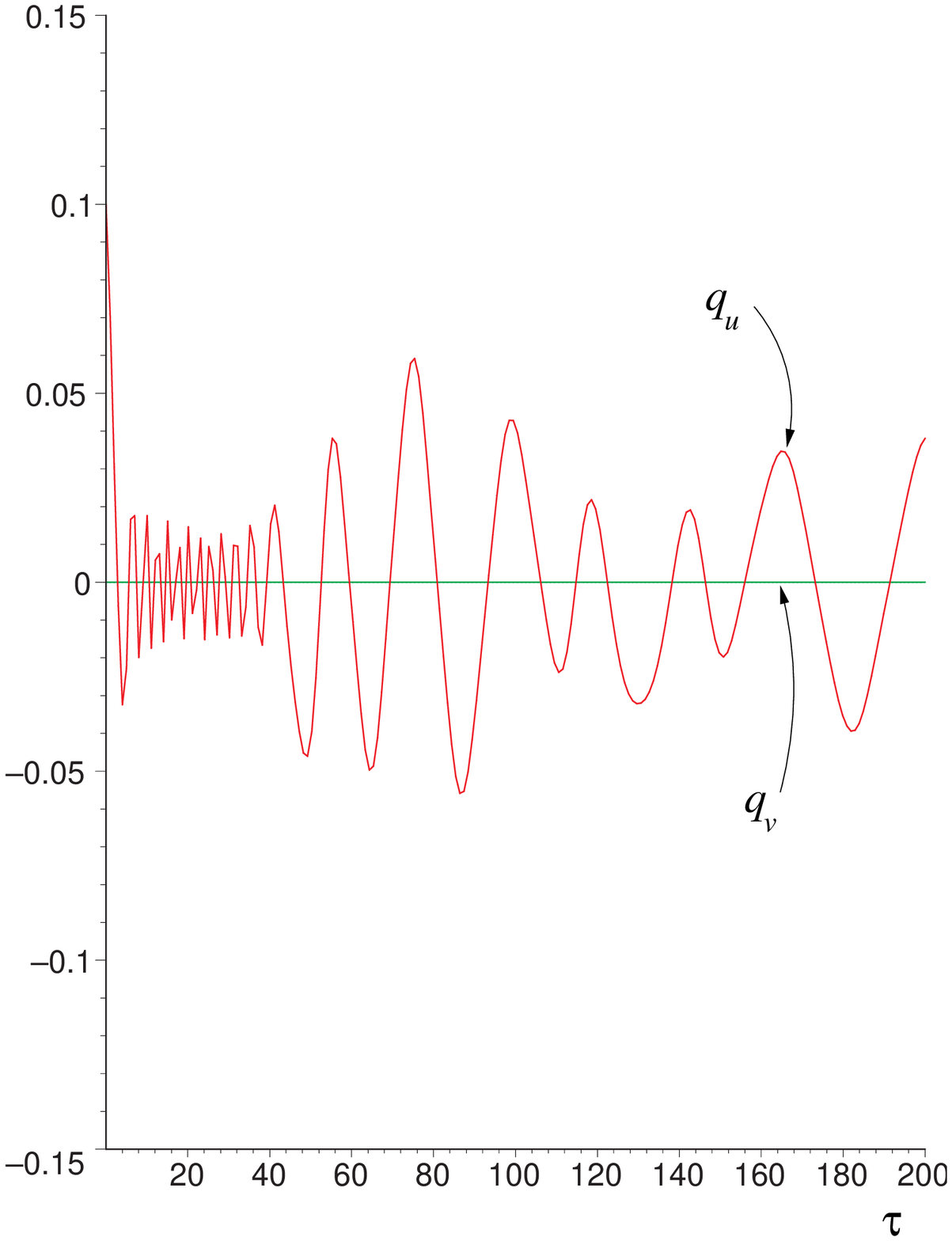} \, \, 
\epsfxsize=0.45\textwidth
\epsffile{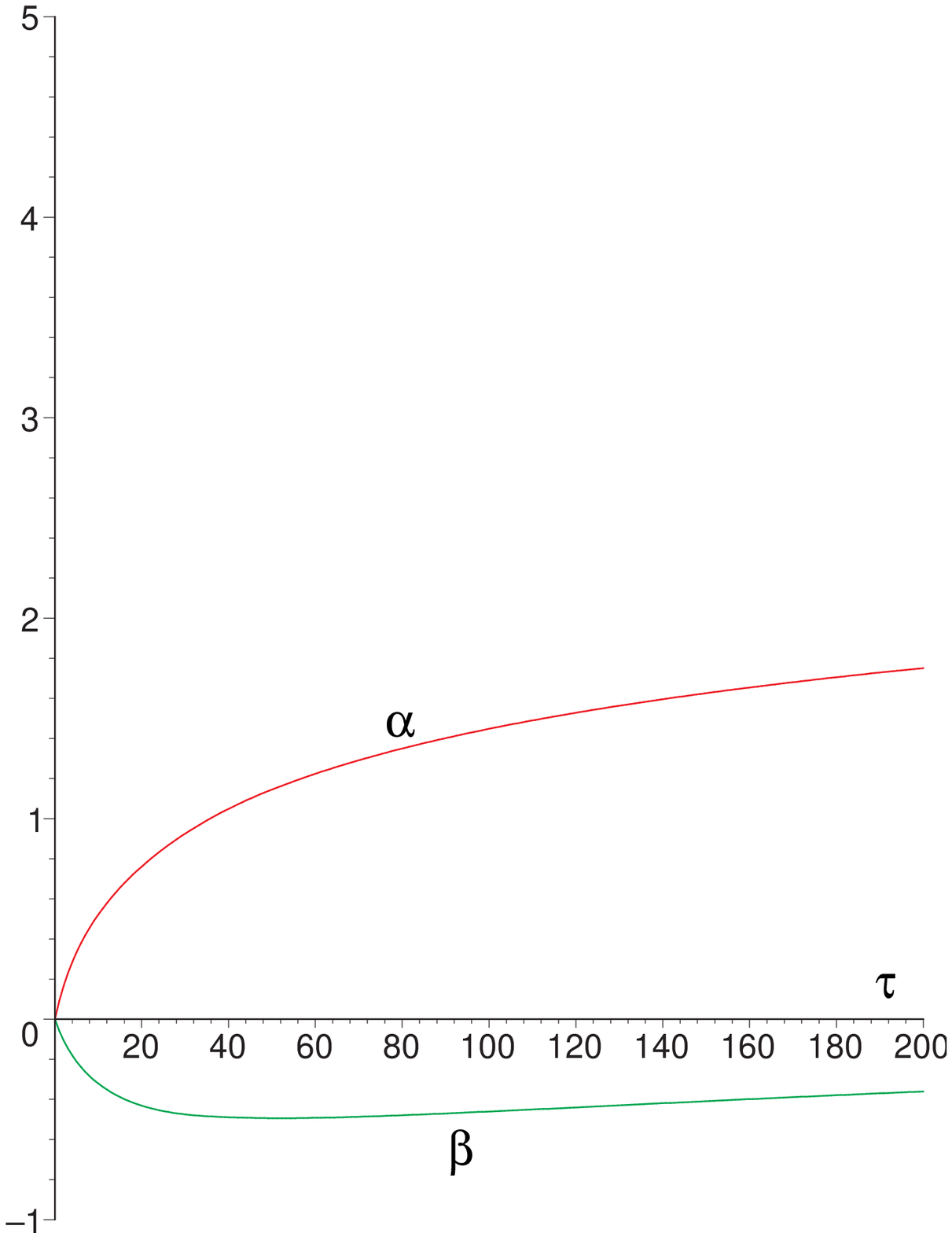}
\end{center}
\parbox[c]{\textwidth}{\caption{\label{acht}{\footnotesize The complete numerical solution of eqs. (\ref{2.31}), (\ref{2.32}) and (\ref{2.33}) starting with the initial values ``${\rm b}^{\prime}$'' given in Table \ref{t.3}. These are the same as for the Out-picture solution ``b'' except for the non-trivial transition state $q_u(0) = 0.1$. Starting from the upper left, the plots depict the dynamics of the vector multiplet scalars $U(\tau), W(\tau)$, the CY volume $V(\tau)$, the transition states $q_u(\tau), q_v(\tau)$ and the logarithmic scale factors $\alpha(\tau), \beta(\tau)$, respectively. After some initial period the CY volume stabilizes, the logarithmic scale factors increase monotonically, and $q_u(\tau)$ oscillates around the fixed plane value $q_u = 0$.}}}
\end{figure}
 We observe that the vector multiplet moduli $U(\tau)$ and $W(\tau)$ oscillate around the flop line $U = W$ instead of approaching the boundary $b_1$. 
The CY volume $V(\tau)$ is obtained by substituting the numerical solution into eq. (\ref{2.20}).  We find that when 
picking the initial value $\dot{V}(0) = 0.1$, as for the Out-picture solution ``b'', the volume increases monotonically. This is completely analogous to the corresponding Out-picture solution. The difference between the two pictures is, however, that the Out-picture volume undergoes an accelerated increase while the  volume in the In-picture shows decelerated increase.

The initial values for the transition states were chosen such that $q_v(\tau)$ is frozen to be zero. The non-trivial initial value for $q_u$ results in $q_u(\tau)$ oscillating around $q_u = 0$. The frequency of oscillations depends on the difference $|U(\tau) - W(\tau)|$ in the sense that a large difference induces rapid oscillations while a small difference, i.e., being close to the flop, corresponds to a low frequency. 
After some initial period the logarithmic scale factors $\alpha(\tau)$ and $\beta(\tau)$  become almost constant. This differs from the Out-picture where $\alpha(\tau) \rightarrow -\infty$, $\beta(\tau) \rightarrow - \infty$ as the solution approaches the boundary $b_1$.

So, comparing these examples of In- and Out-picture solutions we find that the inclusion of the charged transition states drastically modifies the behavior of the solution. In particular, the In-picture solution does not run into a boundary where the solution becomes singular.

The complete solution of our second example ``i'' is shown in Fig. \ref{neun}.
\begin{figure}[p!]
\renewcommand{\baselinestretch}{1}
\begin{center}
\leavevmode
\epsfxsize=0.45\textwidth
\epsffile{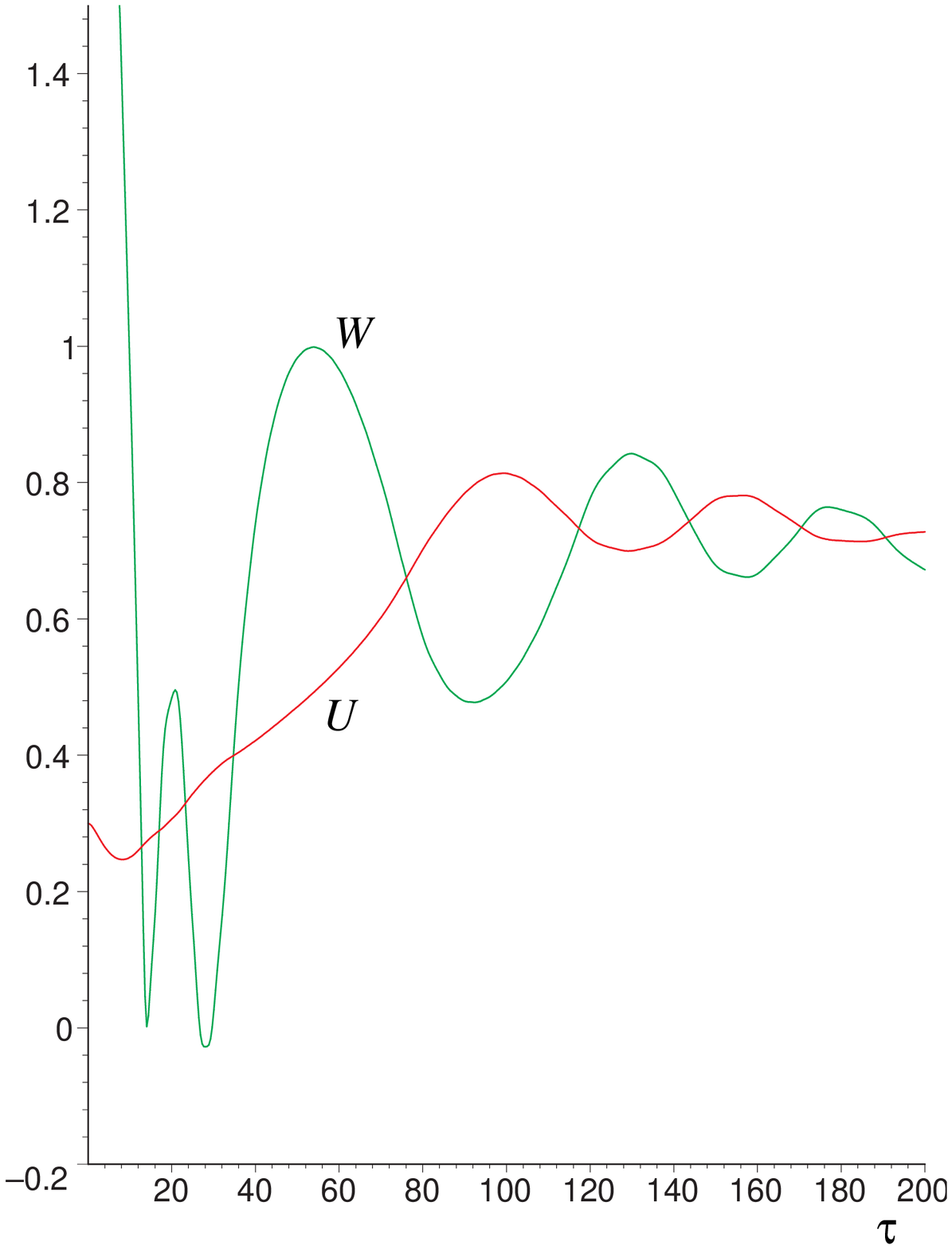} \, \, 
\epsfxsize=0.45\textwidth
\epsffile{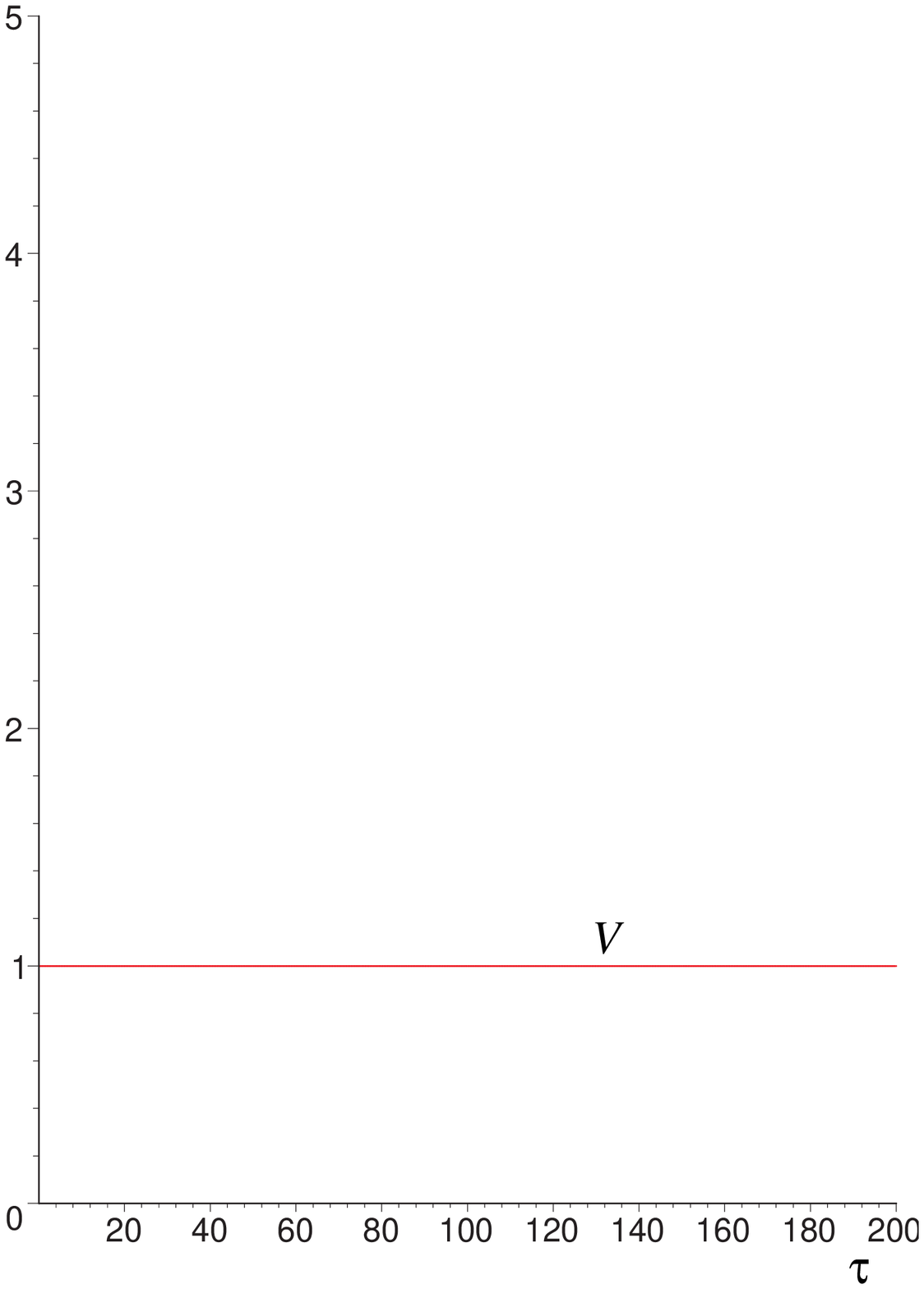}

\leavevmode
\epsfxsize=0.45\textwidth
\epsffile{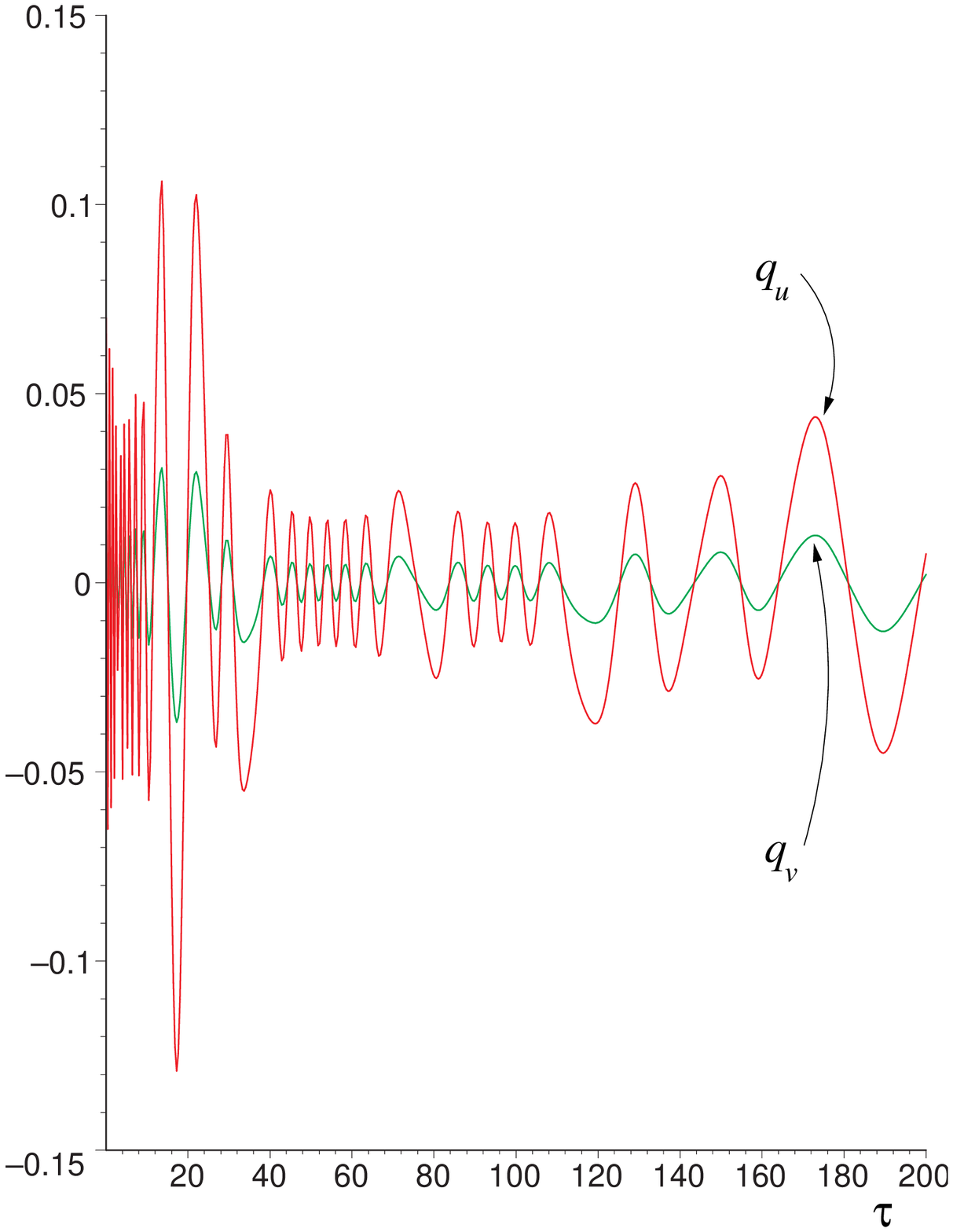} \, \, 
\epsfxsize=0.45\textwidth
\epsffile{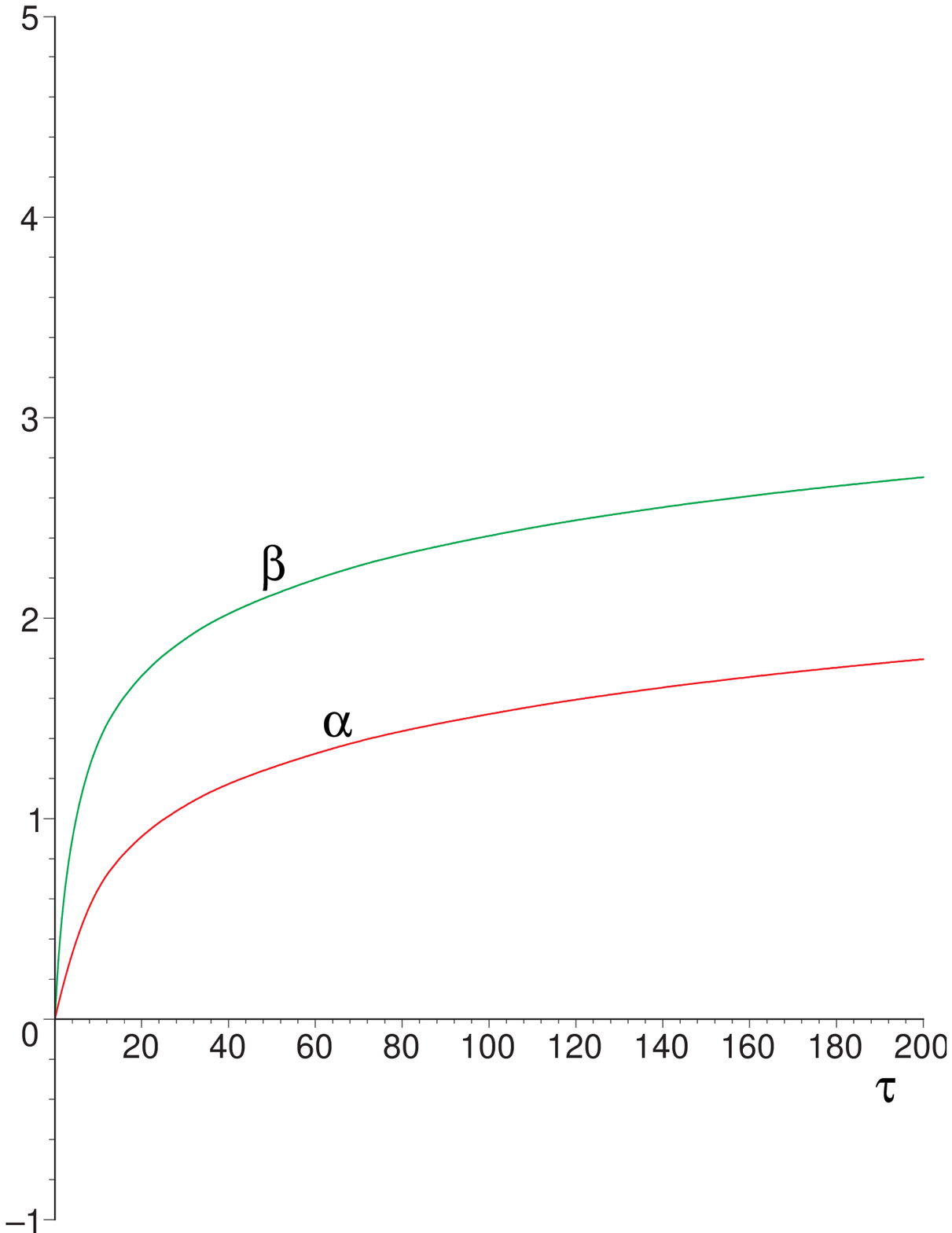}
\end{center}
\parbox[c]{\textwidth}{\caption{\label{neun}{\footnotesize The complete numerical solution of eqs. (\ref{2.31}), (\ref{2.32}) and (\ref{2.33}) starting with the initial values ``i'' given in Table \ref{t.3}. As in Fig. \ref{acht}, the diagrams show the dynamics of the vector multiplet scalars $U(\tau), W(\tau)$, the CY volume $V(\tau)$, the transition states $q_u(\tau), q_v(\tau)$ and the logarithmic scale factors $\alpha(\tau), \beta(\tau)$, respectively.  The vector multiplet scalars are repelled when approaching the boundary $b_2$. The transition states $q_u(\tau)$, $q_v(\tau)$ oscillate with exactly the same frequency.}}}
\end{figure}
Here we have taken the vector multiplet scalar $W$ to start ``far away'' from the flop line and  the initial conditions for $Q_u$ and $Q_v$ were chosen such that the CY volume stays constant, $V(\tau) = 1$.

The scalars $U(\tau), W(\tau)$ start in a  region of high potential so that the solution first rolls down to the region of low potential before it begins to oscillate around the flop line. The first turning point in $W(\tau)$ and the second turning point in $U(\tau)$ correspond to points where the solution is repelled by the boundary $b_2$.
 In the region where $|U(\tau) - W(\tau)|$ is large the moduli $q_u(\tau)$ and $q_v(\tau)$ exhibit rapid oscillations around the zero line. The frequency of the oscillations is identical for both $q_u(\tau)$ and $q_v(\tau)$.
The logarithmic scale factors $\alpha(\tau)$ and $\beta(\tau)$ increase monotonically with time. In fact, a closer look on their  dynamics in subsection 4.5 reveals several short periods of accelerated expansion, although these do not significantly influence the behavior of $\alpha(\tau)$ and $\beta(\tau)$. 
 
For the solution ``g'' shown in Fig. \ref{neun_a} we have chosen our initial values (see Table \ref{t.3}) for $U$ and $W$ to be on the flop line. 
\begin{figure}[t]
\renewcommand{\baselinestretch}{1}
\epsfxsize=0.45\textwidth
\begin{center}
\leavevmode
\epsffile{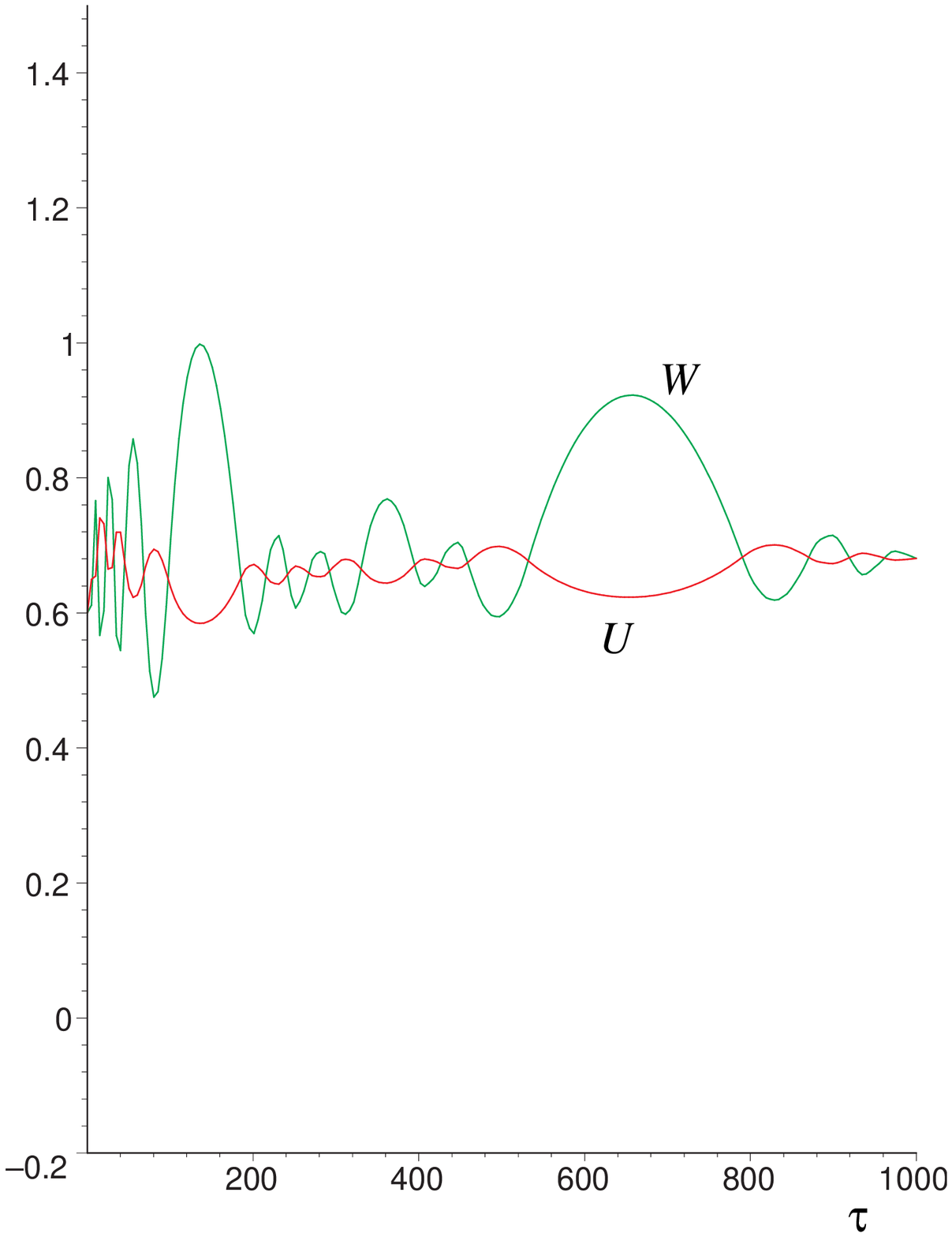} \, \, 
\epsfxsize=0.45\textwidth
\epsffile{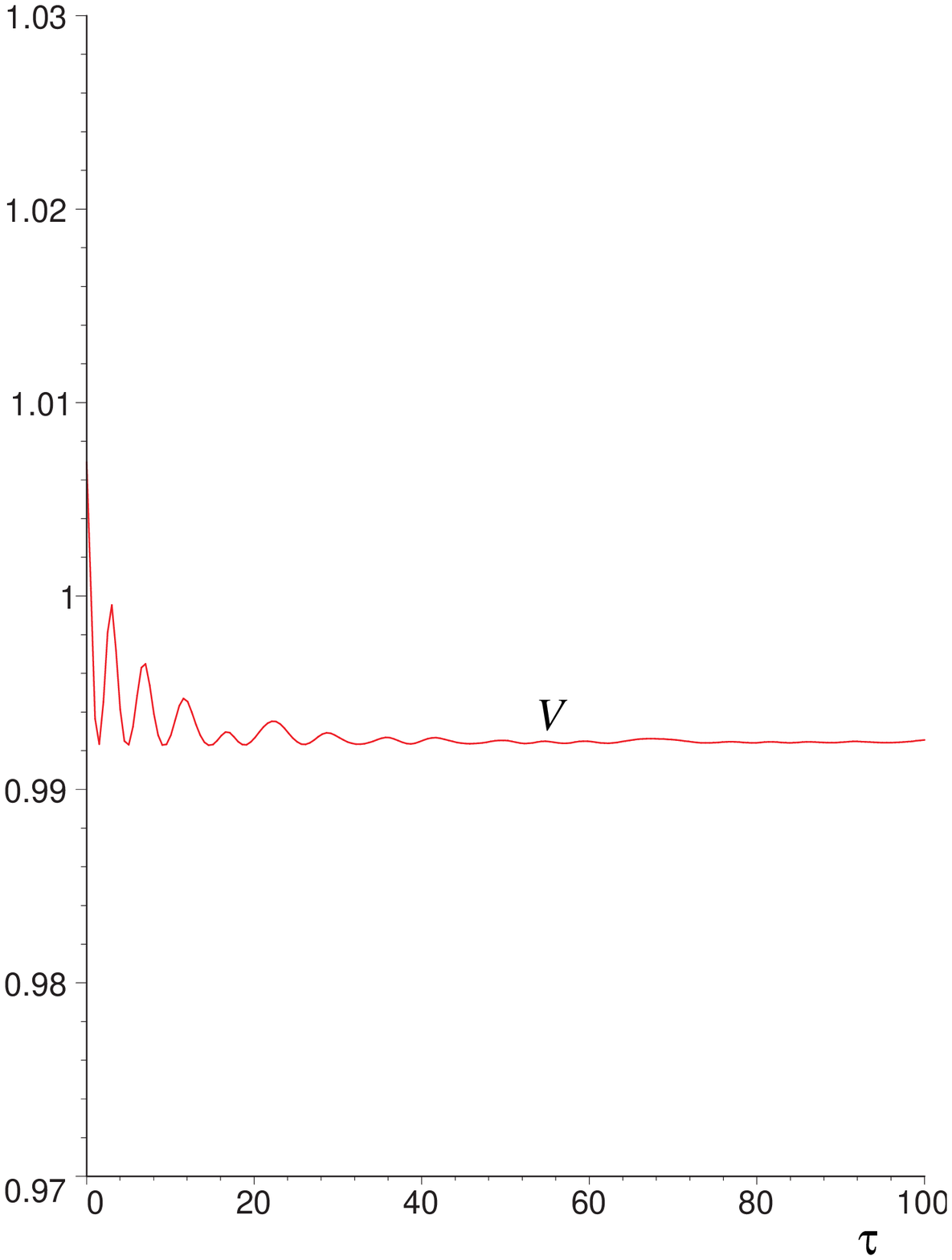}
%
%\epsfxsize=0.45\textwidth
%\epsffile{i1_q_labels.eps} \, \, 
%\epsfxsize=0.45\textwidth
%\epsffile{i1_alpha_labels.eps}
\end{center}
\parbox[c]{\textwidth}{\caption{\label{neun_a}{\footnotesize The left and right diagram show $U(\tau)$, $W(\tau)$ and the CY volume $V(\tau)$ arising from the initial conditions ``g'' given in Table \ref{t.3}. The solution starts at the flop line but then evolves away and oscillates around $U=W$. Furthermore, $V(\tau)$ is no longer monotonic but oscillates before stabilizing.}}}
\end{figure}
However, the solution does not stay at the flop but starts oscillating around it. This result illustrates again that the flop line is neither a stable 
nor an unstable manifold.
Another notable  feature of this solution is  that the CY volume $V(\tau)$ is no longer monotonic, but oscillates before settling to a finite value. This behavior is caused by giving non-trivial initial values to the transition states. 

A common property of all the solutions presented in this subsection is that if we evolve them backwards in time, they become singular. Here the magnitude of $\ad(\tau)$ and $\bd(\tau)$ becomes increasingly large, while the moduli $q^X, \phi^x$ are still inside their respective manifolds. As this happens, our numerical solutions are no longer reliable and we do not obtain any conclusive results about the origin of this singularity.

\end{subsection}
\begin{subsection}{Behavior of solutions at the boundaries of moduli space}
As in subsection 3.3, it is also interesting to discuss the behavior of solutions close to the In-picture boundaries of the vector multiplet scalar manifold.\footnote{Note that in such an analysis we investigate the properties of the 
In-picture Lagrangian far away from the flop line. Thus these results do not necessarily reflect the full M-theory behavior.}
For this purpose it is convenient to introduce
\be\label{4.1}
\epsilon := 12 (U+W) - 17 U^4 - 14 U^3 W - 6 U^2 W^2 + 4 U W^3 + W^4 \, .
\ee
This quantity is positive inside the vector multiplet scalar manifold displayed in Fig. \ref{eins} and vanishes at the boundary $b_2$. In this sense $\epsilon$ measures the distance away from the boundary $b_2$.

We observe that the In-picture equations of motion (\ref{2.31}), (\ref{2.32}), and (\ref{2.33}) become singular as $\epsilon \rightarrow 0$. This is due to $\det(g_{xy}^{\rm (In)}) = 0$ for $\epsilon = 0$. This behavior reflects itself in the Christoffel symbols of the vector multiplet sector and in the scalar potential, which both diverge as $\frac{1}{\epsilon}$. The last observation implies that for non-zero transition states the potential term appearing in the vector multiplet equations of motion diverges as $\frac{1}{\epsilon^3}$. Assuming that $\dot{U}(\tau)$ and $\dot{W}(\tau)$ are finite, this term completely dominates  eq. (\ref{2.33}) close to $b_2$. The shape of the potential shown in Fig. \ref{fuenf} then implies that this term drives the solution {\it away} from the boundary.\footnote{This observation justifies the assumption that $\dot{U}(\tau)$ and $\dot{W}(\tau)$ are finite as the potential term decreases these quantities close to the boundary.} 

The inclusion of non-zero transition states thus provides a new mechanism preventing the solutions from reaching the boundaries of the vector multiplet scalar manifold where the metric degenerates. An analysis along the lines of subsection 3.3 fails, however, due to the complicated nature of the corresponding differential equations close to the boundary. But the second diagram of Fig. \ref{sieben} gives a numerical example for this mechanism when the solution is repelled from the boundary $b_2$.

At the boundary $b_1$ we observe that the potential is finite for $W > 0$. This implies that at this boundary the Christoffel symbols of the vector multiplet sector are the only quantities which become singular. The corresponding behavior of the solutions should be similar to the Out-picture solutions of type 1) and 2) found in subsection 3.3. The case 3), however, is modified by the strong repulsive potential as $W(\tau)$ becomes large.  
\end{subsection}
\begin{subsection}{The search for inflation}
In this subsection we investigate whether inflation is possible in the In-picture model. Eq. (\ref{3.9}) shows that the Out-picture does not admit accelerated expansion in the $\vec{x}$-directions, while in the $y$-direction the occurrence of acceleration depends on the actual value of $\ad$. However, contrary to the Out-picture, the In-picture contains a scalar potential, which, in principle, can give rise to an inflationary phase.

\subsubsection*{General analysis of our model}
The starting point for our analysis are the Einstein equations (\ref{2.31}) 
written in terms of the cosmological time $\tau$. Setting $\nu = 0, \nd = 0$, these take the form:
\bea\label{7.1}
\nonumber 3 \left( \ad^2 + \ad \, \bd \right) & = & T + {\rm g}^2 \,  \cV  \, , \\
2 \ddot{\alpha} + \ddot{\beta} + 2 \ad \bd + 3 \ad^2 + \bd^2  &=& - T + {\rm g}^2 \, \cV \, , \\
\nonumber 3 \left( \ddot{\alpha} + 2 \ad^2 \right) & = & - T + {\rm g}^2 \, \cV \, .
\eea
Eliminating the terms containing $\ad \bd$ and rewriting the second and third equation in terms of $\ddot{a}$ and $\ddot{b}$ given in (\ref{3.8}), we find the following analytic expressions for $\ddot{a}$ and $\ddot{b}$: 
\be\label{7.2}
\ddot{a}   =  \left( \frac{1}{3} \left( {\rm g}^2 \, \cV - T \right) - \ad^2 \right)  \e^{ \alpha} \, , \quad 
\ddot{b}  = \left( - \left( T + \frac{1}{3} {\rm g}^2 \, \cV \right) + 3 \ad^2 \right) \, \e^{\beta} \, .
\ee
These equations show that  while $\cV$ opposes accelerated expansion in the extra-dimension, it also enables accelerated expansion in the $\vec{x}$-directions. Acceleration in the three-space
 occurs if the potential dominates over the kinetic term $T$ and the
$\ad$-contribution. In principle, eq. (\ref{7.2}) allows for de Sitter like solutions, which correspond to $\cV > 0$, $\ad$ positive and approximately constant, and $T$ small compared to $\cV$. However, since the ground state of our model is Minkowski, we cannot realize these de Sitter like solutions. The best approximation we can obtain are solutions which behave de Sitter like for a limited period of time. This can be realized when the scalar fields roll slowly at a non-vanishing value of the potential. 
\subsubsection*{Slow-Roll conditions}
Having found that our model admits inflationary phases in principle, we need
to investigate whether there is a region in our parameter space where our
potential satisfies the slow-roll conditions. 

%%%
We consider the consistency conditions given in terms of the
slow-roll parameter $\epsilon$ \cite{LL}, generalized to non-linear
sigma models (see for example \cite{Stewart}). Since the existence
of a slow-roll regime is a property of the scalar potential
$\cV(\phi, q)$, we will drop the $(3+1)$-split of our space-like
directions and set $\alpha(\tau) = \beta(\tau)$.
We also introduce the standard Hubble parameter $H=\dot{\alpha}$.
In terms of the covariant derivative with respect to the scalar
field metrics, $D_\tau \dot{\phi}^{\Lambda} =  \ddot{\phi}^{\Lambda} +
\Gamma^{\Lambda}_{\;\;\Sigma \Xi} \dot{\phi}^{\Sigma} \dot{ \phi}^{\Xi}$,
the equations of motion (\ref{2.31}), (\ref{2.32}), and (\ref{2.33})
imply
\be\label{7.11}
H^2  = \frac{1}{6} \left( T + {\rm g}^2 \, \cV \right) \, , \quad
D_\tau \dot{\phi}^\Sigma + 4 H \dot{\phi}^{\Sigma} = - \, {\rm g}^2 \, g^{\Sigma \Lambda} \frac{\partial \cV}{\partial \phi^\Lambda} \, .
\ee
Here $\left( \phi^{\Sigma} \right) = \left( \phi^x, q^X \right)$ and $\Gamma^{\Lambda}_{\;\;\Sigma \Xi}$
is the Christoffel connection with respect to the combined scalar 
metric $g_{\Lambda \Xi} = g_{xy} \oplus g_{XY}$. Slow-roll inflation
 implies the following conditions:
\be
T \ll g^2 \cV \;, \;\;\;
D_\tau \, \dot{\phi}^{\Sigma} \ll 4 H \dot{\phi}^{\Sigma} \;.
\ee
In the slow-roll regime we can use the eq. (\ref{7.11}) to
express $\epsilon := \ft{T}{g^2 \cV}$ entirely in terms of the 
potential,
\be\label{7.17}
\epsilon \approx \frac{3}{16 \cV^2} \, g^{\Sigma \Lambda} \, \frac{\partial \cV}{\partial \phi^\Sigma} \, \frac{\partial \cV}{\partial \phi^{\Lambda}} \, .
\ee
Consistency of the slow-roll conditions requires 
$\epsilon \ll 1$.
%%%
By substituting $\alpha(\tau) = \beta(\tau)$ into eq. (\ref{7.2}) and eliminating $\ad(\tau)$, we find that an accelerated expansion needs $3 T < {\rm g}^2 \cV $. Hence the period of inflation ends when $\epsilon = \frac{1}{3}$.

We then use this generalized slow-roll parameter $\epsilon$ (\ref{7.17}) to investigate the possibility of slow-roll inflation for the In-picture scalar potential $\cV( \phi, q)$.  An extended numerical check shows that the condition for slow-roll inflation, i.e., the RHS of eq. (\ref{7.17}) being less than one third, is never satisfied.  Thus we conclude that our scalar potential does not allow for a phase of slow-roll inflation. 
\subsubsection*{Numerical examples}
Even though our model does not admit slow-roll inflation,
it nevertheless has solutions 
with short periods of accelerated expansion. This is not in conflict with the results above, as the conditions leading to (\ref{7.17}) may be violated so that the approximation which we used to show that $\epsilon$ is bigger than one third breaks down. 

Let us return to the numerical solutions ``${\rm b}^\prime$'' and ``i''
studied in subsection 4.3. The functions
$\ddot{a}(\tau)$ and $\ddot{b}(\tau)$ for these solutions are shown in the left
and right diagram of Fig. \ref{zwoelf}, respectively.
\begin{figure}[t]
\renewcommand{\baselinestretch}{1}
\epsfxsize=0.45\textwidth
\begin{center}
\leavevmode
\epsffile{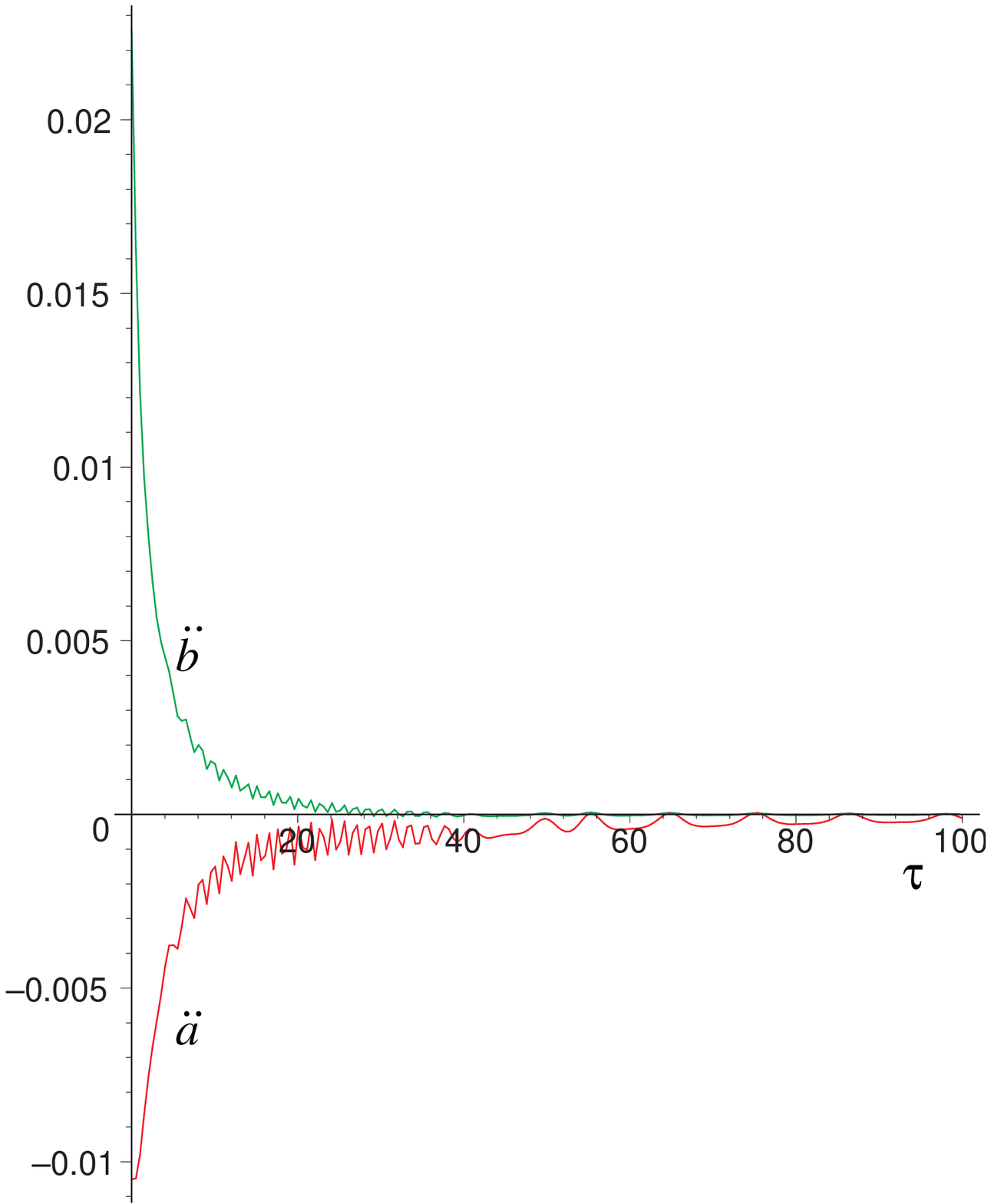} \, \, 
\epsfxsize=0.45\textwidth
\epsffile{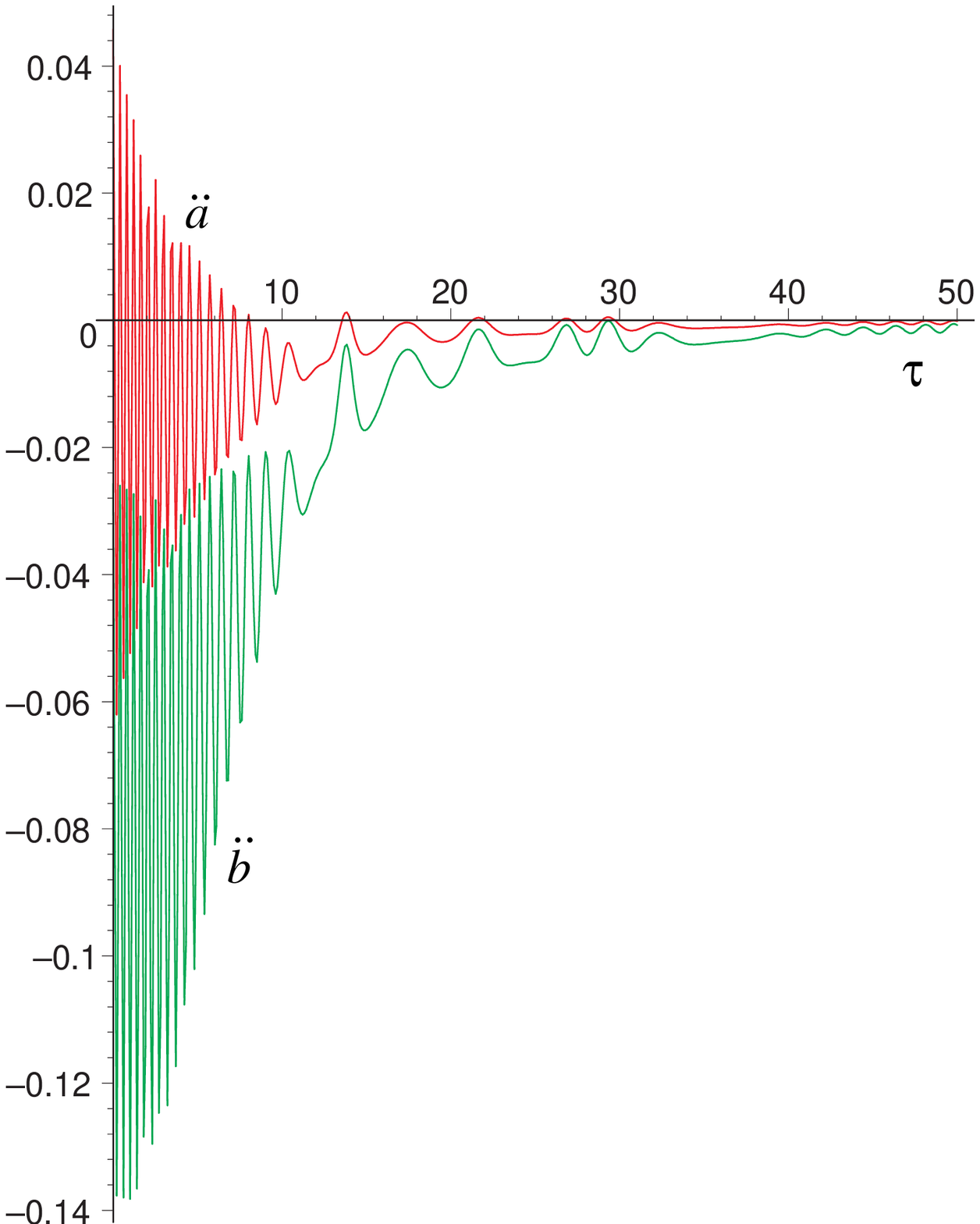}
\end{center}
\parbox[c]{\textwidth}{\caption{\label{zwoelf}{\footnotesize The factors $\ddot{a}(\tau)$ and $\ddot{b}(\tau)$ along the solutions ``${\rm b}^\prime$'' and ``i'' given in Table \ref{t.3}. Both solutions yield multiple short periods of accelerated expansion in three-space. The late time maxima are slightly positive.}}}
\end{figure}
In both cases $\ddot{a}(\tau)$ and $\ddot{b}(\tau)$ show  an 
oscillatory behavior. But the most prominent feature is that
both solutions have short periods of acceleration.
Looking at the solution ``${\rm b}^{\prime}$'' we see that $\ddot{b}(\tau)$ is positive for an extended period. This corresponds to a decelerated contraction of the $y$-direction which later turns into expansion (see Fig. \ref{acht}). In the three-dimensional $\vec{x}$-space we observe several short periods of accelerated expansion at late times. For the solution ``i'' these periods are more pronounced. Here  $\ddot{b}(\tau)$ is initially negative and oscillates strongly whereas at later times we see small fluctuations. A common feature of the scale factors of both solutions is that after an initial period the average value of acceleration is small and negative, showing preference for a decelerating universe.

This behavior can also be understood from eq. (\ref{7.2}). From there it is obvious that accelerated expansion in the three-space requires $\cV(\phi(\tau), q(\tau))$ to dominate over $T(\phi(\tau), q(\tau))$ and $\ad(\tau)^2$ for some $\tau$. For our potential these requirements are most easily met if the transition states are non-zero and the vector multiplet moduli take values far away from the flop line. But there the transition states oscillate rapidly, and each time they pass through zero the potential vanishes, killing inflation. From eq. (\ref{7.2}) we also observe that $\ad$ non-zero generically decreases $\ddot{a}$. Thus there are two mechanisms working against sustained accelerated expansion. 

These effects are also displayed by our numerical solutions. Looking at 
Fig. \ref{acht} shows that solution ``${\rm b}^\prime$'' initially has
a large value of $\dot{\alpha}(\tau)$, which for early times suppresses
acceleration in $a(\tau)$ and induces acceleration in $b(\tau)$. For later
times $\dot{\alpha}(\tau)$ decreases and the value of $\ddot{a}(\tau)$ is controlled
by $T$ and $\cV$. During this phase we observe decelerated expansion when $T$ 
dominates, while accelerated expansion corresponds to $\cV$ 
dominating. The fluctuations in $\ddot{a}(\tau)$ reflect the 
fluctuations of the transition states. The potential
vanishes if the transition states are zero, so that at the corresponding 
points we automatically
obtain decelerated expansion. On the other hand, if the numerical values of the transition states are large, the value of $\cV$ also becomes large. These points are the most likely ones for $\cV$ dominating over $T$ and give rise to the observed short phases of accelerated expansion.
For the solution ``i'' the effect of accelerated expansion is enhanced by choosing a
small initial value of $\dot{\alpha}$ and $|U-W|$ being large, leading to large values of the potential
and hence of $\ddot{a}$. The fluctuations in $\ddot{a}(\tau)$ are induced
by the oscillations of the transition states, which cause
oscillations in the value of the potential. 

Note that the mechanism which leads to accelerated expansion
in the above examples is rather generic. Acceleration is maximal at a collective turning point, where $T$ momentarily vanishes and the moduli fields turn from running ``uphill'' the potential to running ``downhill'' the potential.\footnote{It was pointed out in \cite{emp}, that essentially the same mechanism is responsible for transient accelerating phases in cosmologies of hyperbolic and flux (S-brane) compactifications of string and M-theory.} The farer the system is away from this point, the smaller is the resulting acceleration. As we have seen, the  phases of acceleration in our model
are not strong enough to induce an inflationary
growth of the scale factors $a,b$. This is consistent with our earlier
result that slow-roll inflation cannot be realized.  
\end{subsection}
\end{section}
\begin{section}{Discussion and Outlook} 
In this paper we studied cosmological solutions of M-theory
compactified on a Calabi-Yau threefold which dynamically
passes through a particular topological phase transition,
a flop. The solutions were analyzed in both the Out-picture,
where the extra light states arising in the transition 
are integrated out, and  in the In-picture, where they are
dynamical. In the latter case one has a positive
semi-definite scalar potential, which is completely
determined by microscopic M-theory physics.
This potential drastically modifies 
the behavior of the cosmological solutions, with 
consequences for the two important problems of moduli stabilization 
and inflation. 

Concerning the moduli stabilization we have seen that the usual 
picture of run-away behavior can be highly misleading. As soon
as we allow all light states to be excited the moduli are
dynamically confined to the transition region.
Thus the ``almost singular'' manifolds close to a topological 
phase transition are dynamically preferred. 
This is somewhat surprising, because the potential has still   
many unlifted flat directions, so that there is no energy barrier 
which prevents the 
system from running away. 
Therefore this effect cannot be predicted by just analyzing the critical points of the superpotential.

The behavior of our cosmological solutions can be qualitatively understood from  a thermodynamic analogy. Around the flop line 
additional degrees of freedom can be excited. Generically,
the available energy is then distributed equally
among all the light modes (``thermalization'').
Once this has happened it becomes 
very unlikely (though it remains possible in principle) 
that the system ``finds''
the flat directions and ``escapes'' from the flop region (``entropy beats
energy''). This picture is consistent with all the
numerical solutions we have
looked at: irrespective of the 
initial conditions the system finally settles 
down in a state where all the fields either approach constant finite
values or oscillate around the transition region with comparable and small amplitudes. For long simulation time one
sees ``fluctuations from equilibrium'', i.e., some mode picks up
a bigger share of the energy for a while, but the system 
eventually thermalizes 
again. This fits nicely with the non-hyperbolic character of the 
fixed point manifold, which implies that linearized solutions
describe oscillations.

In a realistic scenario of moduli stabilization one would of course
prefer that the system is damped, so that the moduli are attracted
to fixed point values. This is possible for hyperbolic fixed points,
which are the generic case in dynamical systems. Obviously, the non-generic
feature of our system is the existence of a 
degenerate family of supersymmetric vacua. It would be interesting to investigate if the lifting of the flat directions of the LEEA  makes the fixed point hyperbolic. This would open the possibility
of attractor behavior, but also carries the risk of reintroducing
run-away  behavior.  

Let us now discuss what can be learned about inflation. Our In-picture solutions generically exhibit several periods of transient acceleration. Both the analysis of the slow-roll conditions and the study of numerical solutions shows that the amount of acceleration is much too small to account for an inflationary expansion of the early universe. However, our mechanism might still be relevant for the moderate acceleration suggested by current observations. 

Again, the behavior of the solutions can be understood qualitatively in terms of properties of the scalar potential. The point is that the potential is only flat along the unlifted directions parametrized by the moduli. Thus it is either flat, but vanishing, or non-vanishing, but steep. This also explains why hybrid inflation \cite{Hybr1,Hybr2} is not realized in our model, despite the ingredients are present, namely several scalar fields and both flat and steep directions in the scalar potential. To get a considerable amount of inflation, one needs to 
lift the flat directions gently without making them to steep.

Flux and hyperbolic compactifications are two ways for obtaining positive semi-definite potentials from string and M-theory compactifications. The inclusion of transition states in singular Calabi-Yau compactifications provides an alternative. As was pointed out in \cite{emp,Tow} it is a common feature of the potentials in the two former cases that cosmological solutions exhibit epochs of accelerated expansion, which generically are not pronounced enough to describe primordial inflation. We observe this feature is shared by the potentials induced in the presence of transition states.

Our construction also avoids the no-go theorem \cite{Gib,MalNun} which excludes de Sitter vacua in ten- or eleven-dimensional supergravity compactified on time-independent, smooth and compact internal spaces. In both the Out- and the In-picture we compactified on a time-dependent manifold. Moreover, in the In-picture we have gone beyond eleven-dimensional supergravity by including states of wrapped M2-branes in order to have a sensible theory when the internal space becomes singular. But as we have seen, the mere fact that the no-go theorem is circumvented does not automatically lead to the existence of de Sitter solutions or inflation. Nevertheless it opens up the possibility that combining our approach with additional effects, which do not necessarily violate the no-go theorem by themselves, might lead to de Sitter solutions.

In summary we see that the dynamics of the transition states is
interesting and relevant, but can only be part of the solution
of the problems of moduli stabilization and inflation. 
The first step to extend our work is
to consider more general gaugings of our five-dimensional model. 
Here the detailed study of Kasner solutions performed in this
paper will be helpful, because they show which kinds of contributions
to the equations of motion are needed in order to enhance 
the periods of acceleration and to stabilize the moduli. 
%
%For example, one option is to look
%for a mechanism which damps the oscillatory behaviour of the
%acceleration in solutions of the type `i' in Fig. 11,
%so that one obtains one long period of acceleration.
%We also saw that the accelaration asymptotically oscillates
%around a slightly negative mean value. It might be possible to obtain 
%inflation by gently lifting the potential to make this value
%positive. Finally, some of the solutions in the Out-picture
%show that the non-standard kinetic terms sometimes lead to
%a very slow motion of scalar fields through moduli space.
%This might be useful in cases where the scalar potential
%is naively too steep for inflation.
%
Once gaugings which lead to interesting cosmological solutions
are found, one should  clarify whether these can
be derived from string or M-theory where they correspond to adding fluxes or branes. This framework also allows to address supersymmetry breaking. 

Another direction
is to work out to which extent our results can hold independently
of the detailed form of the hypermultiplet metric. In particular,
one can use the fact that a Calabi-Yau manifold has a flop transition
in order to constrain the hypermultiplet metric in the In-picture.
Yet another direction is to  consider topological transitions 
other than flops, where one also gets non-abelian gauge symmetry enhancement
or additional flat directions (``Higgs branches'') of the potential.
Finally, one should investigate four-dimensional cosmologies, for example
in the context of type II compactifications on Calabi-Yau threefolds.
Here the main new complication is the vector multiplet sector, which is no
longer controlled by a simple cubic polynomial, but by a general
holomorphic prepotential. Ultimately, all these directions need to
be put together to find realistic four-dimensional cosmologies
in compactifications which include the effects of flux, branes, and
transition states.

We have also obtained several nice results concerning the relation between the geometry of the moduli space and space-time. In the Out-picture we found that there are three types of boundaries which we labeled type I, II, and III. Type III boundaries are internal boundaries corresponding to topological phase transitions involving finitely many transition states. For these boundaries we proved in a model-independent way that Kasner cosmological solutions are smooth. For type II boundaries where the vector multiplet metric degenerates, we observed that our solutions can reach these boundaries in a finite time. We also observed that these boundaries are not related to a space-time singularity. At type I boundaries where the moduli space metric diverges, we found a mechanism which prevents the solutions from reaching this boundary in a finite time. We conjecture that the behavior of our solutions at the type I and II boundaries is general, since we found that close to such boundaries it is completely determined by the singularity occurring in the scalar field metric. Finally we have seen that solutions in 
the Out-picture are non-singular, as long as the moduli take values
inside the K\"ahler cone.

All these observations have their 
counterparts in static BPS solutions, such as black holes, black strings,
and domain walls  \cite{GMMS,KMS,TMProc}. This shows that there is a systematic connection
between the geometry of the internal Calabi-Yau space and the geometries
of various space-time geometries of the compactified theory. 
In particular it seems that the K\"ahler cone acts as a cosmic censor. 
These aspects also deserve further study.

\end{section}
\newpage
\subsection*{Acknowledgments}
We would like to  thank
K. Behrndt
and M.N.R. Wohlfarth
for useful discussions. This work is supported by
the DFG within the `Schwerpunktprogramm Stringtheorie.' 
F.S. acknowledges a scholarship from the 
`Studienstiftung des deutschen Volkes'. L.J. was also supported by the Estonian Science Foundation Grant No 5026.
\begin{appendix}
\begin{section}{Initial conditions for numerical solutions}
\renewcommand{\textfraction}{0}
\renewcommand{\floatpagefraction}{1}
\begin{table}[h!]
\begin{tabular*}{\textwidth}{@{\extracolsep{\fill}} cccccccccc} \hline \hline
Traj. & $U$ & $\dot{U}$ & $W$ & $\dot{W}$ & $V$ & $\dot{V}$ & $\alpha$ & $\dot{\alpha}$ & $\beta$ \\ \hline
a & $0.6$ & $ -0.1$ & $0.2$ & $0.1$ & $1$ & $0.1$  & $0$ & $0.1$ & $0$ \\ 
b & $0.8$ & $ -0.1$ & $0.5$ & $0.1$ & $1$ & $0.1$  & $0$ & $0.1$ & $0$ \\ 
c & $0.85$ & $ -0.1$ & $0.7$ & $0.1$ & $1$ & $0$  & $0$ & $-0.2$ & $0$ \\ 
d  & $0.1$ & $ -0.1$ & $0.8$ & $0.1$ & $1$ & $0.2$  & $0$ & $0.2$ & $0$ \\ 
e  & $0.1$ & $ -0.1$ & $1.45$ & $0.1$ & $1$ & $0.2$  & $0$ & $0.1$ & $0$ \\
f  & $0.7$ & $ -0.1$ & $1.5$ & $0.1$ & $1$ & $-0.2$  & $0$ & $-0.05$ & $0$ \\
L  & $0.8$ & $ -0.1$ & $1$ & $0$ & $1$ & $-0.2$  & $0$ & $0.1$ & $0$ \\
\hline \hline
\end{tabular*}
\renewcommand{\baselinestretch}{1}
\parbox[c]{\textwidth}{\caption{\label{t.2}{\footnotesize
Initial conditions for the numerical solutions of the Out-picture equations of motion discussed in subsection 3.2.}}}
\end{table}  
\begin{table}[h!]
\begin{tabular*}{\textwidth}{@{\extracolsep{\fill}} cccccc} \hline \hline
boundary & $U$ & $\dot{U}$ & $W$ & $\dot{W}$ & $\dot{\alpha}$ \\ \hline
$b_1$ & $0.07$ & $ -0.1$ & $0.55$ & $-0.05$  & $0.2$ \\ 
$b_1$ & $0.07$ & $ -0.13$ & $0.55$ & $0.15$  & $0.2$ \\ 
$b_1$ & $0.07$ & $ -0.1$ & $0.2$ & $0.1$  & $0.2$ \\ \hline 
$b_2$ & $0.76$ & $ 0.05$ & $0.01$ & $-0.1$  & $0.1$ \\ 
$b_2$ & $0.765$ & $ 0.05$ & $0.01$ & $-0.1$  & $0.1$ \\
$b_2$ & $0.77$ & $ 0.05$ & $0.01$ & $-0.1$  & $0.1$ \\
$b_2$ & $0.775$ & $ 0.05$ & $0.01$ & $-0.1$  & $0.1$ \\ \hline
$b_3$ & $0.87$ & $ 0.05$ & $0.505$ & $0.05$  & $0.1$ \\
$b_3$ & $0.87$ & $ 0.05$ & $0.515$ & $0.05$  & $0.1$ \\
$b_3$ & $0.87$ & $ 0.05$ & $0.525$ & $0.05$  & $0.1$ \\
$b_3$ & $0.87$ & $ 0.05$ & $0.535$ & $0.05$  & $0.1$ \\
\hline \hline
\end{tabular*}
\renewcommand{\baselinestretch}{1}
\parbox[c]{\textwidth}{\caption{\label{t.4}{\footnotesize
Initial conditions for the numerical solutions close to the boundaries $b_1$, $b_2$ and $b_3$ shown in Fig. \ref{vier}. We further take $V = 1$, $\dot{V} = 0$, $\alpha = 0$ and $\beta = 0$. The column ``boundary'' indicates to which diagram the solution belongs.}}}
\end{table}  
\begin{table}[h!]
\begin{tabular*}{\textwidth}{@{\extracolsep{\fill}} ccccccccccc} \hline \hline
Traj. & $U$ & $\dot{U}$ & $W$ & $\dot{W}$ & $Q_u$ & $\dot{Q}_u$ & $Q_v$ & $q_u$ & $q_v$ & $\dot{\alpha}$   \\ \hline
${\rm b}^{\prime}$ & $0.8$ & $ -0.1$ & $0.5$ & $0.1$ & $0$ & $-0.05$  & $0$ & $0.1$ & $0$ & $0.1$  \\ 
i & $0.3$ & $ 0$ & $3$ & $0$ & $0$ & $0$  & $0$ & $0.07$ & $0.02$ &  $0.1$  \\ 
g & $0.6$ & $0$ & $0.6$ & $0$ & $-0.2$ & $0$ & $0.5$ & $-0.05$ & $0.3$ & $0.2$ \\ %
\hline \hline
\end{tabular*}
\renewcommand{\baselinestretch}{1}
\parbox[c]{\textwidth}{\caption{\label{t.3}{\footnotesize
Initial conditions for the numerical solutions of the In-picture equations of motion discussed in subsection 4.3. Additionally we take $\dot{q}_u = \dot{q}_v = \dot{Q}_v = 0$, $\alpha = \beta = 0$ for all three solutions.}}}
\end{table}  
\end{section}
\clearpage
\end{appendix}

\end{document}